%% file: ms.tex
\documentclass[3p]{elsarticle}

\usepackage{pifont, geometry, fleqn, graphicx}
\usepackage[colorlinks,bookmarksopen,bookmarksnumbered,citecolor=red,urlcolor=red]{hyperref}
\RequirePackage[labelfont={rm,md,up},format=hang]{subfig}

\usepackage{epstopdf}
\usepackage{colortbl}
\usepackage{doi}
\usepackage{xcolor}
\usepackage{hhline}
\usepackage[USenglish]{babel}
\usepackage{amsmath}
\usepackage{amsfonts}
\usepackage{amssymb}
\usepackage{mathrsfs}
\usepackage{multirow}
\usepackage{lipsum}  
\usepackage[]{natbib}

\newcommand\BibTeX{{\rmfamily B\kern-.05em \textsc{i\kern-.025em b}\kern-.08em
T\kern-.1667em\lower.7ex\hbox{E}\kern-.125emX}}

\journal{Computer Methods in Applied Mechanics and Engineering}

\begin{document}

\begin{frontmatter}

\title{Optimization and projection of coated structures with orthotropic infill material}

\author[dep1]{Jeroen P. Groen\corref{cor1}\fnref{ft1}}
\author[dep2]{Jun Wu}
\author[dep1]{Ole Sigmund}

\fntext[ft1]{E-mail: jergro@mek.dtu.dk}
\cortext[cor1]{Correspondence to: J. P. Groen, Department of Mechanical Engineering, Solid Mechanics, Technical University of Denmark, Nils Koppels All\'{e}, Building 404, 2800 Kgs. Lyngby, Denmark}

\address[dep1]{Department of Mechanical Engineering, Solid Mechanics, Technical University of Denmark}
\address[dep2]{Department of Design Engineering, Delft University of Technology}

\input{Abstract.tex}


\end{frontmatter}

\input{Introduction.tex}
\input{Hcoat.tex}
\input{NumExp_TO.tex}

\input{Proj.tex}

\input{NumExp_proj.tex}
\input{Conclusion.tex}

\bibliography{biblio}   
\bibliographystyle{elsarticle-harv}
\biboptions{authoryear}

\end{document}

%% file: Abstract.tex
\begin{abstract}
The purpose of this work is two-fold. First, we introduce an efficient homogenization-based approach to perform topology optimization of coated structures with orthotropic infill material. By making use of the relaxed design space, we can obtain designs with complex microstructures on a relatively coarse mesh. 
In the second part of this work, a method is presented to project the homogenization-based designs on a fine but realizable scale. A novel method to adaptively refine the lattice structure is presented to allow for a regular spacing of the infill. Numerical experiments show excellent behavior of the projected designs, which perform almost identical to the homogenization-based designs. Furthermore, a reduction in computational cost of at least an order of magnitude is achieved, compared to a related approach in which the infill is optimized using a density-based approach.

\end{abstract}

%% file: Introduction.tex
\section{Introduction}
\label{Sec:Intro}

Topology optimization is recognized as an important design method, with numerous applications in industry and academia. Furthermore, the flexibility offered by additive manufacturing (AM) methods makes topology optimization the ideal design method for this rapidly growing field. In recent years a large number of works have considered incorporation of constraints posed by the AM process directly into the optimization framework. Examples are geometric constraints to restrict the overhang angle~\citep{Bib:LangelaarOverhang,Bib:QianOverhang,Bib:GaynorGuest2016,Bib:AllaireOverhang}, and methods to restrict the length-scale of the design as well as to make them robust against manufacturing variations, see~\citep{Bib:FilterMethodsLazarov} for a detailed review of such methods. 
For a global overview on the state of the art and future trends in topology optimization for additive manufacturing, the reader is referred to~\citep{Bib:TOReview_AM}. 

Most additive manufacturing processes, such as Fused Deposition Modeling (FDM) work with a solid shell to represent the surface, reinforced by porous infill. The reason these so-called coated structures are considered instead of solid structures are high strength-to-weight ratio, good energy absorption characteristics, and high thermal and acoustic insulation properties~\citep{Bib:GibsonAshby}. Furthermore, porous structures can alleviate thermal hot-spots that are prone to cause large stresses in printed designs~\citep{Bib:AllaireHeatAM,Bib:RanjanHeatAM}. Compared to their solid counterparts, coated structures with porous infill can obtain significantly increased buckling stability~\citep{Bib:ClausenBuckling}, as well as a better performance w.r.t. unpredicted loading conditions and material deficiency~\citep{Bib:WuUniformInfill}, at the cost of slightly increased compliance. We do remind the reader that truss or lattice-like infill is inferior to close-walled cell structures when only considering stiffness, as e.g. discussed in~\citep{Bib:SigmundMichell}. However, here we consider the 2D-case where the differentiation between open- and closed-walled structures does not directly apply.

Traditionally, the porous infill in coated structures consisted of repetitive infill patterns (e.g., triangles and hexagons). However,~\citet{Bib:WuUniformInfill} proposed a density-based method to design bone-inspired microstructures as porous infill. In another approach~\citet{Bib:ClausenCoating2015,Bib:ClausenCoating3D} presented a method to optimize a coated structure, using a solid shell and an isotropic base material that can be interpreted as a uniform porous infill. Here, density-based optimization is applied and using successive filtering operations a clear distinction between coating and infill material could be made. In a natural subsequent step these methods have been combined to concurrently design both the coating and the infill~\citep{Bib:WuClausenSigmund2017}.  Even more recently, this approach has been extended to dynamic loading problems and multi-material coatings~\citep{Bib:AageCoating2018}. 
Similar to density-based methods, level-set methods have been used to design for material interface properties~\citep{Bib:Vermaak2014, Bib:WangKangLevelCoat}. This approach was recently extended to first design a coating layer, and subsequently an infill~\citep{Bib:Dapogny2018}. Besides, the computational mechanics community, the computer graphics community has recently proposed many works on the optimization of porous structures, e.g.~\citep{Bib:Lu2014ToG,Bib:Martinez2016ToG,Bib:Wu2016CAD}.

\begin{figure}[h!]
\centering
\subfloat[Homogenization-based topology optimization, resolution: $300\times100$ elements.]{\includegraphics[width=0.40\textwidth]{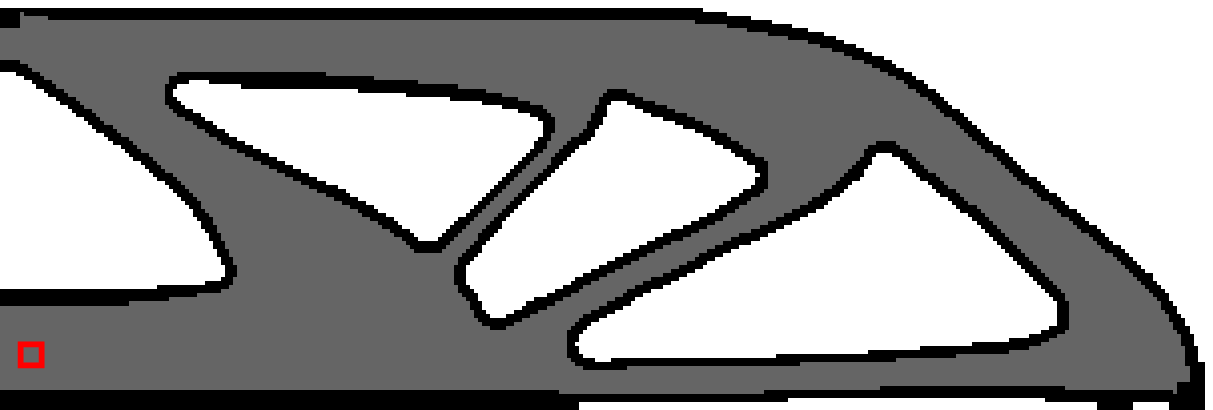}} \quad
\subfloat[Microstructure]{\includegraphics[width=0.15\textwidth]{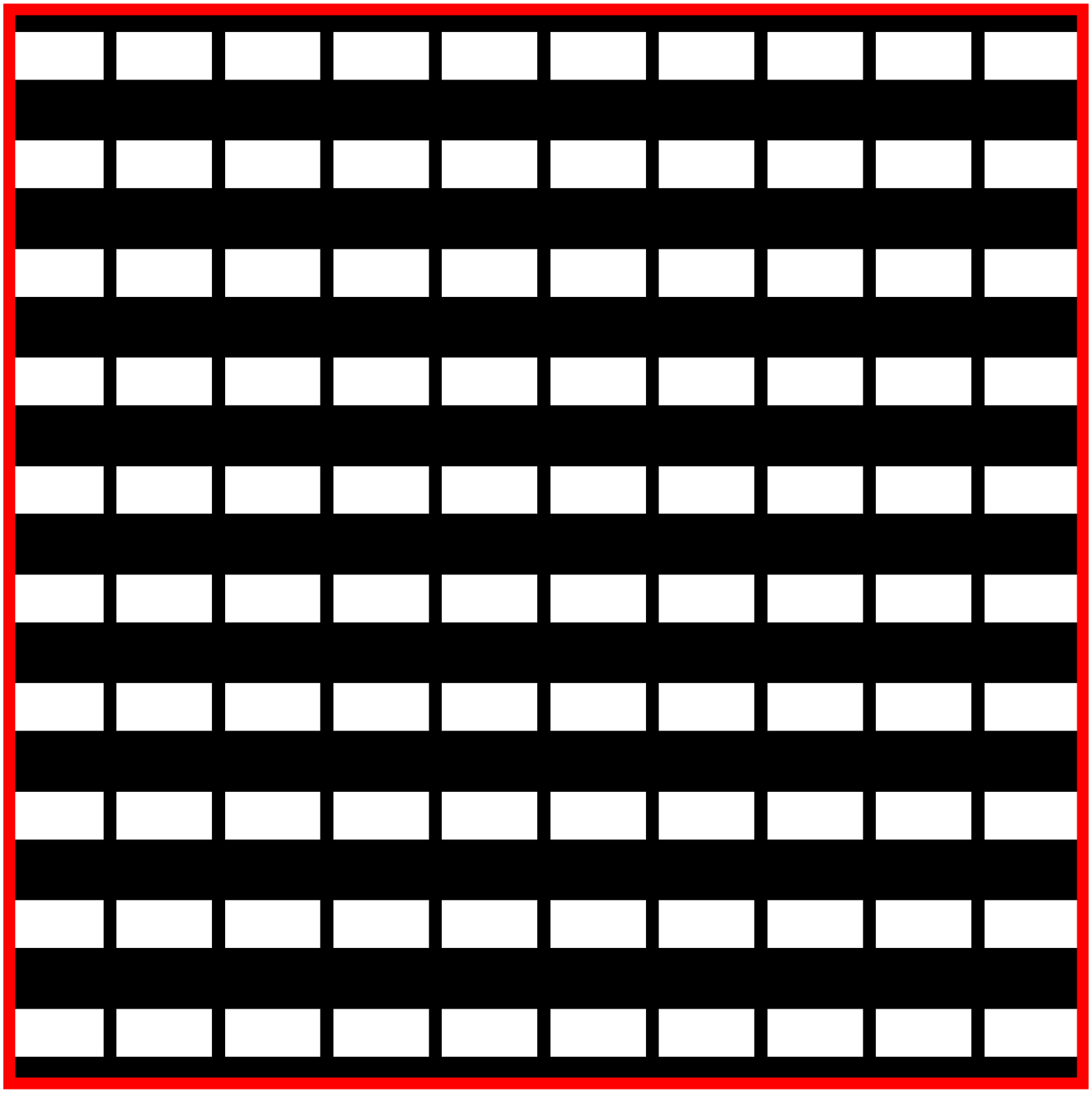}} \quad
\subfloat[Projection of composite design on a fine mesh, resolution: $3000\times1000$ elements.]{\includegraphics[width=0.40\textwidth]{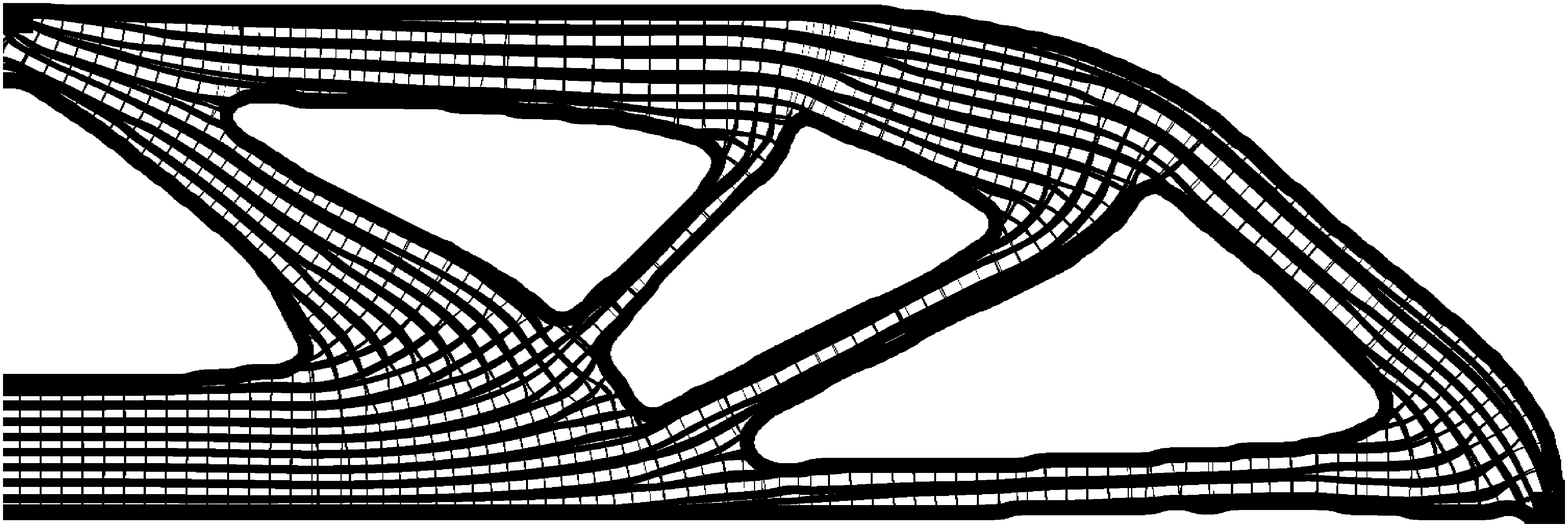}} 
\caption{Overview of the proposed methodology to obtain high-resolution coated designs, with composite orthotropic infill.} 
\label{Fig:Intro.1}
\end{figure}

In this work we extend the method of~\citet{Bib:ClausenCoating2015} to design coated structures with a macroscopically varying orthotropic infill material. By exploiting the relaxed design space, we can obtain a detailed description of the infill behavior on a relatively coarse mesh, e.g. $300\times100$ elements as is shown in Figure~\ref{Fig:Intro.1}(a). To describe the periodic infill composite we make use of the well-known square unit-cell with a rectangular hole \citep{Bib:BendsoeKikuchi}, shown in Figure~\ref{Fig:Intro.1}(b). Recent studies \citep{Bib:GroenSigmund2017,Bib:GDondersAllairePantz2018} inspired by \citet{Bib:Pantz1,Bib:Pantz2} have shown that microstructures with a rectangular hole can be projected on a much finer mesh to obtain manufacturable design, abandoning the separation of scales. Our work is continuing on the projection procedure presented in \citep{Bib:GroenSigmund2017,Bib:GroenFrame}, where an explicit constraint is used to align the projected microstructure with the directions of lamination. To keep the unit-cell spacing as regular as possible we introduce a novel scheme that adaptively refines the periodicity as can be seen in Figure~\ref{Fig:Intro.1}(c), where a resolution of $3000\times1000$ elements is used.

The paper is organized in two parts. The first part encompasses homogenization-based topology optimization for coated structures. The theory is introduced in Section~\ref{Sec:Hcoat}, while numerical experiments regarding performance of the optimized results and comparisons to the use of isotropic infill material are discussed in Section~\ref{Sec:ExpHcoat}. The second part of this article deals with the projection method to obtain high-resolution manufacturable structures on a fine mesh. The theory and implementation will be discussed in Section~\ref{Sec:Proj}, while numerical experiments on the performance and numerical efficiency of the projected structures are discussed in Section~\ref{Sec:NumExp.proj}. Finally, Section~\ref{Sec:Conclusion} contains the most important conclusions of this study.

%% file: Hcoat.tex
\section{Homogenization-based topology optimization for coated structures}
\label{Sec:Hcoat}
A coated structure consists of two material phases, a coating and an infill as can be seen in Figure~\ref{Fig:Intro.1}(a). 
\citet{Bib:ClausenCoating2015} introduced an elegant method to obtain a coated structure using only a single field as design variable. Successive filter and projection operations allowed for a clear distinction between coating, isotropic infill and void. The use of composite infill material, introduced in this work, will add extra design variables; nevertheless the method to distinguish between infill and coating remains almost the same. 

\subsection{Successive filter operations}
The procedure to distinguish between coating and infill makes use of two well-established filter methods in topology optimization. The first is a smoothing operation using the density filter. And the second is a projection step to force the smoothed values on the interval $[0,1]$ towards either $0$ or $1$.

\subsubsection{Smoothing}
As basis for subsequent projection we use the Helmholtz-type PDE-based density filter~\citep{Bib:LazarovDensityPDE},
\begin{equation} \label{Eq:HcoatFilter.1}
-\Big(\frac{R}{2\sqrt{3}}\Big)^{2} \nabla^{2} \tilde{\phi} + \tilde{\phi} = \phi.
\end{equation}
Here scalar $R$ corresponds to the radius of the convolution kernel used in the filter operation. $\phi$ is the unfiltered field, while $\tilde{\phi}$ is the filtered field. Homogeneous Neumann boundary conditions are applied at the boundary of the filter domain. A discussion on undesired boundary effects of standard filter methods will be given later in this section.

\subsubsection{Projection}
Projection methods have been successfully applied in topology optimization to obtain black-and-white designs \citep{Bib:GuestHeaviside,Bib:SigmundFilters}. Here we use the formulation for the smoothed Heaviside projection proposed by~\citet{Bib:WangRobustTopopt},
\begin{equation} \label{Eq:HcoatFilter.2}
\bar{\tilde{\phi}} = \frac{\text{tanh}(\beta\eta) + \text{tanh}(\beta(\tilde{\phi}-\eta))}{\text{tanh}(\beta\eta) + \text{tanh}(\beta(1-\eta))}.
\end{equation}
Here $\bar{\tilde{\phi}}$ is the projected field. $\beta$ determines the steepness of the projection, $i.e.$ when $\beta \rightarrow \infty$ a sharp step is modeled. In general, a continuation approach is used for $\beta$, hence a low value is used during the first iterations, after which $\beta$ is gradually increased. Furthermore, $\eta\in [0,1]$ is the threshold parameter, $\eta > 0.5$ corresponds to an erosion operation, while $\eta < 0.5$ corresponds to a dilation operation. 

\subsubsection{Combining the filters to obtain a coated structure}
\begin{figure}[ht!]
\centering
\includegraphics[width=1.0\textwidth]{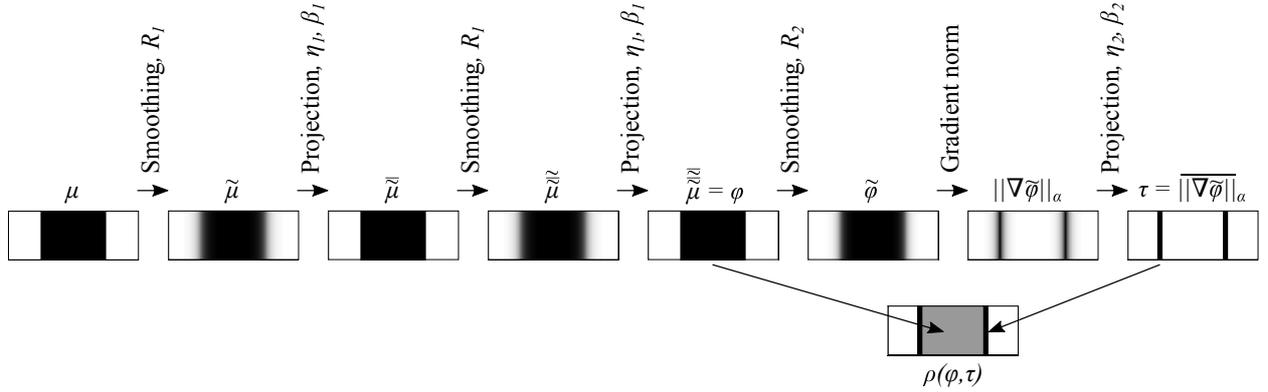}
        \caption{Subsequent filtering steps, allowing to separate the base structure $\varphi$ and the coating $\tau$.}
		\label{Fig:HcoatFilter.1}
\end{figure}

Analogous to \citet{Bib:ClausenCoating2015} we use successive filter operations to obtain a parameter describing the base structure $\varphi$ and a parameter describing the coating $\tau$. Additional variables, which will be introduced in Section~\ref{Sec:Hcoat.mat}, are used to describe the shape of the infill. However, these variables do not affect the distinction between coating $\Omega_{c}$, infill $\Omega_{l}$ or void $\Omega_{v}$ regions.
\begin{equation}\label{Eq:HcoatFilter.3}
\textbf{x} \in
\begin{cases}
\Omega_{v}	& \quad \text{if } \quad \varphi(\textbf{x}) = 0 	\quad \text{and if } \tau(\textbf{x}) = 0,\\
\Omega_{l}	& \quad \text{if } \quad \varphi(\textbf{x}) = 1 	\quad \text{and if } \tau(\textbf{x}) = 0,\\
\Omega_{c}	& \quad \text{if } \quad \tau(\textbf{x}) = 1.\\
\end{cases} 
\end{equation}
The successive smoothing projection and gradient operations used in this work can be seen in Figure~\ref{Fig:HcoatFilter.1}.

The design field is filtered and projected twice using filter radius $R_{1}$, and projection parameters $\beta_{1}$ and $\eta_{1}$, which indirectly control the length-scale of $\varphi$. The base structure which is now either 0 or 1, is smoothed again using $R_{2} < R_{1}$, such that the resulting field $\tilde{\varphi}$ has smooth boundaries. The coating layer can then be defined by taking the Euclidean norm of the spatial gradients of $\tilde{\varphi}$. $\left\vert\left\vert\nabla\tilde{\varphi} \right\vert\right\vert$ is then normalized such that the largest possible gradient norm corresponds to $1$. For this we use a normalization factor $\alpha$, which is related to $R_{2}$ using~\citep{Bib:ClausenCoating2015}, 
\begin{equation}\label{Eq:HcoatFilter.4}
\alpha = \frac{R_{2}}{\sqrt{3}}.
\end{equation}
The normalized gradient $\left\vert\left\vert\nabla\tilde{\varphi} \right\vert\right\vert_{\alpha}$ is subsequently projected using $\beta_{2}$ and $\eta_{2}$ to define a clear coating $\tau = \overline{\left\vert\left\vert\nabla\tilde{\varphi} \right\vert\right\vert}_{\alpha}$. In~\cite{Bib:ClausenCoating2015} an analytical relation is shown between $R_{2}$ and the maximum coating thickness. This is used to select $R_{2}$ for a user-specified coating thickness $t_{ref}$ as,
\begin{equation}\label{Eq:HcoatFilter.5}
R_{2} =  \frac{\sqrt{3}}{\text{ln}(2)} t_{ref} \approx 2.5 t_{ref}.
\end{equation}

The motivation to filter and project the design field twice is that numerical experiments using a single smoothing and projection step showed that the possibility that $\varphi$ and $\tau$ do not converge exactly to $0$ or $1$.  A similar observation can be made from Figure~13 in \cite{Bib:ClausenCoating2015}, where it is difficult to distinguish between infill and void. To circumvent this undesired effect we use the double filter approach as proposed by \cite{Bib:ChristiansenDouble}. More details on the effect of this double filtering approach versus the approach from~\cite{Bib:ClausenCoating2015} will be given in Section~\ref{Sec:ExpHcoat}.

\subsubsection{Note on the filter boundary conditions}
\begin{figure}[ht!]
\centering
\includegraphics[width=0.8\textwidth]{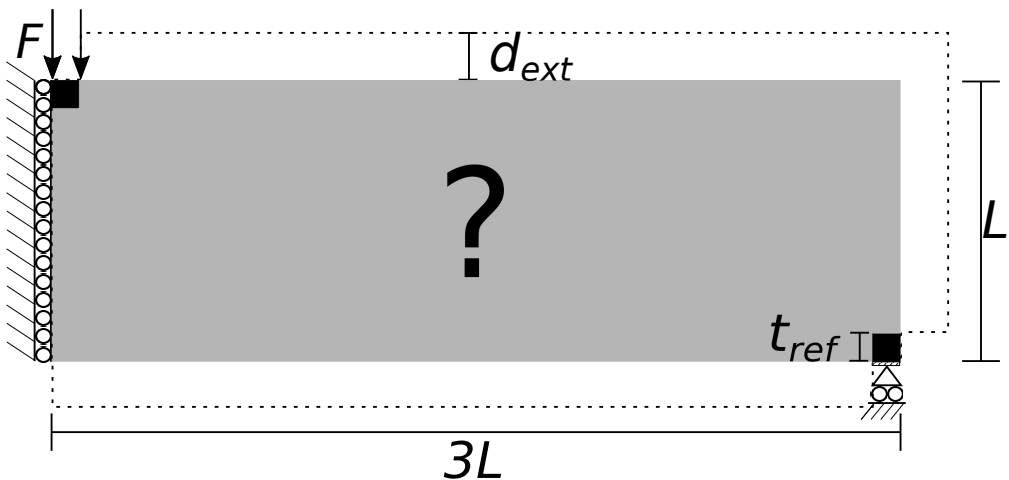}
        \caption{Design domain $\Omega$ and boundary conditions of the MBB-beam example. $\Omega$ is indicated using gray material, while the extended filter domain is bounded by the dotted line.}
		\label{Fig:Hcoat.Filter.1}
\end{figure}

It is well-known that the use of homogeneous Neumann boundary conditions used in the smoothing operation of Equation~\ref{Eq:HcoatFilter.1} causes artifact on the optimized design near the domain boundary. To circumvent this undesired effect, we make use of the domain extension approach proposed by~\cite{Bib:ClausenAndreassenFilterBC}. In this approach the physical domain is padded using void elements $(\mu = 0)$, except at boundaries at which the displacement field is constrained. 
The extension distance $d_{ext}>R_{1}$  is chosen large enough, such that the homogeneous Neumann filter boundary conditions, do not affect the final design.
All filtering operations, finite element analysis and objective and constraint calculations are performed on this extended domain. An overview of the design domain $\Omega$ and boundary conditions for the MBB-beam example including the extended domain can be seen in Figure~\ref{Fig:Hcoat.Filter.1}.

Solid blocks of material are used at the two boundary conditions. This guarantees a minimum feature size of $2t_{ref}$, and reinforces the structure thus preventing load concentrations. Furthermore, the load is applied in a distributed fashion over the complete top of the solid block, while the displacement constraint is applied in an average sense over the bottom of the lower block. This is done to prevent load concentrations when the optimized structure is mapped on a finer mesh, as will be discussed later on in this paper.

The domain extension approach works well for robust topology optimization problems or three-field SIMP optimization problems (density filtering, followed by projection), using $\eta = 0.5$. However, in the context of coated structures, it is possible that a part of the coating $(t_{ref}/2)$ can exceed the design domain. To make sure that the optimized structure is within the bounds of $\Omega$, the elasticity tensor in the padded domain is multiplied by penalization parameter $q <1$, while $q=1 \in \Omega$. Numerical experiments have shown that the use of $q = 0.2$ effectively restricts optimized structures to $\Omega$.

\subsection{Interpolation of elastic properties and density}
\label{Sec:Hcoat.mat}
\begin{figure}[ht!]
\centering
\includegraphics[width=0.6\textwidth]{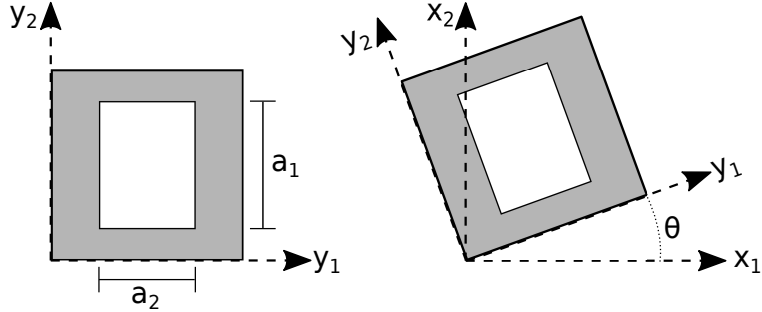}
        \caption{Layout of the unit-cell with a rectangular hole, in local (y$_{1}$,y$_{2}$), and global (x$_{1}$,x$_{2}$) coordinate system.}
		\label{Fig:Hcoat.mat.1}
\end{figure}

For the numerical examples we restrict ourselves to coating and infill made from the same material, with Young's modulus $E^{0} = 1$, Poisson's ratio $\nu^{0} = 0.3$, and mass density $m^{0} = 1$; however, extension to different coating and infill material is trivial. As infill we chose to use the square unit-cell with rectangular hole introduced by \citet{Bib:BendsoeKikuchi}, shown in Figure~\ref{Fig:Hcoat.mat.1}. 

The constitutive properties which are close in performance to the optimal rank-2 laminate are obtained using numerical homogenization. A database of the constitutive properties in the local frame for a large number of different combinations for the height $(a_{1})$ and width $(a_{2})$ of the hole has been created. Afterwards, we can interpolate the effective properties $\textbf{E}^{H}$ and sensitivities for any combination of $a_{1}$ and $a_{2}$. The corresponding elasticity tensor of the infill in the global frame of reference $\textbf{E}^{I}$ can be calculated as, 
\begin{equation}\label{Eq:Hcoat.mat.1}
\textbf{E}^{I}(\theta,a_{1},a_{2}) = \textbf{R}(\theta)\textbf{E}^{H}(a_1,a_2)\textbf{R}(\theta)^{T},
\end{equation}
with $\textbf{R}$ being the well-known frame rotation matrix. The mass fraction of the microstructure $m^{I}$ is given as,
\begin{equation}\label{Eq:Hcoat.mat.2}
m^{I} = 1-a_{1}a_{2}.
\end{equation}
With the properties of the coating and infill known, we can define the interpolation of the density $\rho$ and elasticity tensor $\textbf{E}$ throughout the design domain. To this we use material indicator $\varphi$ and coating indicator $\tau$, such that,
\begin{equation}\label{Eq:Hcoat.mat.3}
\rho(\varphi,\tau,a_{1},a_{2}) = m^{I}(a_{1},a_{2})\varphi + (1 -m^{I}(a_{1},a_{2}) \varphi )  \tau.
\end{equation}
Similarly, the local elasticity tensor $\textbf{E}$ can be defined.
\begin{equation}\label{Eq:Hcoat.mat.4}
\textbf{E}(\varphi,\tau,\theta,a_{1},a_{2}) = 
10^{-9} \textbf{E}^{0}+ q\Big(\textbf{E}^{I}(\theta,a_{1},a_{2}) - 10^{-9} \textbf{E}^{0}\Big) \varphi^{p_{1}}   + q(\textbf{E}^{0} -\textbf{E}^{I}(\theta,a_{1},a_{2})  \varphi^{p_{1}}) \tau^{p_{2}},
\end{equation}
where, $\textbf{E}^{0}$ is the elasticity tensor for the isotropic coating material, and $p_{1}$ and $p_{2}$ are penalty parameters to penalize intermediate values of $\varphi$ and $\tau$. Please note that the model using isotropic infill from~\citet{Bib:ClausenCoating2015} can easily be recovered by substituting $\textbf{E}^{I}$, with $\lambda_{E} \textbf{E}^{0}$, and $m^{I}$ with $\lambda_{m} m^{0}$. 

\subsection{Problem formulation}
We focus on the minimization of a compliance functional $\mathcal{J}$ for plane stress, single load-case problems under the assumption of linear elasticity. We discretize the design domain in bi-linear finite elements; furthermore, the material properties are assumed to be element-wise constant. The microstructures for the infill are described by design vectors $\textbf{a}_{1}$, $\textbf{a}_{2}$, $\boldsymbol{\theta}$, while the coating and base structure are defined by $\boldsymbol{\mu}$. 

The topology optimization problem is solved in nested form.
 For each design iteration the equilibrium equations are satisfied by FE-analysis. As is shown by~\citet{Bib:Pedersen1989,Bib:Pedersen1990}, the optimal orientation of an orthotropic composite coincides with the principal stress directions, hence $\boldsymbol{\theta}$ is aligned accordingly for each minimization step. Subsequently, design vectors $\boldsymbol{\mu}$, $\textbf{a}_{1}$ and $\textbf{a}_{2}$ are updated at each minimization step based on their gradients. The discretized optimization problem can thus be written as,
\begin{equation}\label{Eq:TOForm.1}
\begin{aligned}
 & & \displaystyle \min_{\boldsymbol{\mu},\textbf{a}_{1},\textbf{a}_{2},\boldsymbol{\theta}} & :   \mathcal{J}(\boldsymbol{\mu},\textbf{a}_{1},\textbf{a}_{2},\boldsymbol{\theta},\textbf{U}),     \\
 & & \textrm{s.t.}             & :  \textbf{K}(\boldsymbol{\mu},\textbf{a}_{1},\textbf{a}_{2},\boldsymbol{\theta}) \textbf{U} = \textbf{F}, 								   \\ 
 & & 						     & :   \textbf{v}^{T} \boldsymbol{\rho}(\boldsymbol{\mu},\textbf{a}_{1},\textbf{a}_{2})  - V_{max} A\leq 0,  	\\
 & &                           & :     \textbf{a}_{l}\leq  \textbf{a}_{1},\textbf{a}_{2}  \leq  \textbf{a}_{u},  \\	
 & &                           & :     \textbf{0}  \leq  \boldsymbol{\mu}  \leq  \textbf{1},				
\end{aligned}
 \end{equation}
where  $\textbf{v}$ is the vector containing the element volumes, and $V_{max}$ is the maximum allowed volume fraction of the material in the design domain, where $A$ is the area of $\Omega$. Stiffness matrix $\textbf{K}$ is a function of $\boldsymbol{\mu}$, $\textbf{a}_{1},~\textbf{a}_{2}$, and $\boldsymbol{\theta}$, $\textbf{F}$ describes the loads acting on the domain, and $\textbf{U}$ describes the solution of the equilibrium equation. The vectors $\textbf{a}_{l}$ and $\textbf{a}_{u}$ on interval $[0,1]$ describe the lower and upper bound on the shape of the microstructure, subject to $\textbf{a}_{u} > \textbf{a}_{l}$. For the design update of $\boldsymbol{\mu}$, $\textbf{a}_{1}$ and $\textbf{a}_{2}$ the MATLAB implementation of the Method of Moving Asymptotes (MMA) introduced by~\citet{Bib:MMA} is used.  

The formulation above is the most general formulation. We allow full freedom for the microstructures, within bounds $\textbf{a}_{l}$ and $\textbf{a}_{u}$. To avoid fully void or solid infill material we use $\textbf{a}_{l} \geq 0.1$ and $\textbf{a}_{u} \leq 0.9$, where the exact values are problem dependent. In some optimization examples we do not want to exploit the full design freedom of the microstructures. Therefore, we present in total 4 different problem formulations that will be used throughout this work.

\subsubsection*{Problem 0: Fixed microstructure density, isotropic microstructure\\}
This is the original coating approach as proposed in~\cite{Bib:ClausenCoating2015}. We specify an infill mass fraction $m^{I}$, and choose a stiffness for the isotropic microstructure ${E}^{I}$ that satisfies the upper bound of the Hashin-Shtrikman bounds~\citep{Bib:HS},
\begin{equation} \label{Eq:TOForm.2}
{E}^{I} = 
\frac{m^{I}}{3-2m^{I}}.
\end{equation}

\subsubsection*{Problem 1: Fixed microstructure density, equal widths of hole\\}
This is the simplest problem using the orthotropic microstructure. The infill has a user-defined mass fraction $m^{I}$, and both widths are set equal such that, 
\begin{equation}\label{Eq:TOForm.3}
a_{i} = \sqrt{1-m^{I}},  \quad i = 1,2.
 \end{equation}

 \subsubsection*{Problem 2: Fixed microstructure density, variable widths of hole \\}
Again we have a user-defined mass fraction $m^{I}$. However, $a_{1}$ and $a_{2}$ are allowed to vary such that $m^{I}$ is always satisfied. Hence, we need one design variable to describe the shape of the rectangular hole. For this we use variable $a_{1}$, such that 
\begin{equation}\label{Eq:TOForm.4}
a_{2} = \frac{1-m^{I}}{a_{1}},
\end{equation}
Furthermore, we can set box-constraints $\textbf{a}_{l}$ and $\textbf{a}_{u}$ on $\textbf{a}_{1}$ to restrict the set of allowed microstructures. Here we always use an upper bound of $\textbf{a}_{u} = 0.9$, after which we calculate $\textbf{a}_{l}$ which will be larger than $0.1$. 

\subsubsection*{Problem 3: Variable microstructure density, variable widths of hole\\}
This is the most general optimization problem, which is the problem shown in Equation~\ref{Eq:TOForm.1}, with variables $\boldsymbol{\mu}$, $\textbf{a}_{1}$ and $\textbf{a}_{2}$ updated using the MMA. However, to avoid checkerboard patterns in the infill we do need to regularize design vectors $\textbf{a}_{1}$ and $\textbf{a}_{2}$ into $\tilde{\textbf{a}}_{1}$ and $\tilde{\textbf{a}}_{2}$ describing the physical size of the hole using a density filter with a filter radius just larger than the finite element size.

\subsection{Sensitivity analysis}
The sensitivity of the self-adjoint objective function $\mathcal{J}$ with respect to any design variable $x$ can be derived as,
\begin{equation}\label{Eq:TOSens.1}
\frac{\partial \mathcal{J}}{\partial x} = -\textbf{U}^{T} \frac{\partial \textbf{K}}{\partial x} \textbf{U} = -\textbf{U}^{T} \frac{\partial \textbf{K}}{\partial x} \textbf{U} = -\sum_{j}  \textbf{u}_{j}^{T} \int_{\Omega_{j}} \textbf{B}^{T} \frac{\partial \textbf{E}_{j}}{\partial x}  \textbf{B} \text{d}\Omega_{j} \textbf{u}_{j},
\end{equation}
where $j$ indicates the set of elements for which the elasticity tensor $\textbf{E}_{j}$ is influenced by $x$ due to filter operations. 

\subsubsection{Sensitivities w.r.t. $\mu$}
The sensitivity of $\mathcal{J}$ w.r.t the variable determining the base structure of element $e$ is,
\begin{equation}\label{Eq:TOSens.2}
\frac{\partial \mathcal{J}}{\partial \varphi_{e}} = - \textbf{u}_{e}^{T}\int_{\Omega_{e}}\textbf{B}^{T} \frac{\text{d} \textbf{E}_{e}}{\text{d} \varphi_{e}}  \textbf{B} \text{d}\Omega_{e} \textbf{u}_{e}  
  -\sum_{k}  \textbf{u}_{k}^{T} \int_{\Omega_{k}} \textbf{B}^{T} \frac{\text{d} \textbf{E}_{k}}{\text{d} \tau}  \frac{d  \tau}{d \varphi_{e}}  \textbf{B} \text{d}\Omega_{k} \textbf{u}_{k}.
\end{equation}
Here $k$ indicates the set of elements for which the elasticity tensor $\textbf{E}_{k}$ is influenced by $\varphi_{e}$ due to filter operations. Furthermore,
\begin{equation}\label{Eq:TOSens.3}
 \frac{\text{d} \textbf{E}_{e}}{\text{d} \varphi_{e}}  = 
 qp_{1} \Big( \textbf{E}^{I}(\theta,a_{1},a_{2}) - 10^{-9} \textbf{E}^{c}\Big) \varphi_{e}^{p_{1}-1}   -q p_{1} \textbf{E}^{I}(\theta,a_{1},a_{2})  \varphi_{e}^{p_{1}-1} \tau^{p_{2}},
\end{equation}
\begin{equation}\label{Eq:TOSens.4}
\frac{\text{d} \textbf{E}_{k}}{\text{d} \tau}  =
 qp_{2}(\textbf{E}^{c} -\textbf{E}^{I}(\theta,a_{1},a_{2})  \varphi^{p_{1}}) \tau^{p_{2}-1},
\end{equation}
The derivative of $ \tau$ w.r.t $\varphi_{e}$ is omitted here, but can be found in~\cite{Bib:ClausenCoating2015}. With the derivatives of the objective w.r.t $\varphi$ known, it is easy to get the derivatives w.r.t $\mu$. These are standard chain rule modifications for the two smoothing and two projection steps. These expressions are well-known and can be found in~\cite{Bib:LazarovDensityPDE} and~\cite{Bib:WangRobustTopopt}.

\subsubsection{Sensitivities w.r.t. $a_{i}$}
For an optimization problem of type 3 (full freedom), where both $a_{1}$ and $a_{2}$ are design variables, we can write the sensitivity of the elasticity tensor w.r.t  the filtered variable determining the size of the hole of element $e$ as,
\begin{equation}\label{Eq:TOSens.5}
\frac{\partial \textbf{E}_{e}}{\partial \tilde{a}_{e,1}} = 
 q  \textbf{R}_{e}    \frac{\partial \textbf{E}^{H}_{e}}{\partial \tilde{a}_{e,1}} \textbf{R}_{e}^{T}  \varphi_{e}^{p_{1}}   - q  \textbf{R}_{e}    \frac{\partial \textbf{E}^{H}_{e}}{\partial \tilde{a}_{e,1}} \textbf{R}_{e}^{T}    \varphi_{e}^{p_{1}} \tau^{p_{2}}.
\end{equation}
The derivative of the homogenized elasticity tensor w.r.t. $\tilde{a}_{e,1}$ can be interpolated as discussed above. Furthermore, the standard chain rule modification to get the derivative w.r.t the design variable $a_{1}$ is again trivial. 

For an optimization problem of type 2, where we only have one variable per element to describe the shape of the hole $a_{1}$ we can write,
\begin{equation}\label{Eq:TOSens.6}
\frac{\partial \textbf{E}_{e}}{\partial {a}_{e,1}} = 
 q \textbf{R}_{e} \Big(\frac{\partial \textbf{E}^{H}_{e}}{\partial {a}_{e,1}} -\frac{1-m^{I}}{a_{e,1}^{2}} \frac{\partial \textbf{E}^{H}_{e}}{\partial {a}_{e,2}}   \Big)\textbf{R}_{e}^{T}  \varphi_{e}^{p_{1}}   - q  \textbf{R}_{e} \Big(\frac{\partial \textbf{E}^{H}_{e}}{\partial {a}_{e,1}} -\frac{1-m^{I}}{a_{e,1}^{2}} \frac{\partial \textbf{E}^{H}_{e}}{\partial {a}_{e,2}}   \Big)\textbf{R}_{e}^{T}   \varphi_{e}^{p_{1}} \tau^{p_{2}}.
\end{equation}

\subsection{Parameters to control the optimization problem}
An overview of all parameters that are kept the same in all numerical examples in this work can be seen in Table~\ref{Tab:Hcoat.set.1}. 
\begin{table}[ht!]
\centering
\caption{The parameters that are used in all numerical examples in this work.}
\label{Tab:Hcoat.set.1}
\begin{tabular}{ccc} \hline
$p_{1}$ & 3 & Stiffness penalization of base structure $\varphi$  \\
$p_{2}$ & 1 &  Stiffness penalization of coating indicator $\tau$ \\
$\eta_{1}$ & 0.5 & Threshold parameter used to obtain $\varphi$ \\
$\eta_{2}$ & 0.5& Threshold parameter used to obtain $\tau$ \\
$\beta_{start}$  & 2& Starting steepness parameter for the projection  \\
$\beta_{end}$ & 128 &  Final steepness parameter for the projection \\ 
$q$ & 0.2 & Penalization of elasticity tensor in padded domain \\ \hline
 \end{tabular}
\end{table}

These parameter choices have been extensively motivated using numerical experiments. For example, the choice for not penalizing the coating ($p_{2} = 1)$ might seem counter-intuitive at first sight; however, numerical examples using $p_{2}>1$ yielded solutions that more easily ended up in less optimal local minima. For the steepness parameter of the projection operation we use $\beta_{1} = \beta_{2}$, which we start at $\beta_{start}$ and double every $30-100$ iterations until $\beta_{end}$. 

%% file: NumExp_TO.tex
\section{Numerical examples for topology optimization of coated structures}
\label{Sec:ExpHcoat}
\begin{figure}[ht!]
\centering
\includegraphics[width=0.6\textwidth]{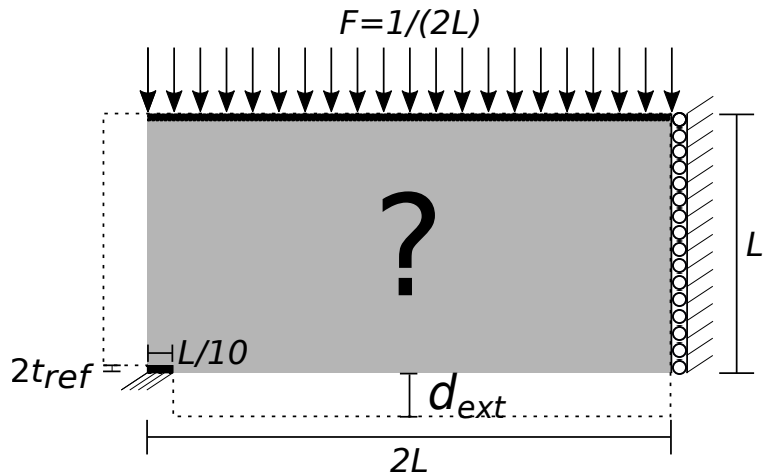}
        \caption{Design domain $\Omega$ and boundary conditions of the bridge example. $\Omega$ is indicated using gray material, while the extended filter domain is bounded by the dotted line.}
		\label{Fig:Hcoat.ExpTO.1}
\end{figure}

For the numerical experiments in this paper we focus on the MBB-beam example shown in Figure~\ref{Fig:Hcoat.Filter.1}. Furthermore, we use the bridge example for which the loads and boundary conditions, including padded domain are shown in Figure~\ref{Fig:Hcoat.ExpTO.1}.

\subsection{Experiments on the MBB-beam example}
For the experiments on the MBB-beam example we use a discretization of $300\times100$ finite elements. A volume constraint of $V_{max} = 0.4$, $R_{1} = 0.075~L$ and $t_{ref}=0.015~L$ is used. The problem formulations with the fixed microstructure density (problems~0-2), have been used to optimize the structure for various infill volume fractions ($m^{I} = 0.4$, $m^{I} = 0.5$, $m^{I}=0.6$, $m^{I}=0.7$, $m^{I}=0.8$ and $m^{I}=0.9$). The corresponding compliance values on the coarse optimization mesh $\mathcal{J}^{c}$, can be seen in Table~\ref{Tab:NumExp.TO.1}. 
\begin{table}[ht!]
\centering
\caption{Compliance $\mathcal{J}^{c}$ for different problem formulations and different infill densities $m^{I}$.}
\label{Tab:NumExp.TO.1}
\begin{tabular}{cccccccc}
\hline
 &  & $m^{I} = 0.4$ & $m^{I} = 0.5$ & $m^{I}=0.6$ & $m^{I}=0.7$ & $m^{I}=0.8$ & $m^{I}=0.9$ \\ \hline
Problem form 0 &$D$  & 362.83 & 325.94  & 297.78  & 279.80 & 256.68 &  237.63 \\ 
Problem form 1 &$D$ & $318.45$ & $291.14$ & $274.92$ & $266.85$  & $251.38$ & $236.37$ \\  
Problem form 2 &$D$ & 267.92 & 247.52 & 234.30 & 227.02 & 219.67 & 217.73 \\  \hline
Problem form 0 &$ND$ & 360.45 & 325.75  & 291.49  & 273.20 & 254.35 &  236.55 \\ 
Problem form 1 &$ND$ & $315.42$ & $292.41$ & $274.16$ & $258.53$  & $245.42$ & $235.16$ \\  
Problem form 2 &$ND$ & 266.02 & 246.83 & 231.44 & 224.44 & 220.19 & 217.84 \\  \hline
 \end{tabular}
\end{table}

Label $D$ indicates that we use the double smoothing and projection approach to obtain $\varphi$ as is shown in Figure~\ref{Fig:HcoatFilter.1}. Label $ND$ means that a single smoothing and projection step has been used, similar to the approach by~\cite{Bib:ClausenCoating2015}. 

\begin{figure}[h!]
\centering
\subfloat[Problem form 0.]{\includegraphics[width=0.3\textwidth]{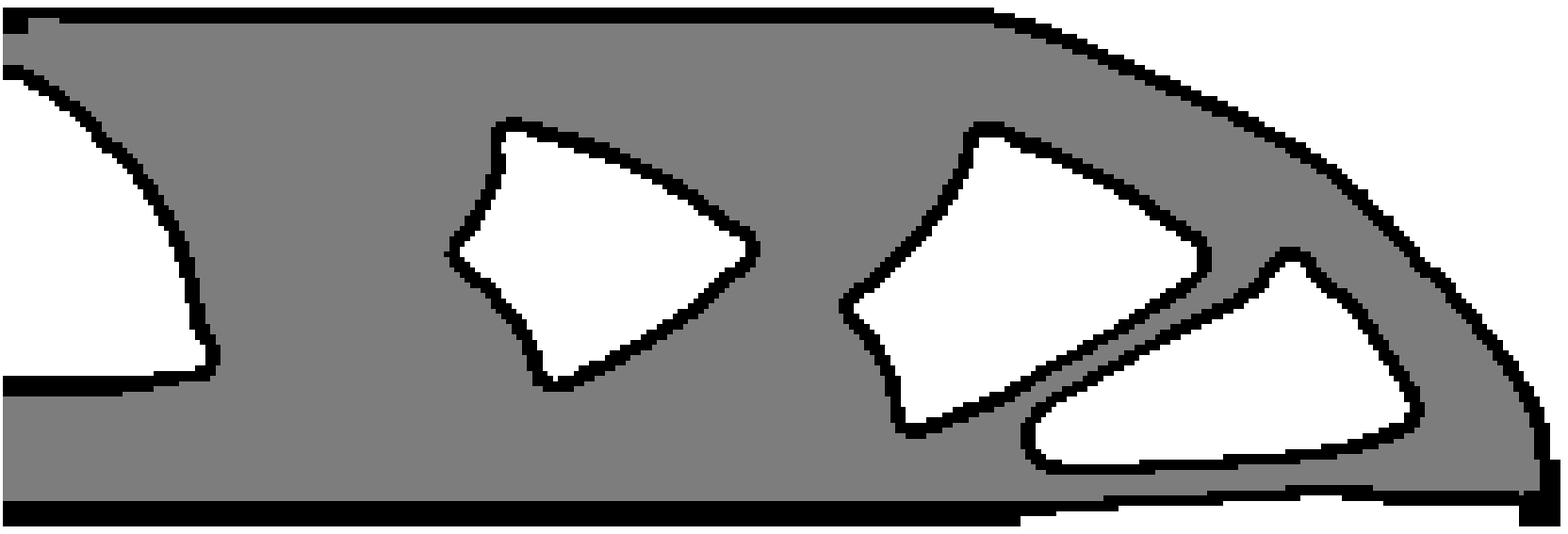}} \quad
\subfloat[Problem form 1.]{\includegraphics[width=0.3\textwidth]{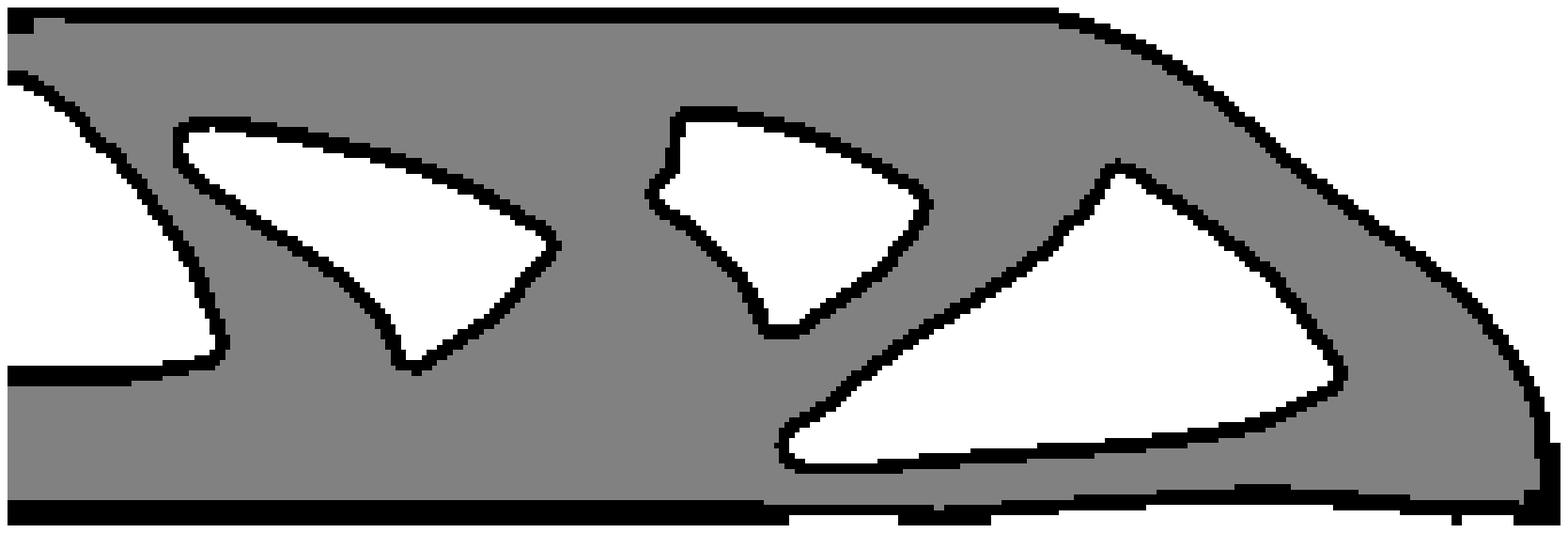}} \quad
\subfloat[Problem form 2.]{\includegraphics[width=0.3\textwidth]{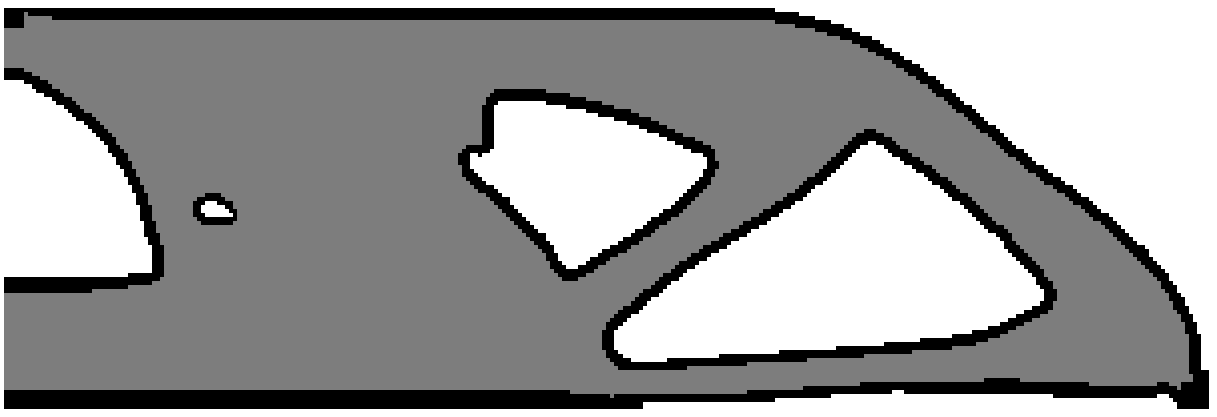}} 
\caption{Density distributions for the MBB-beam example using $m^{I} = 0.5$ and optimized for different problem formulations using the double smoothing and projection approach $(D)$.} 
\label{Fig:NumExp.TO.1}
\end{figure}

It can be seen that the compliances $\mathcal{J}^{c}$ become lower as more freedom is introduced in the microstructures, $i.e.$ going from problem 0 to 2. The corresponding density distributions for optimized results using $m^{I} = 0.5$ are shown in Figure~\ref{Fig:NumExp.TO.1}.

\begin{figure}[h!]
\centering
\subfloat[Problem form 1, $D$, $m^{I} = 0.7.$]{\includegraphics[width=0.3\textwidth]{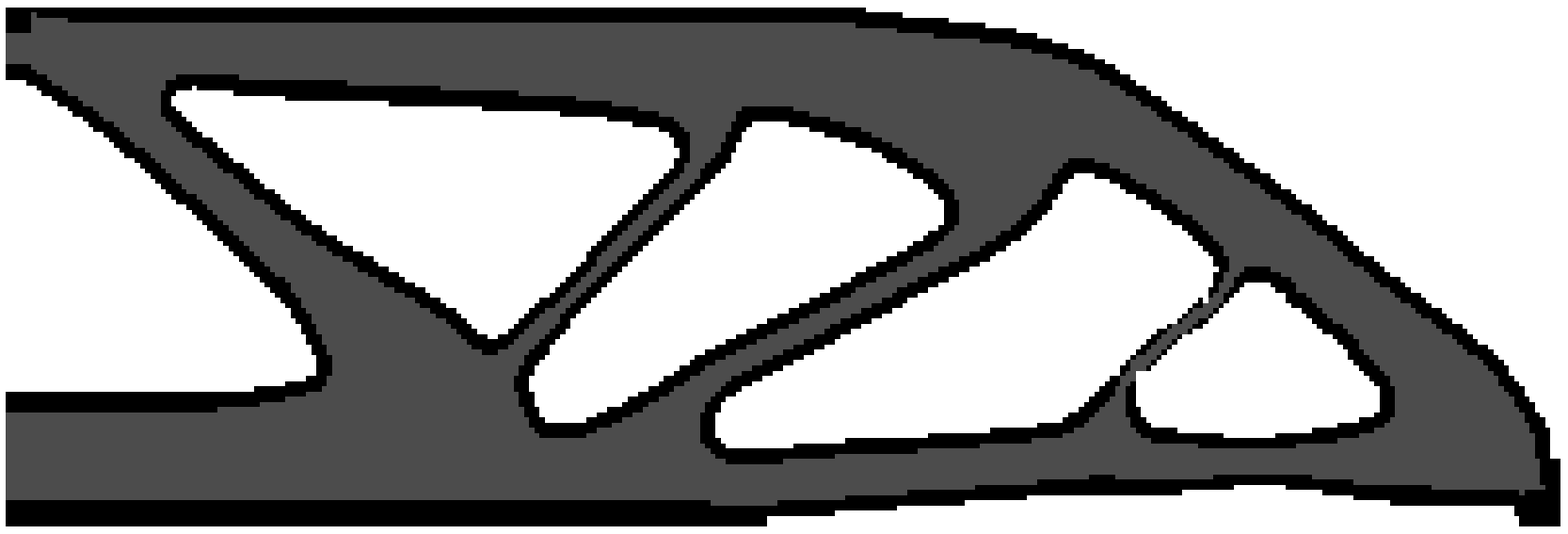}} \quad
\subfloat[Problem form 1, $D$, $m^{I} = 0.9.$]{\includegraphics[width=0.3\textwidth]{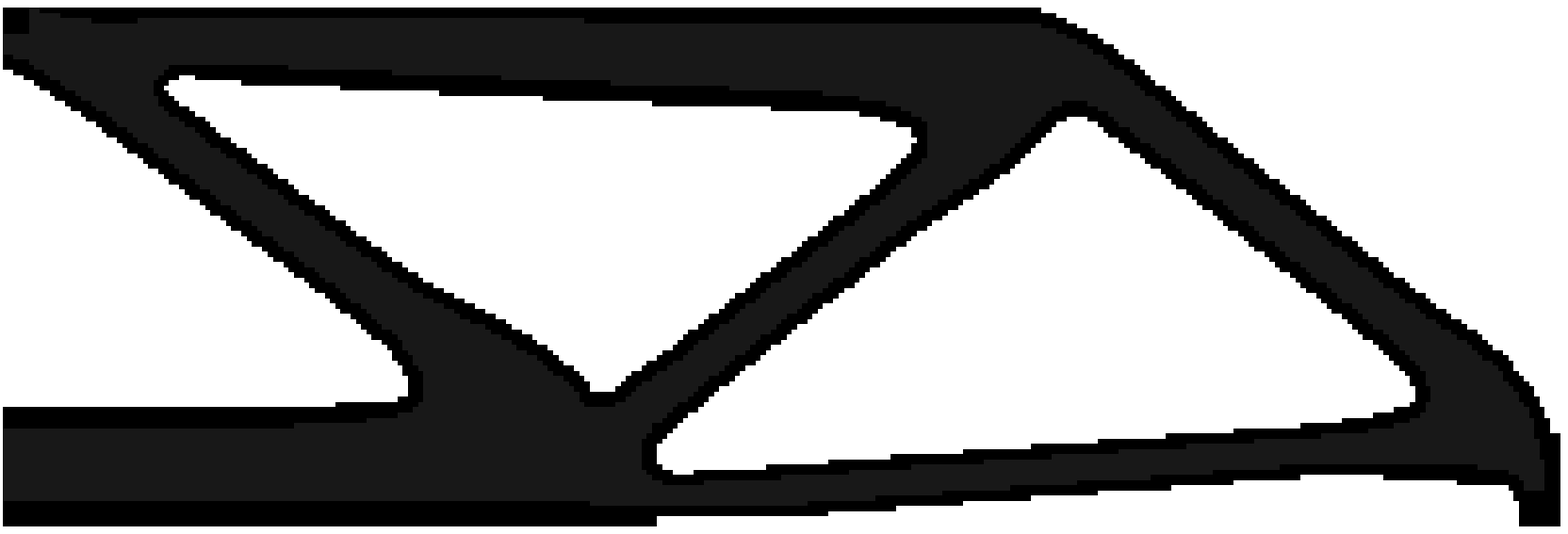}} \quad
\subfloat[Three-field SIMP $m^{I}=1.$]{\includegraphics[width=0.3\textwidth]{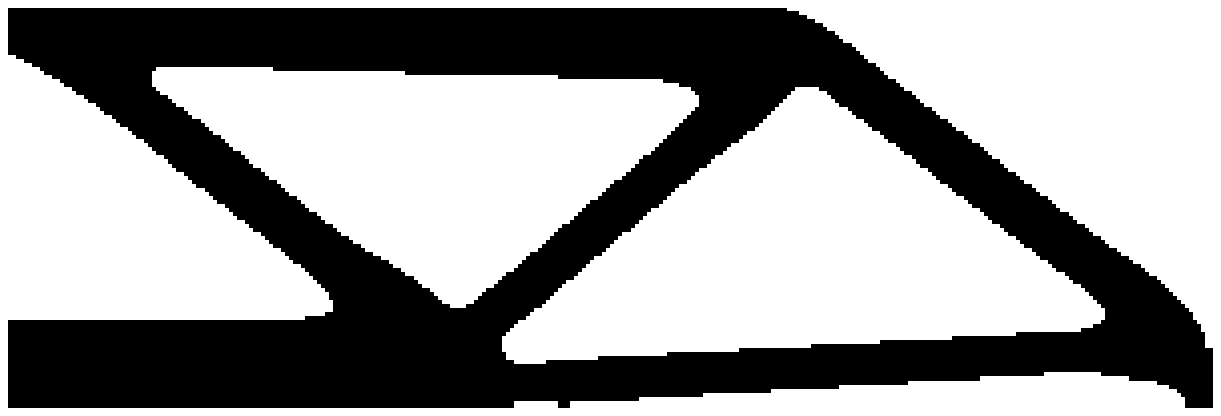}} 
\caption{Density distributions for the MBB-beam example optimized for different infill volume fractions $m^{I}$.} 
\label{Fig:NumExp.TO.2}
\end{figure}

From Table~\ref{Tab:NumExp.TO.1} it is clear that for larger infill densities the structures converge to the results of a three-field SIMP approach (SIMP with smoothing and projection). This can also be seen in Figure~\ref{Fig:NumExp.TO.2}(a) and (b). The MBB-beam optimized using three-field SIMP with these filter settings has a corresponding compliance $\mathcal{J}^{c}=213.57$, and can be seen in Figure~\ref{Fig:NumExp.TO.2}(c).

\begin{figure}[h!]
\centering
\subfloat[Problem form 1, $m^{I} = 0.5$.]{\includegraphics[width=0.4\textwidth]{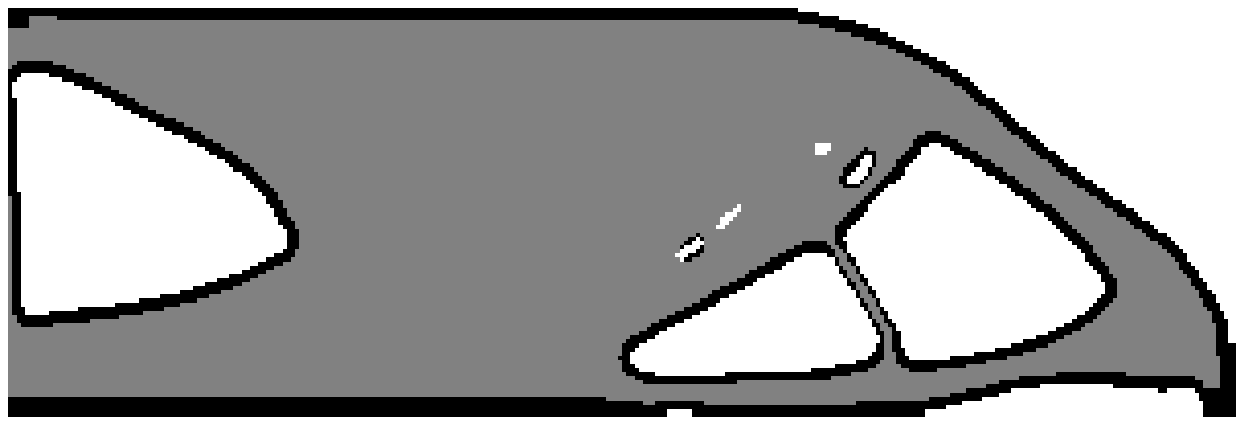}} \quad
\subfloat[Problem form 1, $m^{I} = 0.6$.]{\includegraphics[width=0.4\textwidth]{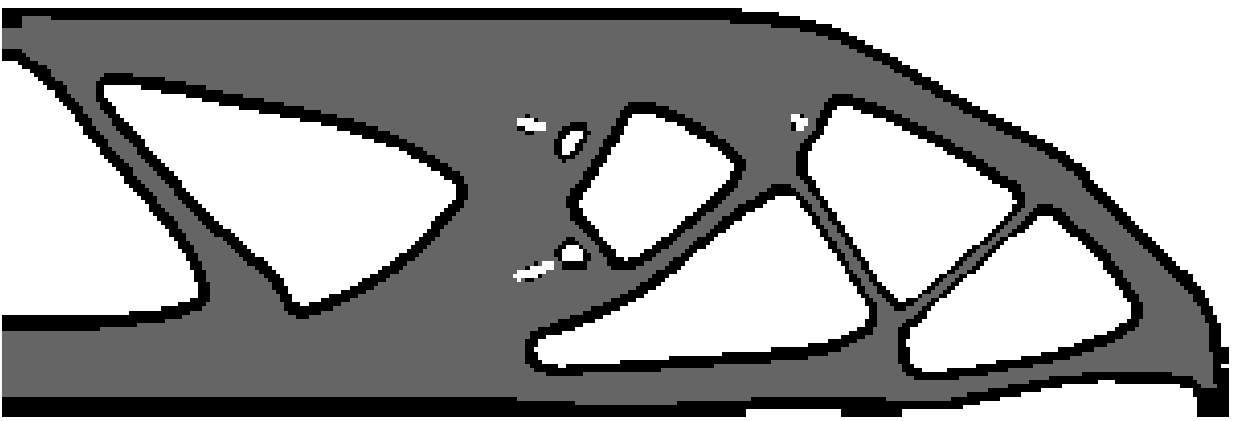}} 
\caption{Density distributions for the MBB-beam example optimized without the double filter approach $(ND)$.} 
\label{Fig:NumExp.TO.3}
\end{figure}

The structures without the double filter approach $(ND)$ seem to have a slightly better compliance. However, this is at the cost of undetermined coating and infill, as seen in Figure~\ref{Fig:NumExp.TO.3}(a) and (b), where small features of void can be observed, that are not bounded by coating. We cannot guarantee that filtering and smoothing $\boldsymbol{\mu}$ twice will always create 0-1 features in $\boldsymbol{\varphi}$; however, numerical experiments have shown promising results in nearly all cases.

Finally, when we use the optimization problem with full freedom (problem 3), an even better performing structure than using standard three-field SIMP can be obtained. In Table~\ref{Tab:NumExp.TO.2} one can find the compliance values corresponding to several lower $a_{l}$ and upper $a_{u}$ bounds on the parameters that describe the shape of the hole.
\begin{table}[ht!]
\centering
\caption{Compliance $\mathcal{J}^{c}$ for problem formulations 3 using double filter approach $D$ and without double filter approach $ND$ for various bounds on the shape of the microstructure.}
\label{Tab:NumExp.TO.2}
\begin{tabular}{ccccccccccc}
\hline
$a_{l}$  & 0.2 & 0.2 & 0.3 & 0.3 & 0.4 & 0.4 & 0.5 & 0.5 & 0.6 & 0.6 \\
$a_{u}$ & 0.8 & 0.9 & 0.8 & 0.9 & 0.8 & 0.9 & 0.8 & 0.9 & 0.8 & 0.9 \\ \hline
$D$& 222.81& 207.12 & 228.70& 212.42 & 236.29& 219.35 & 245.61& 229.48 &259.57& 245.15 \\ 
$ND$ & 226.91 & 207.09 & 231.91 & 212.32 &  238.04& 219.38 & 248.50 & 229.34 & 261.96& 244.97 \\ \hline
 \end{tabular}
\end{table}

\begin{figure}[h!]
\centering
\subfloat[Problem form 3, $D$, $a_{l} = 0.2$ and $a_{u} = 0.9$.]{\includegraphics[width=0.4\textwidth]{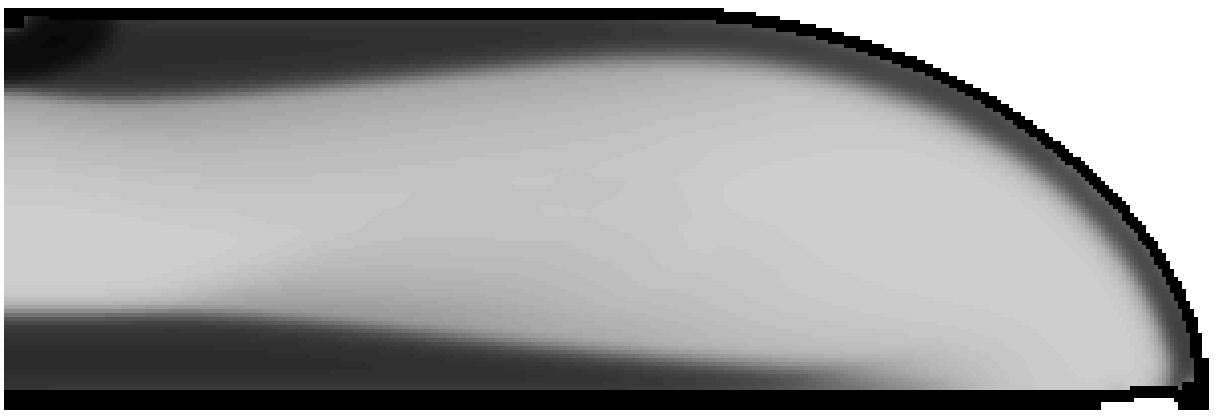}} \quad
\subfloat[Problem form 3, $D$, $a_{l} = 0.3$ and $a_{u} = 0.8$.]{\includegraphics[width=0.4\textwidth]{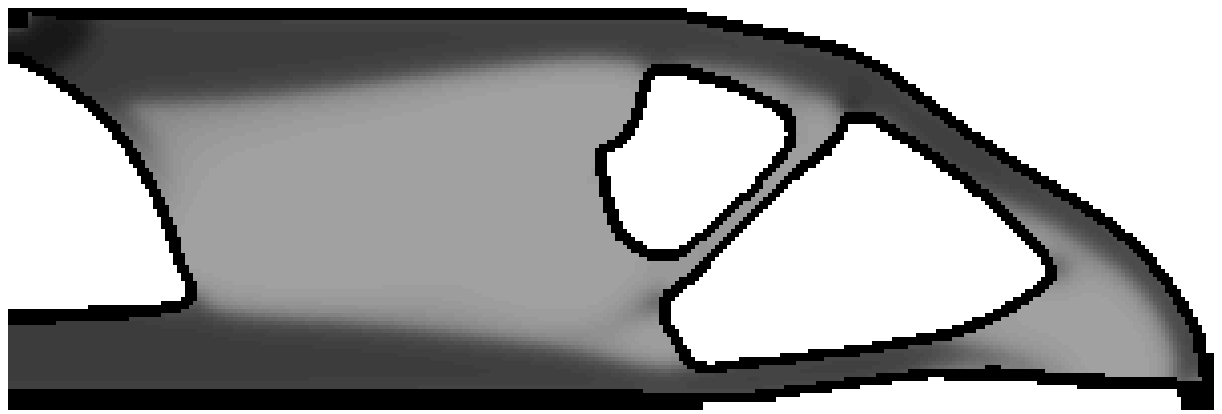}} 
\caption{Density distribution for the MBB-beam example optimized with the double filter approach for the most general optimization problem, shown for various bounds on the shape of the microstructure.} 
\label{Fig:NumExp.TO.4}
\end{figure}

It can be seen that the lowest compliance can be reached when there is the largest freedom for parameters $a_{i}$ to vary. To illustrate this, consider the density distribution shown in Figure~\ref{Fig:NumExp.TO.4} (a). Here the microstructures are allowed to be nearly solid $a_{l} = 0.2$ and also get close to void $a_{u} = 0.9$. Hence, there is no need to create holes. However, if the bounds are set a bit tighter, e.g. $a_{l} = 0.3$ and $a_{u} = 0.8$ holes will originate in the optimized structure, as can be seen in Figure~\ref{Fig:NumExp.TO.4} (b).

\subsection{Bridge design example}
\begin{figure}[h!]
\centering
\subfloat[Problem form 0, $\mathcal{J}^{c} = 34.08$.]{\includegraphics[width=0.4\textwidth]{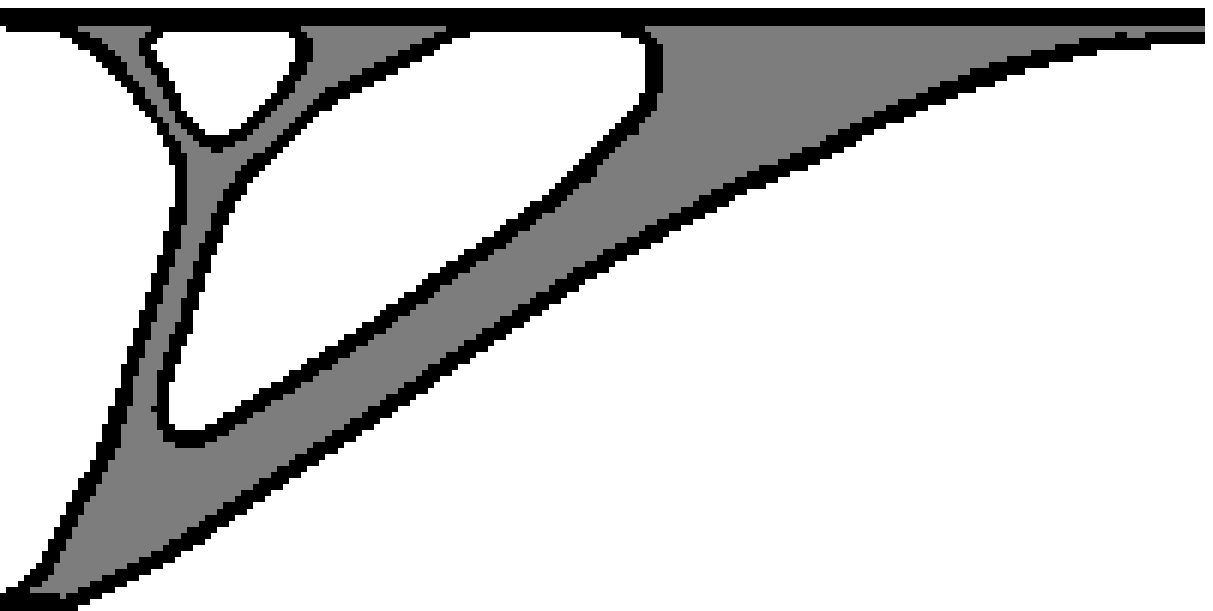}} \quad
\subfloat[Problem form 1, $\mathcal{J}^{c} = 30.63$.]{\includegraphics[width=0.4\textwidth]{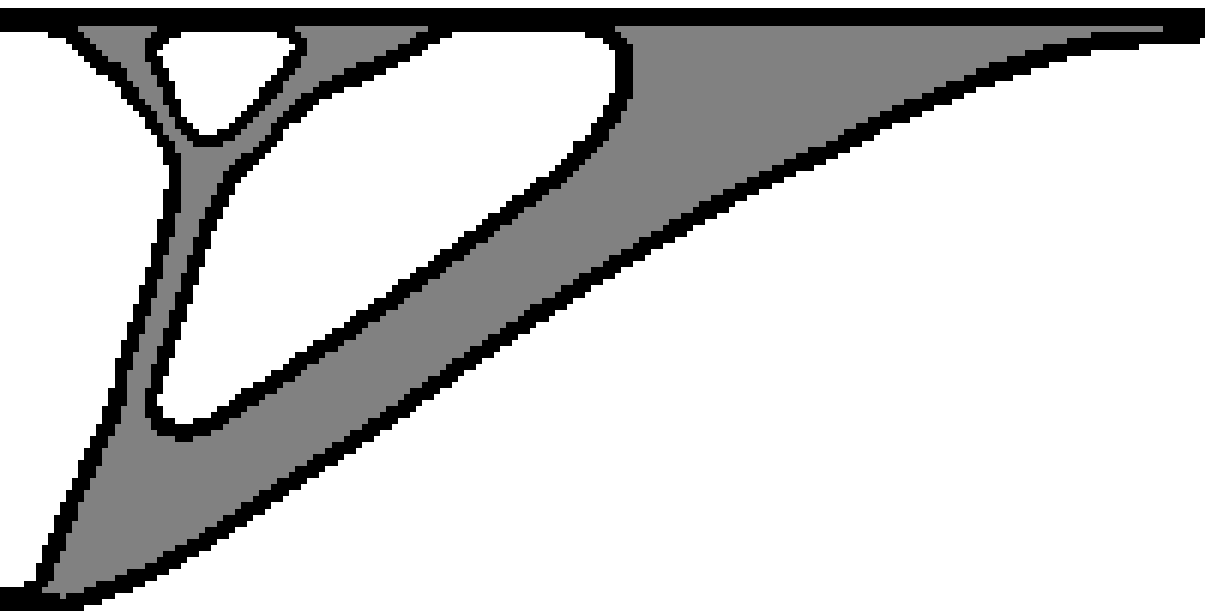}} \\
\subfloat[Problem form 2, $\mathcal{J}^{c} = 27.26$.]{\includegraphics[width=0.4\textwidth]{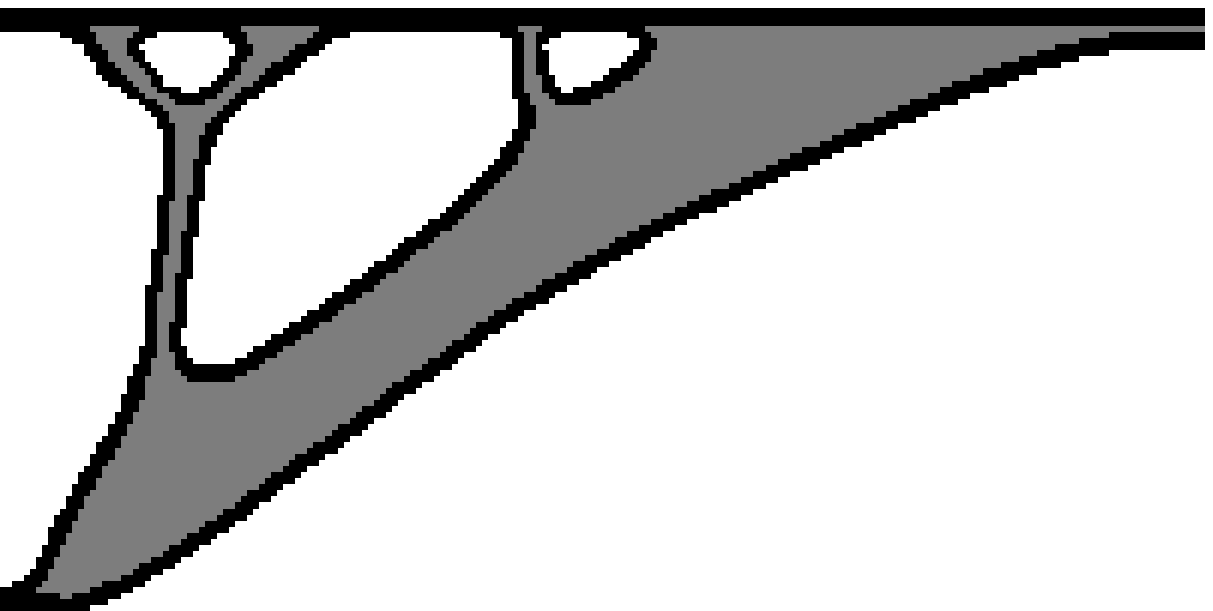}} \quad
\subfloat[Problem form 3, $\mathcal{J}^{c} = 26.13$.]{\includegraphics[width=0.4\textwidth]{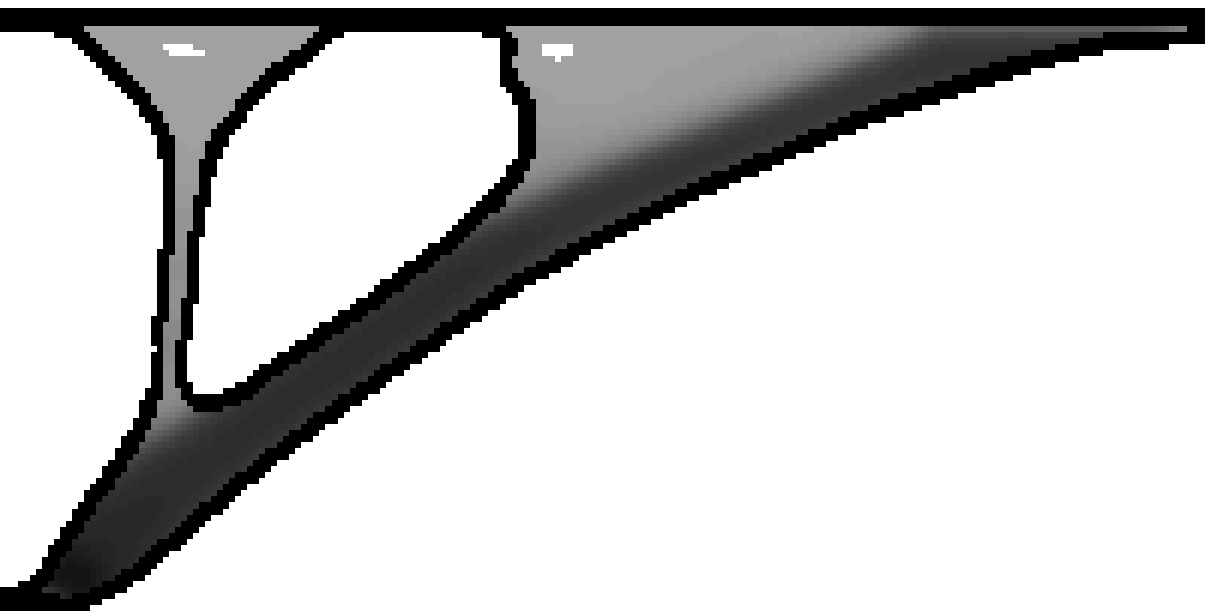}} \\
\caption{Density distribution for the bridge design example, for all 4 types of optimization problems.} 
\label{Fig:NumExp.TO.5}
\end{figure}

The second optimization example considered is the bridge design example, shown in Figure~\ref{Fig:Hcoat.ExpTO.1}. The optimization is performed on a coarse mesh of $200\times100$ elements. A volume constraint of $V_{max} = 0.2$, $R_{1} = 0.075~L$ and $t_{ref}=0.015~L$ is used. The optimized designs and their compliance values, for $m^{I} = 0.5$, and for problem forms 0-2 are shown in Figure~\ref{Fig:NumExp.TO.5} (a)-(c). The design optimized for problem form 3, using $a_{l} = 0.2$ and $a_{u}=0.8$ is shown in Figure~\ref{Fig:NumExp.TO.5} (d).

It is interesting to note that the two small void areas in the top of Figure~\ref{Fig:NumExp.TO.5} (d) are not bounded by coating material. This is a rare case where even the double filter approach cannot guarantee coating everywhere. Although this does not happen in many cases, it is still an undesired effect. A possible remedy can be to apply image analysis on the optimized designs and explicitly enforce coating if required.

%% file: Proj.tex
\section{Projecting coated structures with microstructures as infill}
\label{Sec:Proj}
To project the infill as a smooth and continuous lattice structure two mapping functions $\phi_{1}$ and $\phi_{2}$ have to be determined, representing the two orthogonal layers of the sequence of unit-cells~\citep{Bib:Pantz1,Bib:Pantz2,Bib:GroenSigmund2017}. These mapping functions are then later used to project the composite shape of the microstructures on a fine mesh. We will not go into full detail on the derivation of the mapping functions; instead we focus on changes and improvements compared to the approach presented in~\citet{Bib:GroenSigmund2017}.

\subsection{Projecting a periodic composite shape}
Contrary to previous approaches we do not solve for the mapping function in the void part of $\Omega$, i.e. we create a conformal, albeit regular mesh on $\Omega_{m} = \Omega_{c} \bigcup \Omega_{l}$. Both mapping functions $\phi_{1}$ and $\phi_{2}$ can be obtained independently of each other; therefore, we restrict ourselves to the derivation of $\phi_{1}$. A suitable parameterization of $\phi_{1}$ has to fulfill: 
\begin{enumerate}
\item{} $\phi_{1}$ should be constant in the direction tangential to the layer normal $\textbf{n}_{i}$.
\item{} The spacing between the contour lines of $\phi_{1}$, should be as regular as possible without violating the first requirement.
\end{enumerate}
To solve for $\phi_{1}$ we use the following minimization problem,
\begin{equation}\label{Eq:proj.1}
\begin{aligned}
\displaystyle \min_{\phi_{1}(\textbf{x})} & :  \mathcal{I}(\phi_{1}(\textbf{x}))		      =  \frac{1}{2}\int_{\Omega_{m}} \left \Vert \nabla \phi_{1}(\textbf{x})  -  \textbf{n}_{1}(\textbf{x}) \right \Vert ^{2} \textnormal{d}\Omega_{m},&  &  \\
\textrm{s.t.}             		 & :  \nabla \phi_{1} (\textbf{x})\cdot \textbf{t}_{1}(\textbf{x}) = 0. &  & \\ 
 \end{aligned}
\end{equation}
Here $\textbf{t}_{1}$ is tangential to normal vector $\textbf{n}_{1}$, hence both depend on the local directions of lamination $\theta$,
\begin{equation} \label{Eq:proj.2}
\begin{aligned}
& \textbf{n}_{1}(\textbf{x}) = \textbf{t}_{2}(\textbf{x}) = \begin{bmatrix} -\textnormal{sin}(\theta(\textbf{x})) \\ \textnormal{cos}(\theta(\textbf{x}))
\end{bmatrix},
& & 
\textbf{n}_{2}(\textbf{x}) = \textbf{t}_{1}(\textbf{x}) =\begin{bmatrix} \textnormal{cos}(\theta(\textbf{x})) \\ \textnormal{sin}(\theta(\textbf{x}))
\end{bmatrix}.
& 
\end{aligned}
\end{equation}
It has to be noted that the principal stress directions used to calculate $\theta$ are rotationally symmetric, hence there may be jumps of size $\pi$ in angle field $\theta$. These jumps are identified using connected component labeling and aligned consistently as suggested in~\cite{Bib:GroenSigmund2017}, to allow for a smooth projection using Equation~\ref{Eq:proj.1}.

The mapping functions can then be used to project the optimized shape. As opposed to~\citet{Bib:GroenSigmund2017} that used a cosine we here use a triangle wave function $\mathcal{S}$ using the $\texttt{sawtooth}$ function in MATLAB.
\begin{equation} \label{Eq:proj.3}
\tilde{\rho_{1}}(\textbf{x}) = \frac{1}{2} + \frac{1}{2} \mathcal{S} (P_{1}\phi_{1}(\textbf{x})),
\end{equation}
where $P_{1}$ is a periodicity scaling parameter. The exact widths of the microstructure are then projected using Heaviside function $H$,
\begin{equation} \label{Eq:proj.4}
\rho_{1} (\textbf{x})= H\big(\tilde{\rho}_{1}(\textbf{x})-(1-a_{1}(\textbf{x}))\big).
\end{equation}
After solving for the densities of both layers independently, they can be combined to obtain the density field $\rho$,
\begin{equation} \label{Eq:proj.5}
\rho (\textbf{x})= \text{max}(\rho_{1}(\textbf{x})+\rho_{2}(\textbf{x})+\tau,1).
\end{equation}

The mapping problem is solved using bi-linear finite elements on an intermediate mesh using $h^{i} = h^{c}/2$, on which $\theta$ (after consistent alignment) is interpolated using linear interpolation. The constraint is not enforced explicitly; but in penalty form using penalty parameter $\gamma$. For large values of $\gamma$ the constraint enforces mapping functions $\phi_{1}$ and $\phi_{2}$ to be aligned with $\theta$, at the cost of relaxed periodicity. 

Locally we can identify the spacing of the mapping by making use of the norm of the derivatives of the mapping functions $\left \vert \left \vert \nabla \phi_{i} \right \vert \right \vert$. If the value of $\left \vert \left \vert \nabla \phi_{i} \right \vert \right \vert > 1$, then the corresponding layer distance is locally compressed, similarly if $\left \vert \left \vert \nabla \phi_{i} \right \vert \right \vert < 1$ the corresponding layer distance is locally stretched. In general we would like to impose an average layer distance $\varepsilon$. To do that we can determine periodicity scaling parameter $P_{i}$ as,
\begin{equation}\label{Eq:proj.6}
P_{i} = \frac {2\pi\int_{\Omega_{m}} \text{d}\Omega_{m}}{\varepsilon\int_{\Omega_{m}} ||\nabla \phi_{i}(\textbf{x})||  \text{d}\Omega_{m}}.
\end{equation}

After scaling, the mapping functions are interpolated on a fine mesh where $h^{f} \leq h^{c}/10$, and the microstructure can be projected using Equations~\ref{Eq:proj.3}-\ref{Eq:proj.5}. The normalized gradient norm $\left\vert\left\vert\nabla\tilde{\varphi} \right\vert\right\vert_{\alpha}$ is interpolated using linear interpolation from coarse to fine mesh, afterwards the projection to obtain a clear coating $\tau=\overline{\left\vert\left\vert\nabla\tilde{\varphi} \right\vert\right\vert}_{\alpha}$is performed. 

\begin{figure}[h!]
\centering
\subfloat[$\rho$ on $\mathcal{T}^{c}$.]{\includegraphics[width=0.24\textwidth]{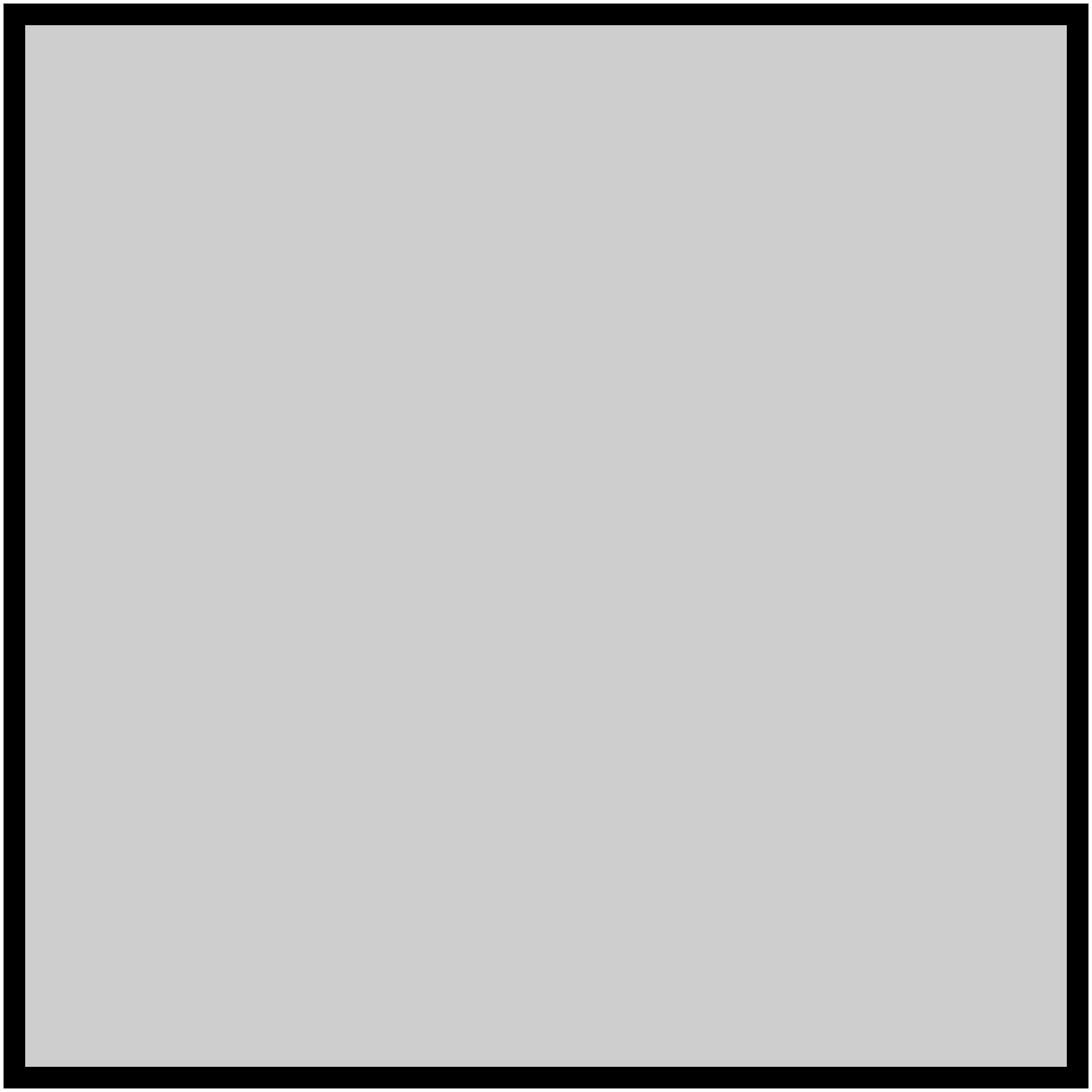}} 
\subfloat[$\theta$ on $\mathcal{T}^{c}$.]{\includegraphics[width=0.27\textwidth]{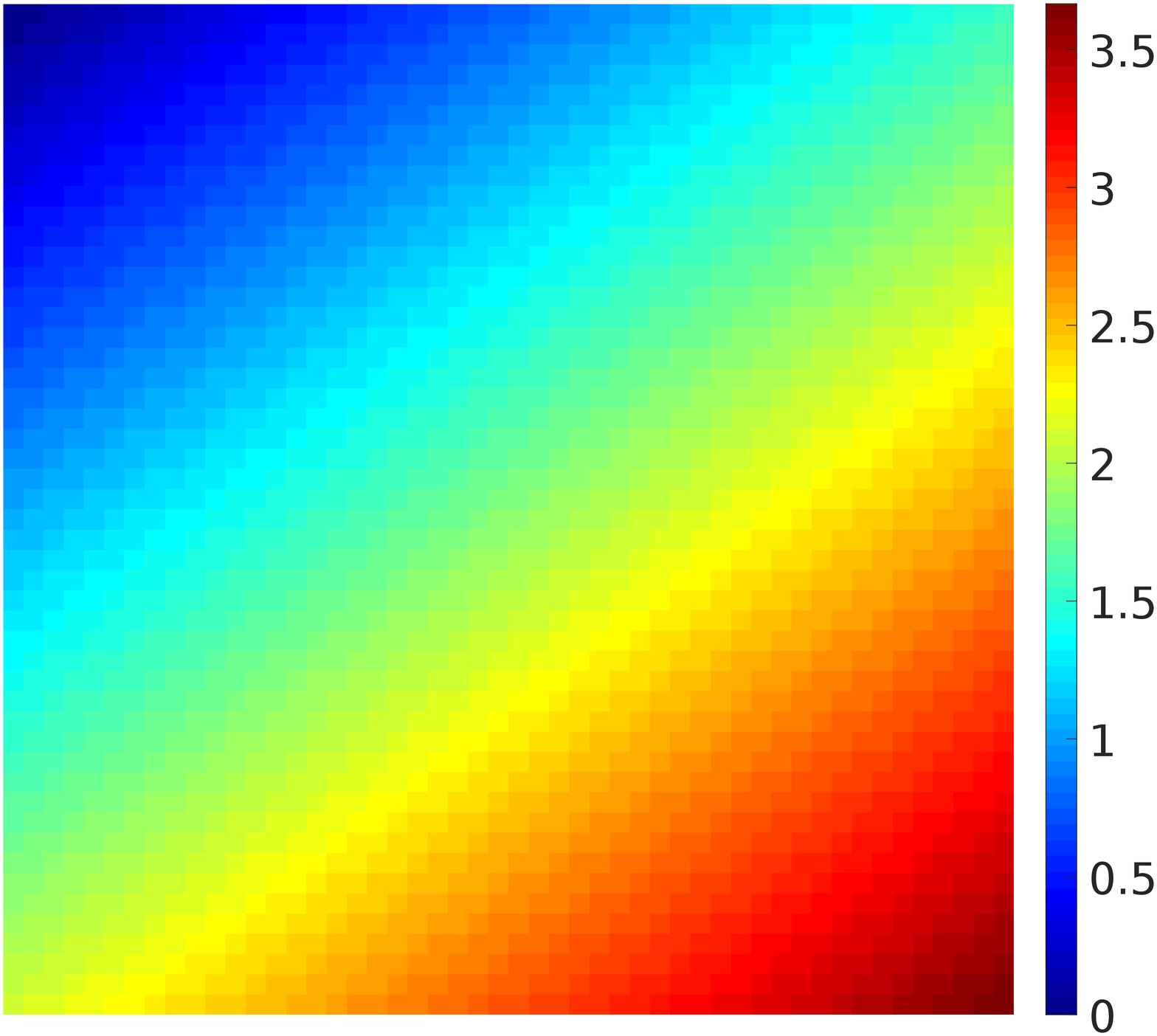}} 
\subfloat[$\phi_1$ on $\mathcal{T}^{i}$.]{\includegraphics[width=0.24\textwidth]{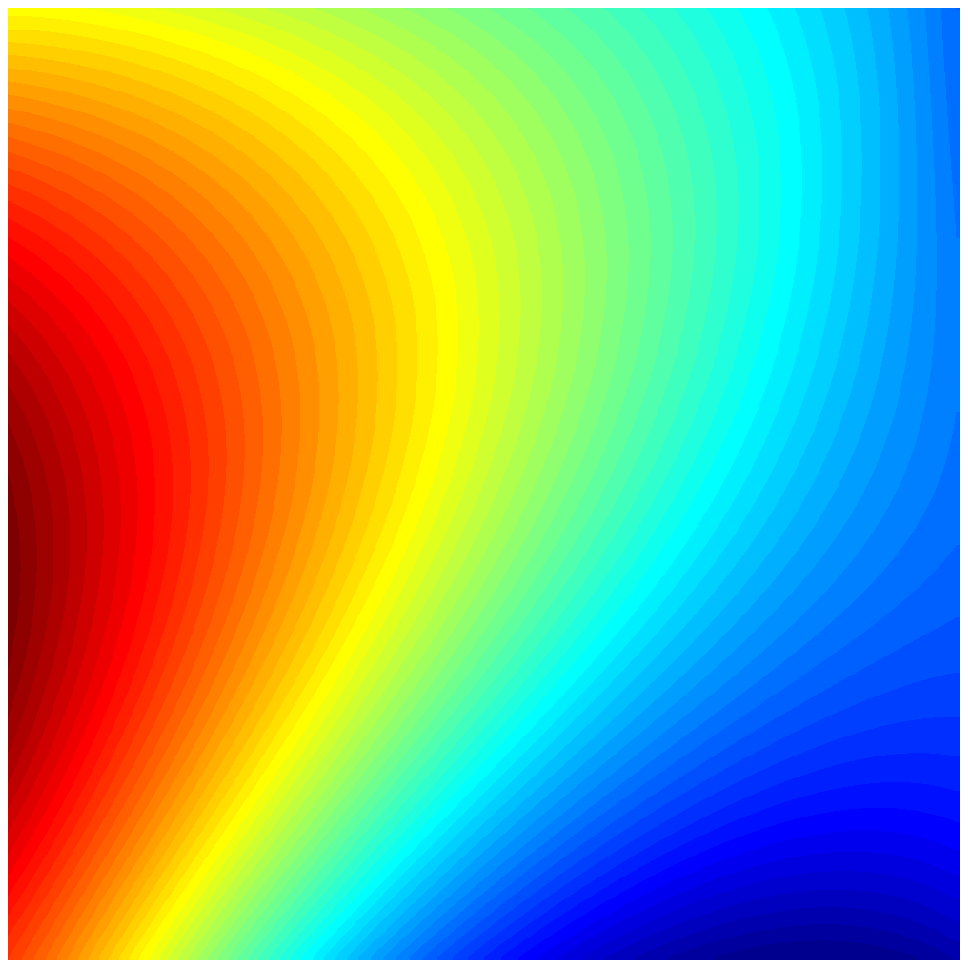}} 
\subfloat[Map on $\mathcal{T}^{f}$.]{\includegraphics[width=0.24\textwidth]{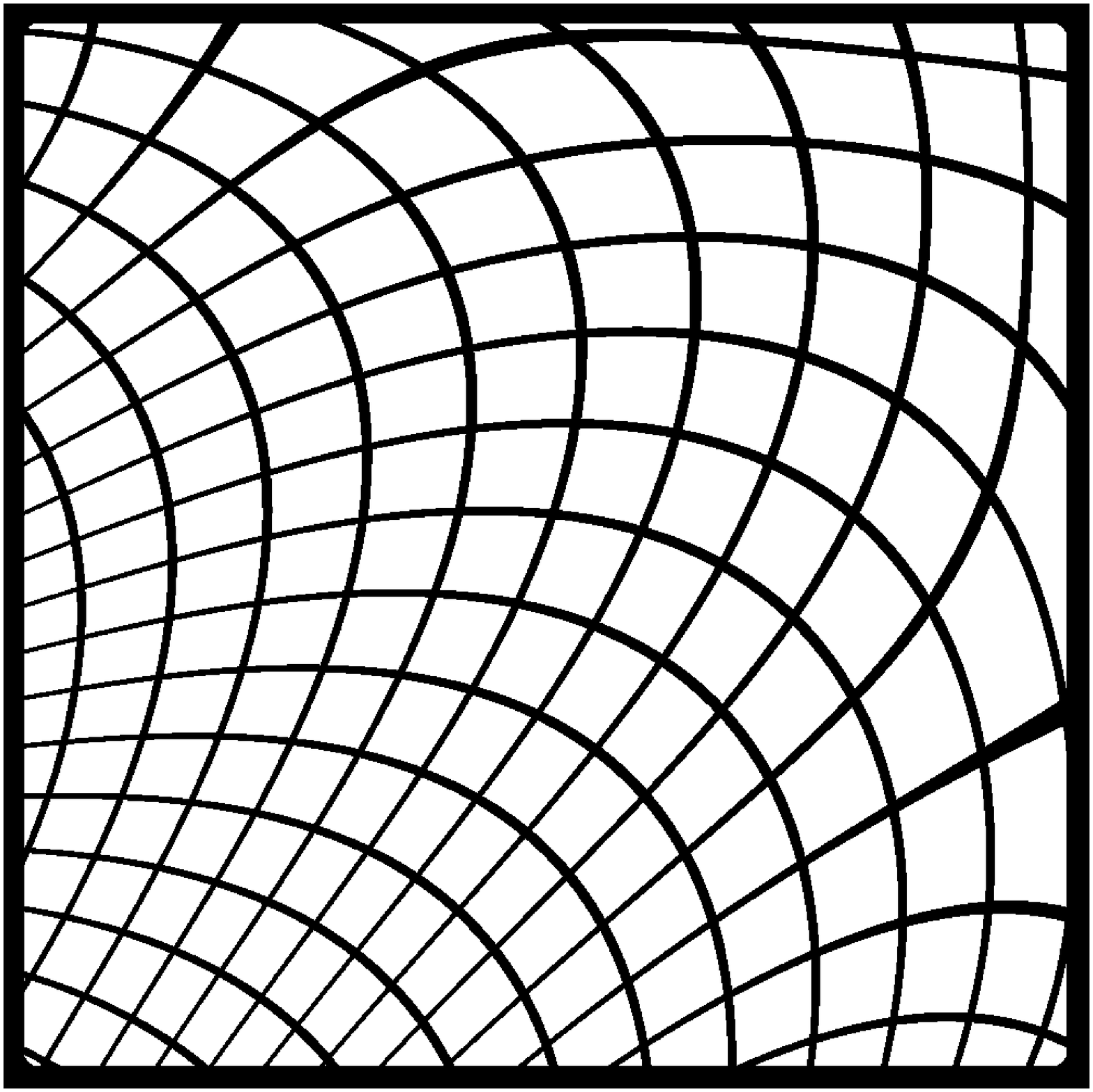}} 
\caption{Example of the mapping procedure, using $\varepsilon = 80~h^{f}$. To demonstrate the use of a coarse $\mathcal{T}^{c}$, intermediate $\mathcal{T}^{i}$ and fine mesh $\mathcal{T}^{f}$.} 
\label{Fig:proj.1}
\end{figure}

To demonstrate the mapping procedure consider the simple test case shown in Figure~\ref{Fig:proj.1}(a). Here we have a coated structure on coarse mesh $\mathcal{T}^{c}$ consisting of $50\times50$ coarse elements, where the coating layer is exactly 1 element wide. The square microstructure has $a_{1} = a_{2} = 0.9$ and the corresponding angle field is shown in Figure~\ref{Fig:proj.1}(b). The mapping is performed on $\mathcal{T}^{i}$ using $100\times100$ elements, where $\phi_{1}$ is shown in Figure~\ref{Fig:proj.1}(c). The corresponding projection on $\mathcal{T}^{f}$ consisting of $1000\times1000$ elements is shown in Figure~\ref{Fig:proj.1}(d).

\subsection{Method to adaptively refine the periodicity}
Figure~\ref{Fig:proj.1}(d) shows a large variation in the layer spacing throughout the domain. However, for performance, and manufacturability of 3D printed structures, a regular spacing is desired. A possible solution to obtain a more regular spacing is to relax the angle constraint by using a small value for $\gamma$; however, this results in a decreased performance. Close inspection of optimization results from~\citet{Bib:WuClausenSigmund2017} (where the infill is optimized using SIMP) shows that structural members tend to split in two to counter increasing spacing between structural members. This observation lead to the idea of locally adapting the periodicity, to have a layer spacing as close to $\varepsilon$ as possible.

\begin{figure}[h!]
\centering
\subfloat[$\lambda_{1}$ on $\mathcal{T}^{f}$.]{\includegraphics[width=0.25\textwidth]{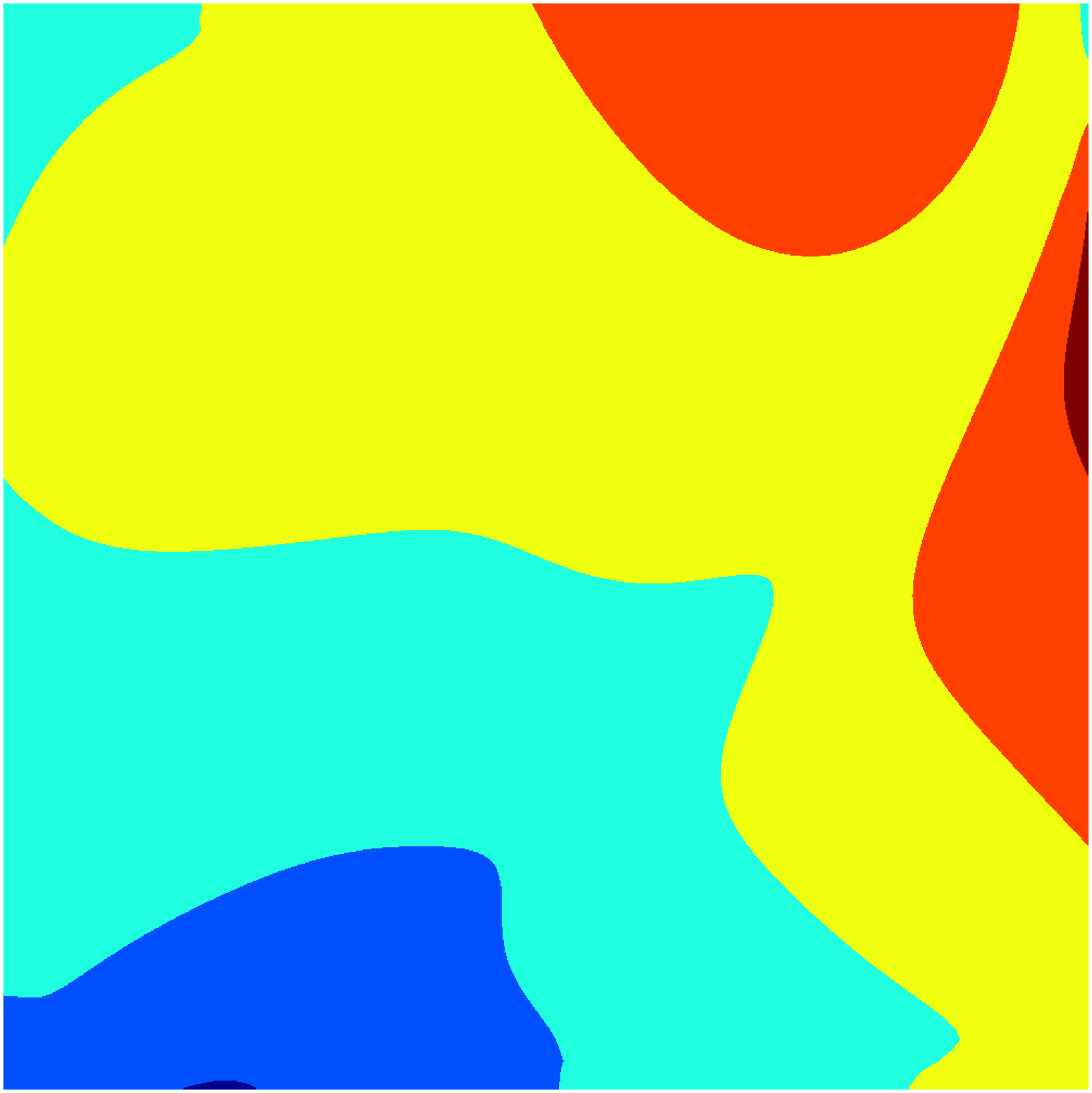}} 
\subfloat[Map on $\mathcal{T}^{f}$.]{\includegraphics[width=0.25\textwidth]{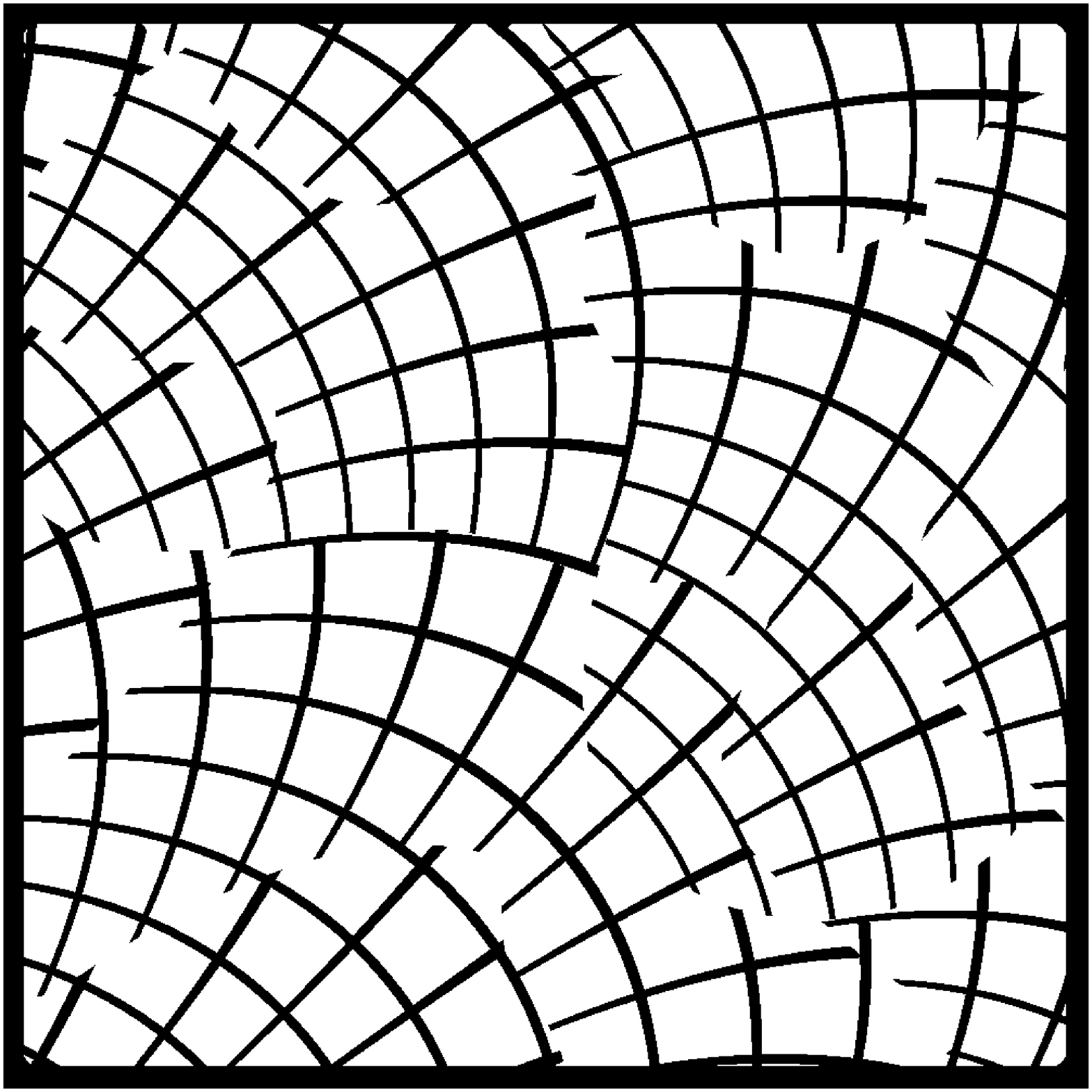}} 
\subfloat[$\tilde{\lambda}_{1}$ on $\mathcal{T}^{f}$.]{\includegraphics[width=0.25\textwidth]{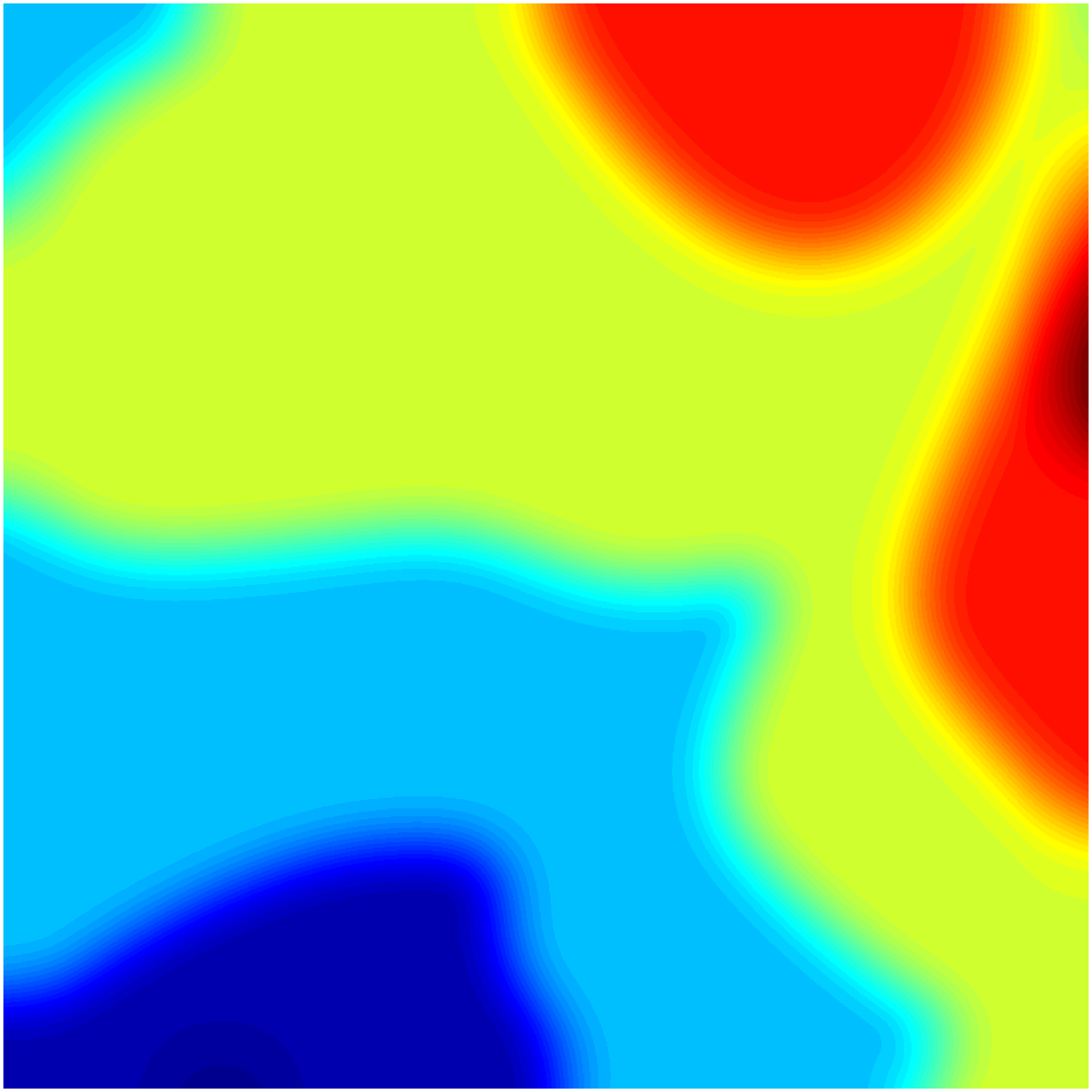}} 
\subfloat[Map and $\psi_{1}$ on $\mathcal{T}^{f}$.]{\includegraphics[width=0.25\textwidth]{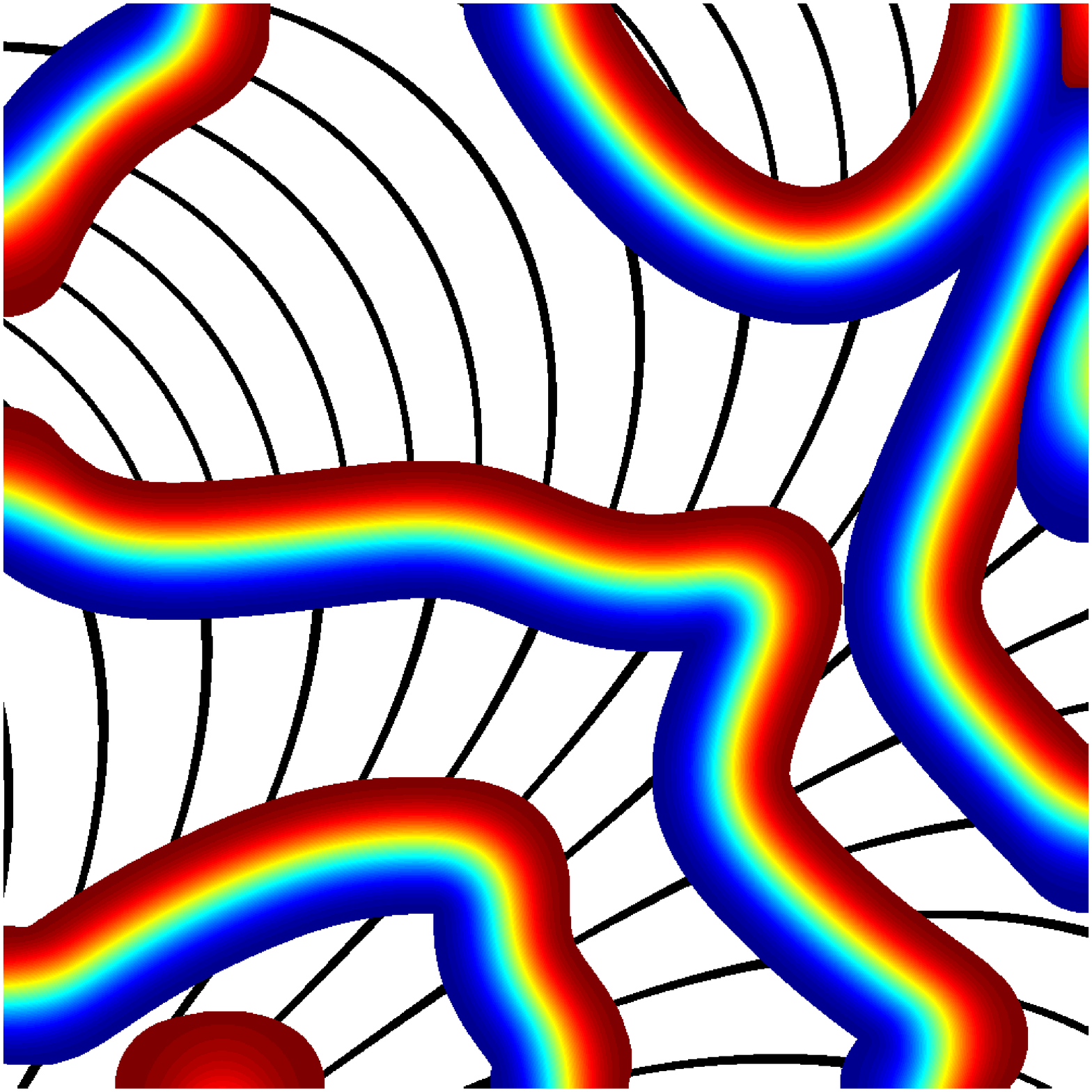}} 
\caption{Visual explanation of adaptive periodicity and required transition zone, to split the structural members for $R^{*} = 0.8\varepsilon$.} 
\label{Fig:proja.1}
\end{figure}

Instead of using Equation~\ref{Eq:proj.3} we use a function such that the  periodicity is adapted using discrete periodicity scaling parameter $\lambda_{i}$, corresponding to the $i$-th mapping function, to obtain 
\begin{equation} \label{Eq:proja.1}
\tilde{\rho_{i}}(\textbf{x},\lambda_{i}(\textbf{x})) = \frac{1}{2} + \frac{1}{2} \mathcal{S} (2^{(\lambda_{i}(\textbf{x})+1)}\pi \phi_{i}(\textbf{x})+\lambda_{i}(\textbf{x})\pi)).
\end{equation}
When $\lambda_{i}=0$, we have the same function as Equation~\ref{Eq:proj.3} with $\varepsilon = 1$. However, when $\lambda_{i} = 1$, we have the same function as when $\varepsilon = 0.5$. Hence, the periodicity is doubled. To get the periodicity in Equation~\ref{Eq:proja.1} as close to $\varepsilon$ as possible we choose $\lambda_{i}$ as,
\begin{equation} \label{Eq:proja.2}
\lambda_{i}(\textbf{x}) = \text{round}\Big(\text{log}\big(\frac{1}{\varepsilon  ||\nabla \phi_{i}(\textbf{x})||}\big)\frac{1}{\text{log}(2)}\Big).
\end{equation}
A plot of $\lambda_{1}$ corresponding to $\phi_{1}$ in Figure~\ref{Fig:proj.1}(c), can be seen in Figure~\ref{Fig:proja.1}(a). The corresponding projection using Equation~\ref{Eq:proja.1} can be seen in Figure~\ref{Fig:proja.1}(b), where it can be seen that the effective lattice spacing is now bounded to the interval $[\varepsilon2^{-1/2}, \varepsilon2^{1/2}]$.

Unfortunately however, the projected structure in Figure~\ref{Fig:proja.1}(b) is now discontinuous. To reconnect the bars a transition zone $\Omega_{T,i}$ is needed. Therefore, we use a convolution operation to obtain $\tilde{\lambda_{i}}$, shown in Figure~\ref{Fig:proja.1}(c), using a linearly decaying convolution kernel with radius $R^{*}$. Using $\tilde{\lambda_{i}}$, we can determine splitting parameter $\psi_{i}$, which is used to identify if we are in the transition zone between two discrete values of $\lambda_{i}$.
\begin{equation} \label{Eq:proja.3}
\psi_{i}(\textbf{x}) = \text{modulo}(\tilde{\lambda_{i}}(\textbf{x}),1).
\end{equation}
If $\psi_{i} = 0$, then we are outside $\Omega_{T,i}$ and we can use Equation~\ref{Eq:proja.1}, as is shown in Figure~\ref{Fig:proja.1}(d), while inside $\Omega_{T,i}$  the value of $\psi_{i}$ is shown. 

\begin{figure}[h!]
\centering
\subfloat[$\tilde{\rho_{i}}$ for $\lambda_{i} = 0$ and $\lambda_{i} = 1$.]{\includegraphics[width=0.25\textwidth]{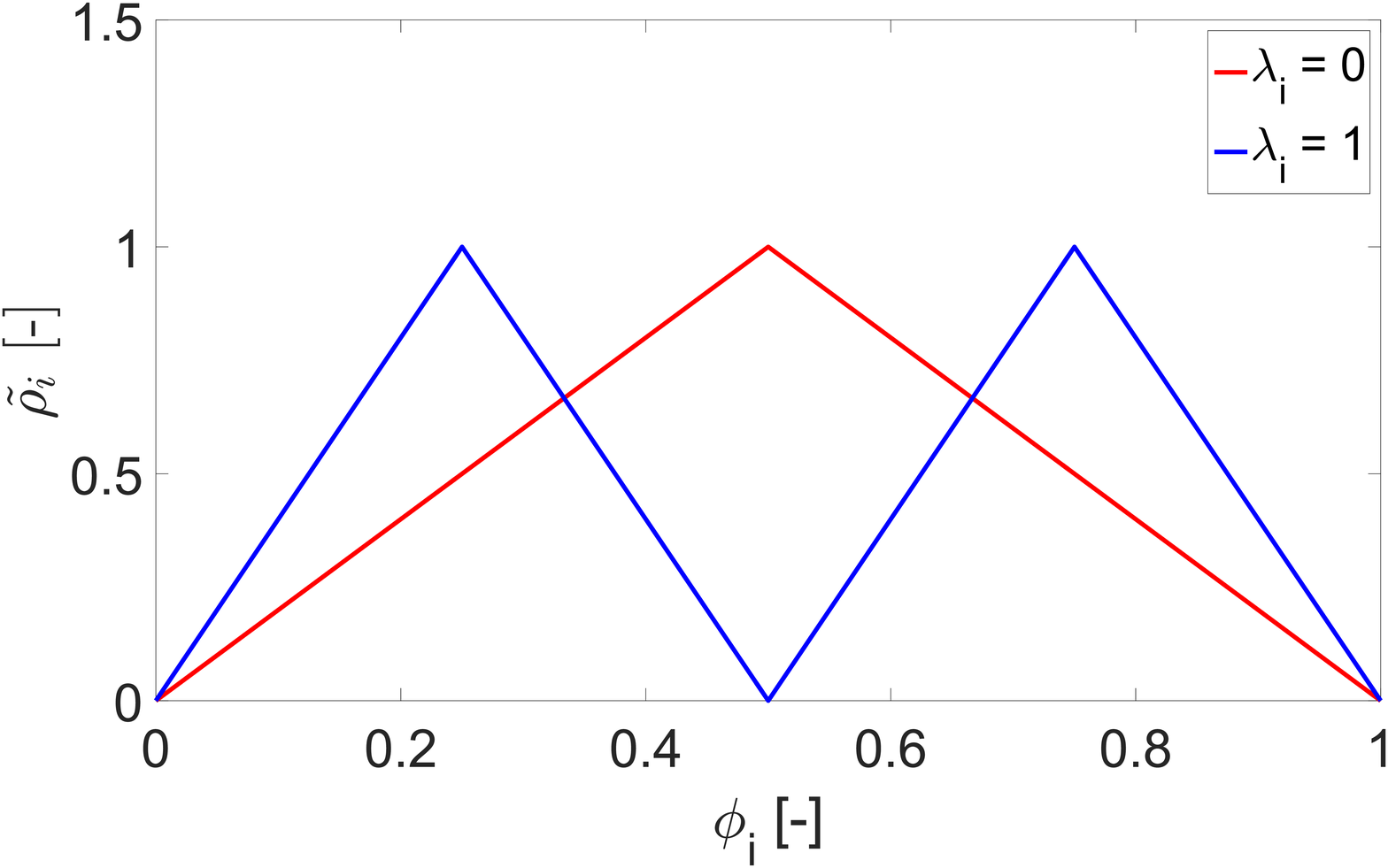}} 
\subfloat[Base minus $1/2\psi_{i}$.]{\includegraphics[width=0.25\textwidth]{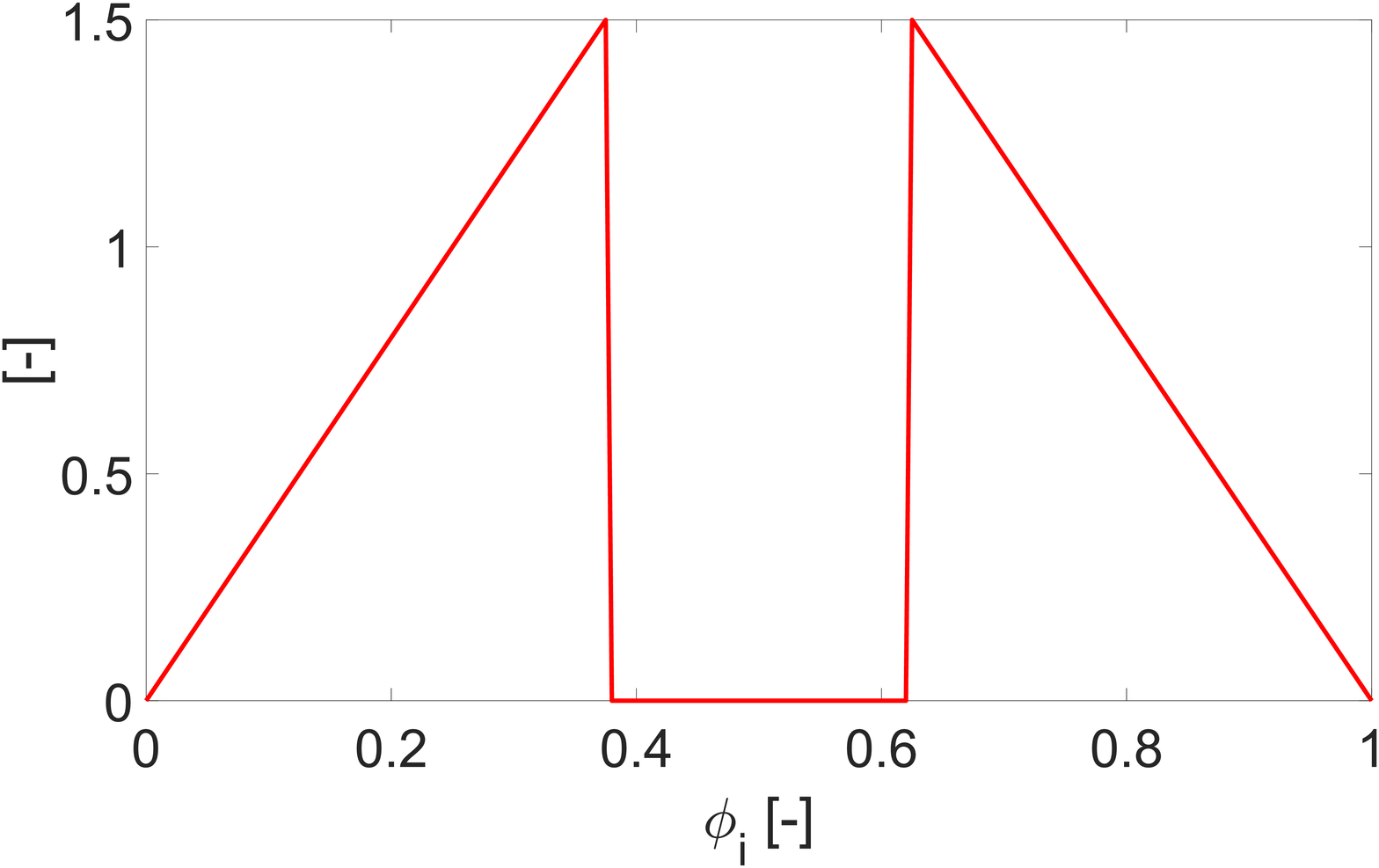}} 
\subfloat[$\mathcal{F}_{i}$ for $\psi_{i} = 0.5$.]{\includegraphics[width=0.25\textwidth]{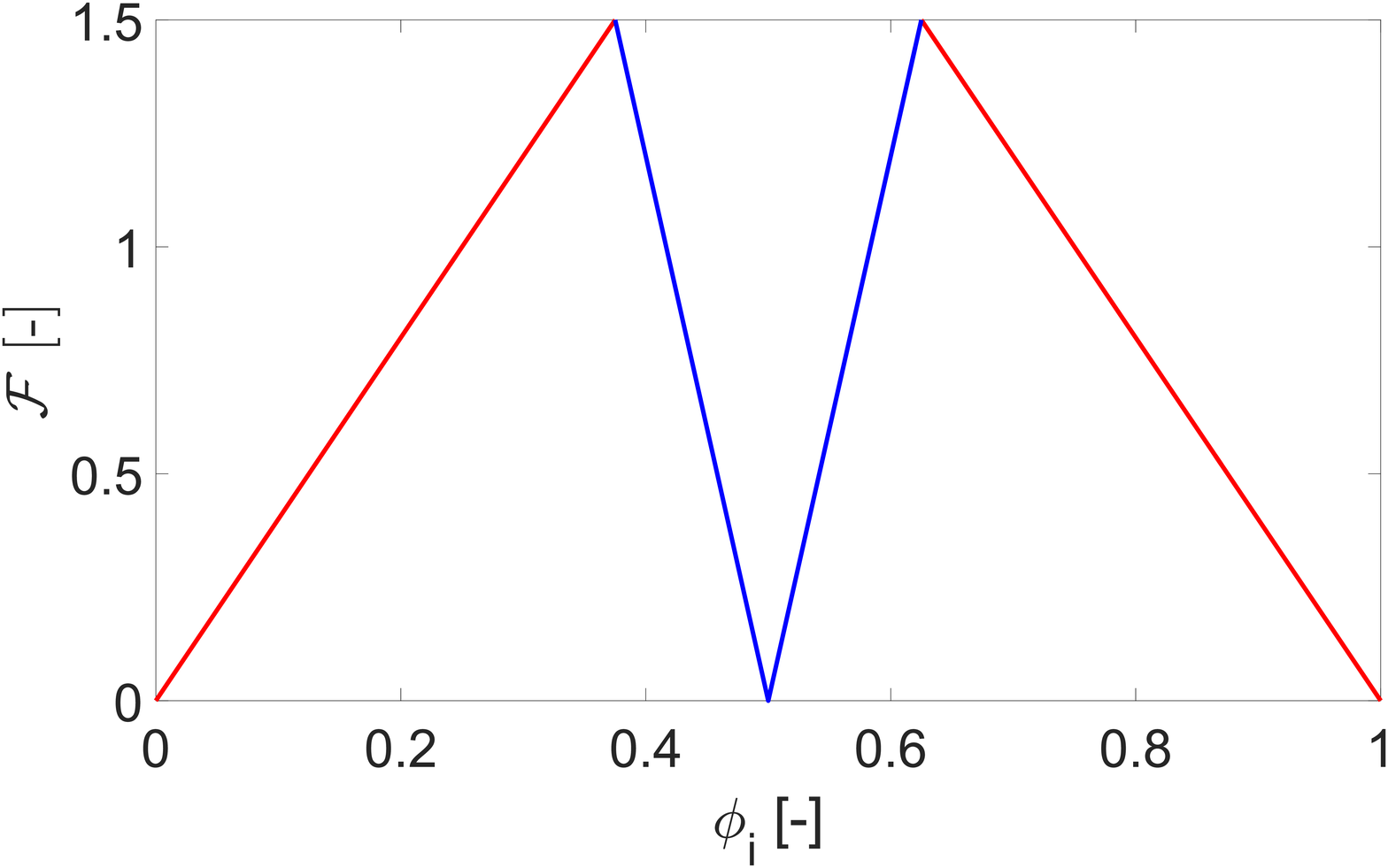}} 
\subfloat[$\mathcal{F}_{i}$ for $\psi_{i} = 0.8$.]{\includegraphics[width=0.25\textwidth]{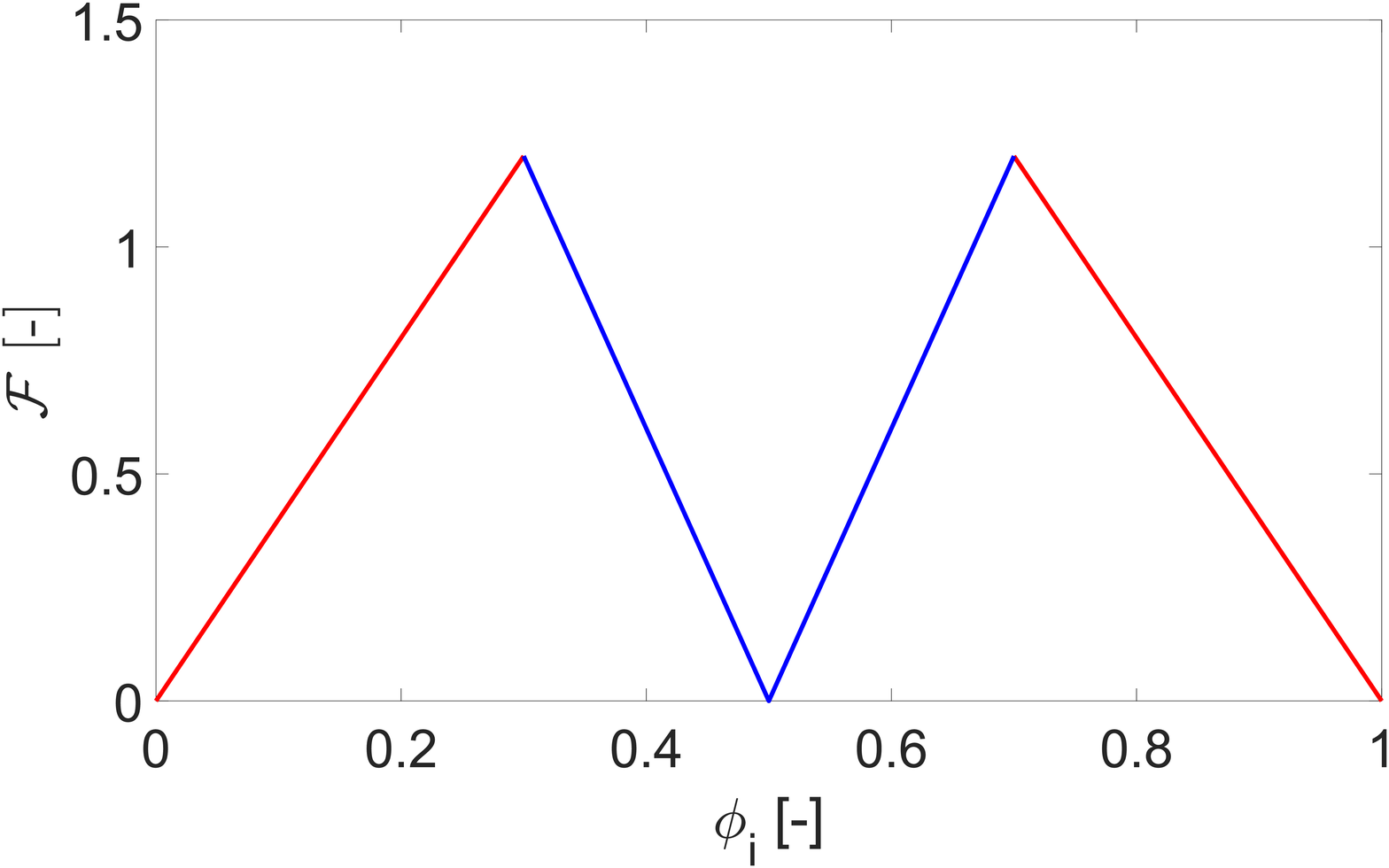}} 
\caption{Functions used in the transition zone to let a structural member split up into two members. } 
\label{Fig:proja.2}
\end{figure}

In $\Omega_{T,i}$ a structural member splits from a low periodicity $(\text{floor}(\tilde{\lambda_{i}}))$ to a higher periodicity $(\text{ceil}(\tilde{\lambda_{i}}))$, where both corresponding functions $\tilde{\rho_i}$ are shown in Figure~\ref{Fig:proja.2}(a). To model this splitting of members inside $\Omega_{T,i}$, we create function $\mathcal{F}_{i}$, which is a function of $\phi_{i}$, $\tilde{\lambda_{i}}$, and $\psi_{i}$. $\mathcal{F}_{i}$ consists of a base wave, on the interval $[0,2]$. This base wave is defined as  $\tilde{\rho_i}$ for a low periodicity $(i.e.~\text{floor}(\tilde{\lambda_{i}}))$ multiplied by 2, from which a part of width $1/2\psi_{i}$ is subtracted as is shown for $\psi_{i}= 0.5$ in Figure~\ref{Fig:proja.2}(b). Afterwards, another sawtooth-like wave is added to create $\mathcal{F}_{i}$. The distance between the peaks in $\mathcal{F}_{i}$ linearly increases with $\psi_{i}$, as can be seen in Figures~\ref{Fig:proja.2}(c)-(d).

\begin{figure}[h!]
\centering
\subfloat[$R^{*} = 0.8\varepsilon$.]{\includegraphics[width=0.3\textwidth]{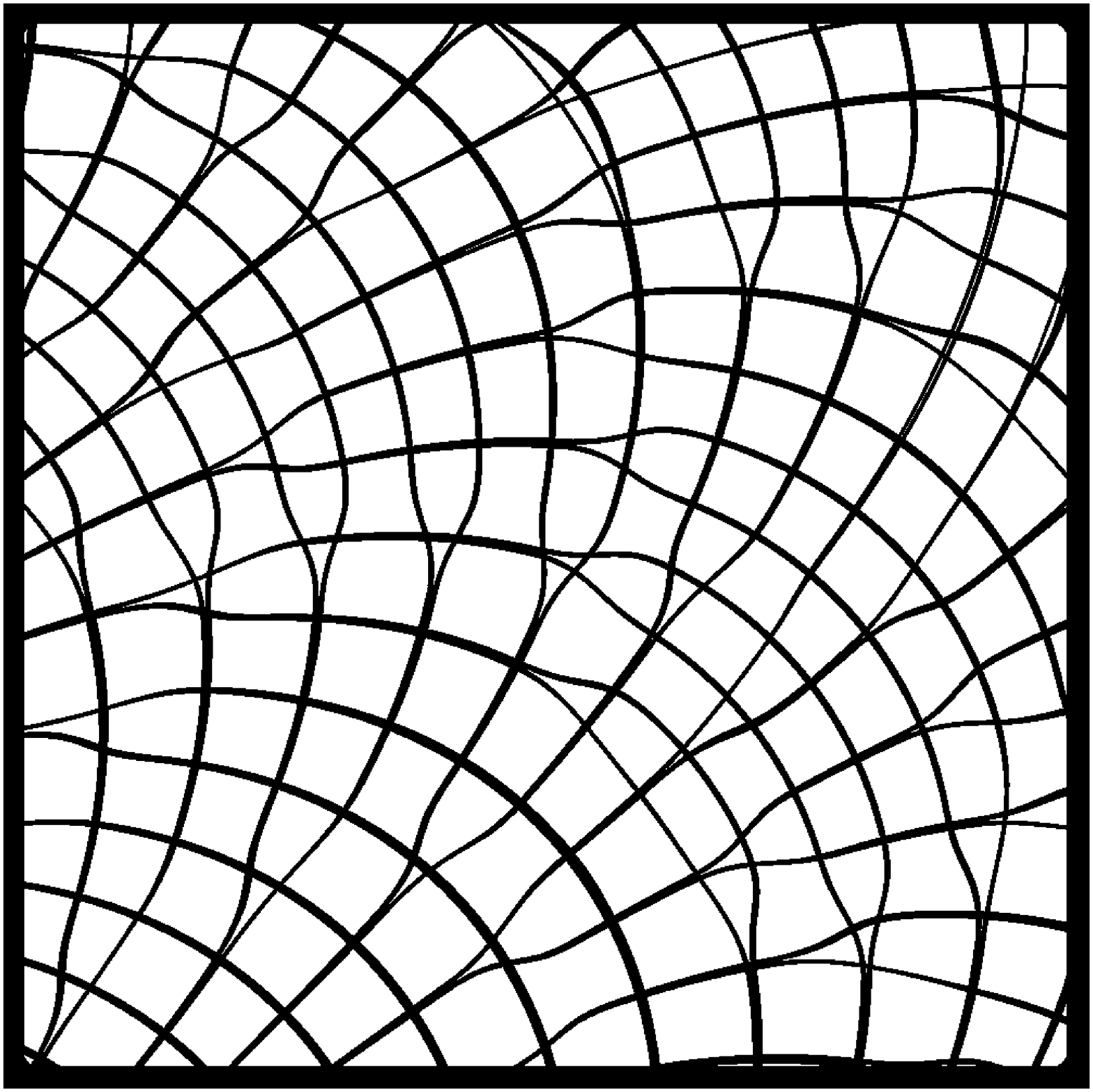}} \quad
\subfloat[$R^{*} = 1.2\varepsilon$.]{\includegraphics[width=0.3\textwidth]{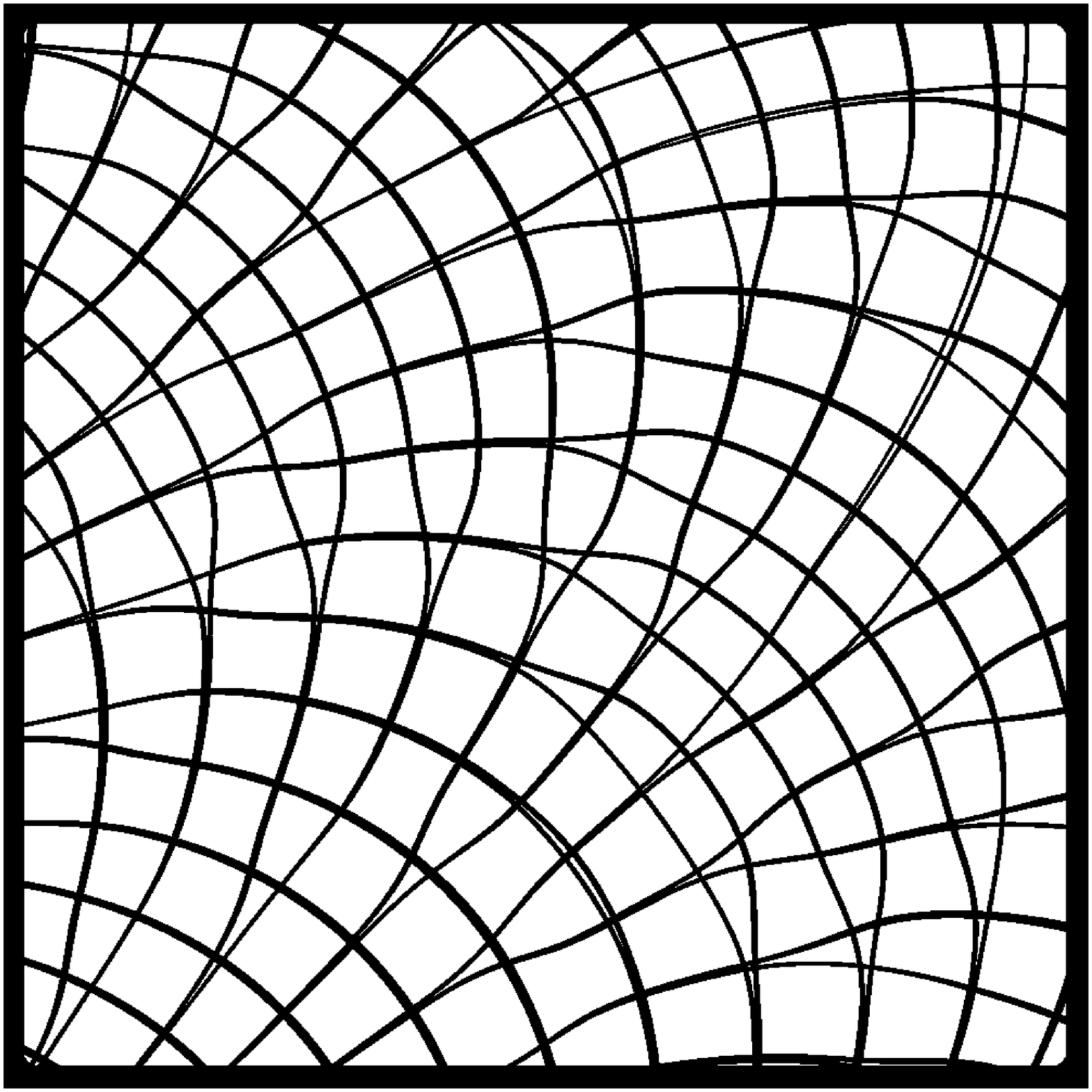}} \quad
\subfloat[$R^{*} = 1.6\varepsilon$.]{\includegraphics[width=0.3\textwidth]{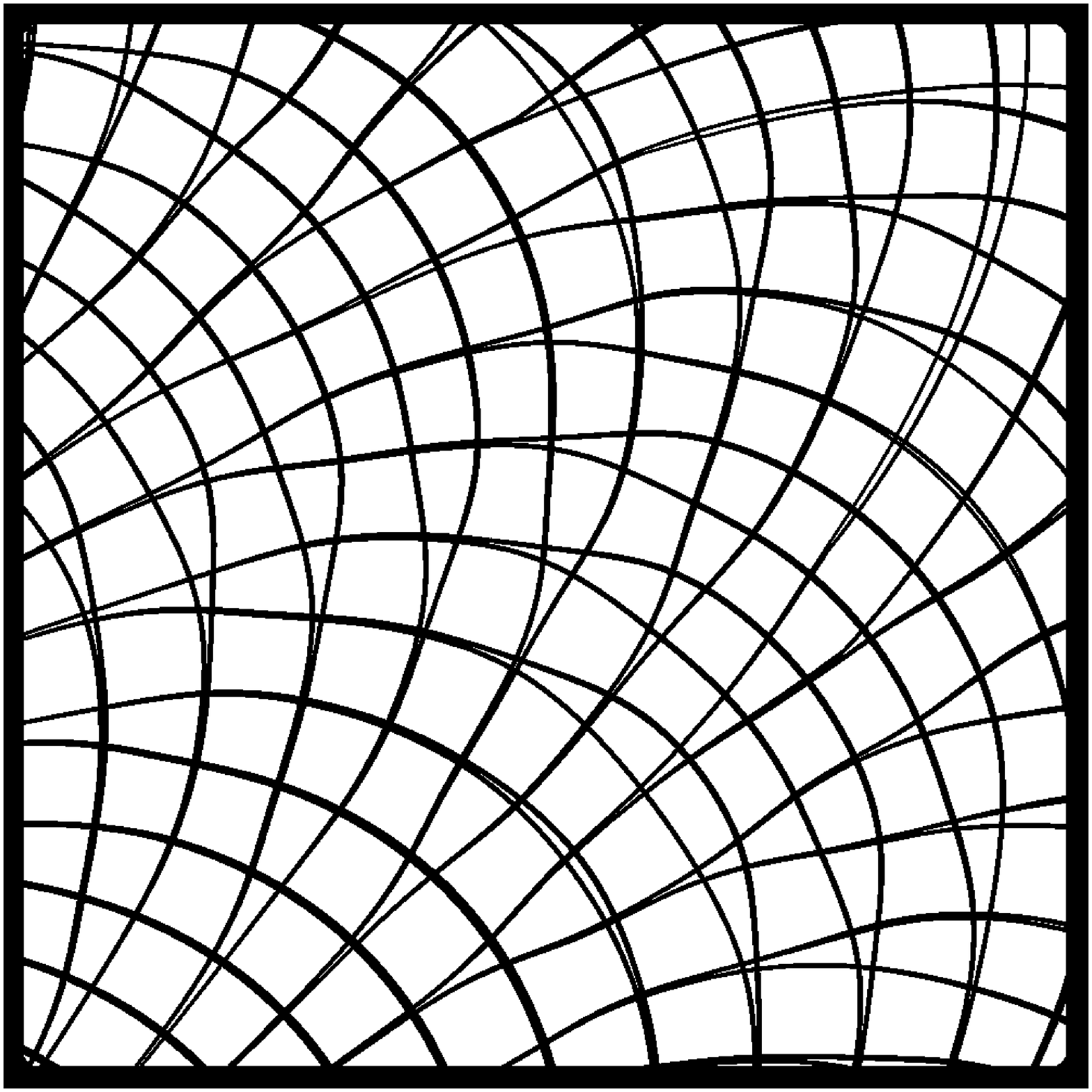}} 
\caption{Projection on a fine mesh $\mathcal{T}^{f}$ using various radii for the convolution kernel $R^{*}$, using $\varepsilon = 80~h^{f}$.} 
\label{Fig:proja.3}
\end{figure}

It has to be mentioned that Equation~\ref{Eq:proj.4} cannot be used to impose the exact widths on $\mathcal{F}_{i}$ to obtain $\rho_{i}$. The reason is that, due to the choice of base wave, the peak of $\mathcal{F}_{i}$ is close to 2 for $\psi \rightarrow 0$. Hence, we need to numerically determine the threshold for the Heaviside function that we use to obtain $\rho_{i}$ from $\mathcal{F}_{i}$, for a given combination of $a_{i}$ and $\psi_{i}$. To do so, a bi-section scheme is used. The projection of the structure from Figure~\ref{Fig:proj.1} using adaptive periodicity scaling can be seen for different values of $R^{*}$ in Figure~\ref{Fig:proja.3}(a)-(c). Here it can be observed that increasing $R^{*}$ leads to a more smooth transition for splitting of structural members. Numerical experiments have shown that $R^{*}=1.6\varepsilon$ gives the best results.

%% file: NumExp_proj.tex
\section{Numerical examples for projection of coated structures}
\label{Sec:NumExp.proj}
The presented projection approach, allows the coarse-scale optimized structure to be interpreted on a much finer mesh $\mathcal{T}^{f}$.
We will demonstrate the performance of the projected designs $\mathcal{J}^{\phi}$, and compare these results to the compliance of the homogenization-based designs on the coarse mesh $\mathcal{J}^{c}$. Three different sources are identified that can lead to a difference between $\mathcal{J}^{\phi}$ and $\mathcal{J}^{c}$: the effect of $h$-refinement, the interpolation of the coating from $\mathcal{T}^{c}$ onto $\mathcal{T}^{f}$, and the projection procedure. First, the effect of the former two, will be discussed. Afterwards, various numerical examples will be used to demonstrate the potential of the projection approach in terms of both performance and computational cost.

\subsection{Effect of mesh refinement and interpolation of coating}

To test the effect of $h$-refinement, we interpolate the designs optimized with problem form 1 from Table~\ref{Tab:NumExp.TO.1}, on a fine mesh ($3000\times1000$ elements) using nearest-neighbor interpolation. The compliance values of the homogenization-based design on the coarse mesh $\mathcal{J}^{c}$, will be compared to the compliance values of the homogenization-based design on the fine mesh $\mathcal{J}^{f}$.  Corresponding results are shown in Table~\ref{Tab:NumExp.proj.1}
\begin{table}[ht!]
\centering
\caption{Compliance on coarse $\mathcal{J}^{c}$, and on fine mesh $\mathcal{J}^{f}$, for the MBB-beam example using the same settings as in Table~\ref{Tab:NumExp.TO.1}, optimized with problem form 1, for several infill densities $m^{I}$. Different interpolation methods for the coating are used.}
\label{Tab:NumExp.proj.1}
\begin{tabular}{cccccccc}
\hline
 & Interpolation & $m^{I} = 0.4$ & $m^{I} = 0.5$ & $m^{I}=0.6$ & $m^{I}=0.7$ & $m^{I}=0.8$ & $m^{I}=0.9$ \\ \hline
$\mathcal{J}^{c}$& - & $318.45$ & $291.14$ & $274.92$ & $266.85$  & $251.38$ & $236.37$ \\  
$\mathcal{J}^{f}$& nearest-neighbor  & 320.17 & 295.52  & 279.40  & 270.27 & 256.75 &  239.82 \\ 
$\mathcal{J}^{f}$& linear & 328.34 & 305.32  & 286.30  & 273.18 & 258.39 &  239.09 \\ 
 \end{tabular}
\end{table}

\begin{figure}[h!]
\centering
\subfloat[nearest-neighbor, $\mathcal{J}^{f}=295.52$.]{\includegraphics[width=0.4\textwidth]{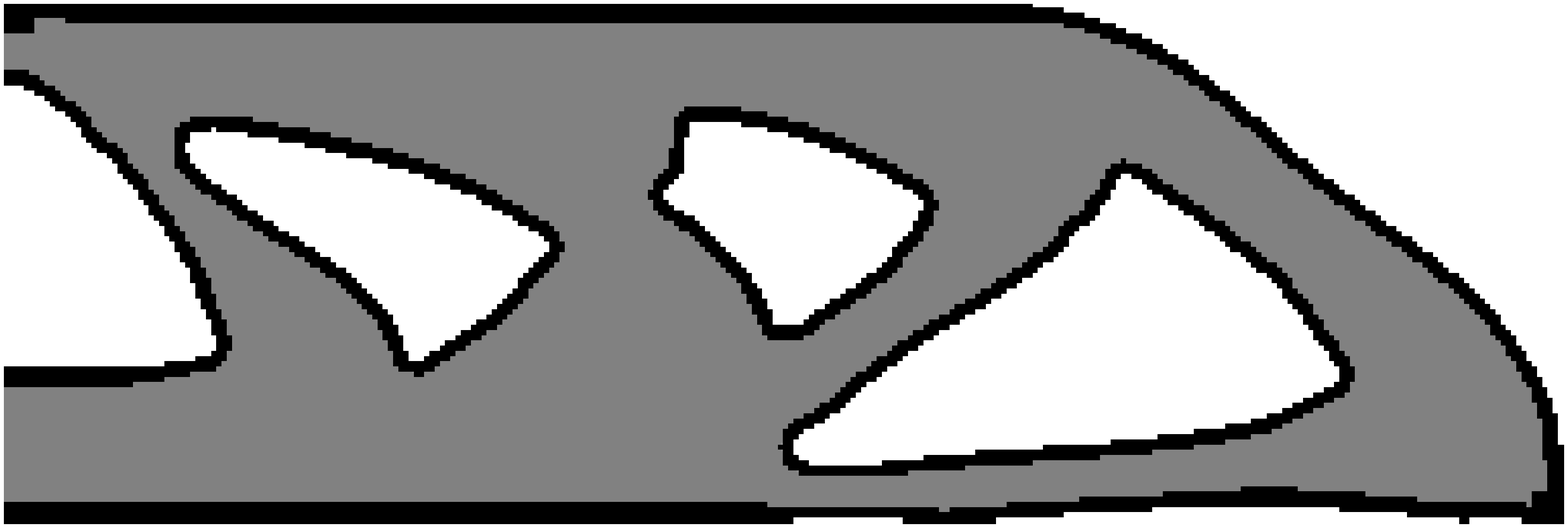}} \quad
\subfloat[linear, $\mathcal{J}^{f}=305.32$.]{\includegraphics[width=0.4\textwidth]{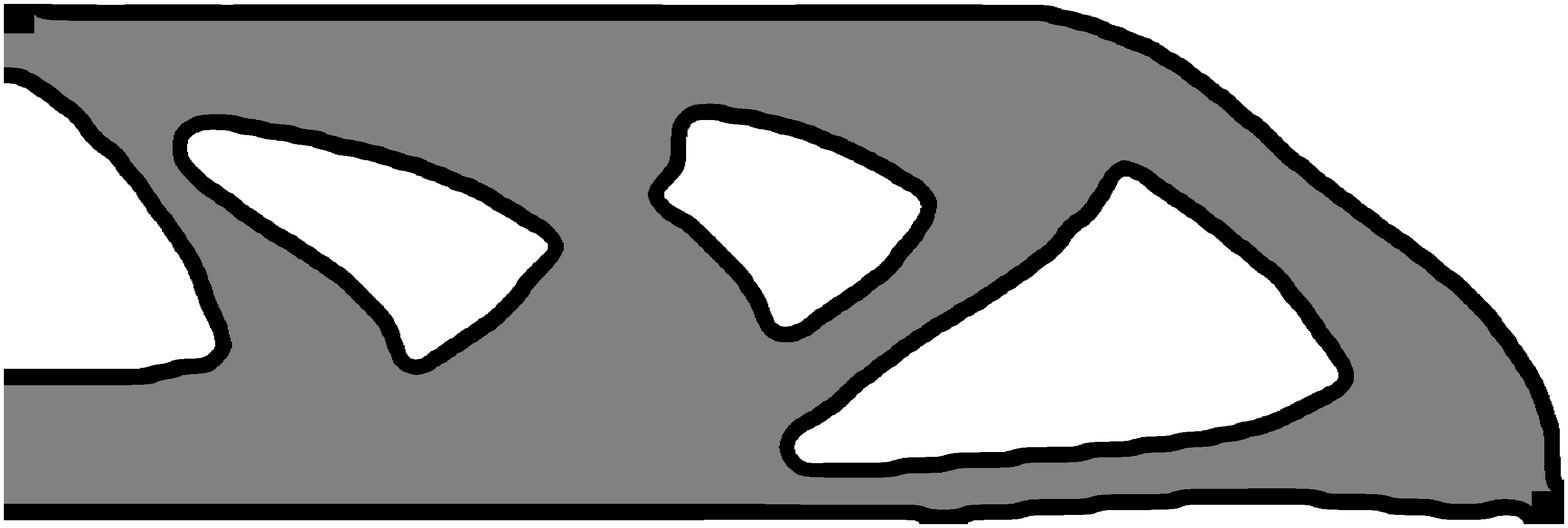}} 
\caption{Density distribution for the MBB-beam example interpolated on a fine mesh of $3000\times1000$ elements, using different interpolation methods for $\left\vert\left\vert\nabla\tilde{\varphi} \right\vert\right\vert_{\alpha}$. The structure optimized using problem form 1, with $m^{I}=0.5$.} 
\label{Fig:NumExp.proj.1}
\end{figure}

As expected $\mathcal{J}^{f}$ is larger than $\mathcal{J}^{c}$, due to more accurate displacement calculation. Furthermore, the effect of the nearest-neighbor interpolation can be seen in Figure~\ref{Fig:NumExp.proj.1}(a), where the jagged edges of the coating correspond to $10$ elements. To avoid these jagged-edges, we use a linear interpolation to map the normalized gradient norm $\left\vert\left\vert\nabla\tilde{\varphi} \right\vert\right\vert_{\alpha}$ onto the fine mesh. The corresponding values of $\mathcal{J}^{f}$ demonstrating the effect of this interpolation are shown in Table~\ref{Tab:NumExp.TO.1} as well.

It can be observed that interpolation of $\left\vert\left\vert\nabla\tilde{\varphi} \right\vert\right\vert_{\alpha}$ using nearest-neighbor interpolation, leads to a better performance than linear interpolation. This seems rather counter-intuitive, since the latter method results in a smooth coating without jagged edges as can be seen in~\ref{Fig:NumExp.proj.1}(b). However, it can be seen that the use of linear interpolation causes somewhat strange jumps close to the bottom right boundary  condition, which is assumed to be the cause of the increased error between $\mathcal{J}^{c}$ and $\mathcal{J}^{f}$.

Despite a slightly increased compliance, linear interpolation is used for the mapping of $\left\vert\left\vert\nabla\tilde{\varphi} \right\vert\right\vert_{\alpha}$ onto $\mathcal{T}^{f}$. The reason is that the reference coating thickness is more uniform using this interpolation method.
A future remedy can be to define the coating and interface on a much finer mesh than the coarse analysis mesh using a multi-resolution approach, e.g.~\citep{Bib:NguyenMTOP1,Bib:GroenFCMTOP}.

\subsection{Effect of the different projection procedures}
\begin{figure}[h!]
\centering
\subfloat[$\varepsilon=20h^{f}$, $\gamma=10$, $\mathcal{J}^{\phi}=315.56$ and $V_{\phi}=0.399$.]{\includegraphics[width=0.3\textwidth]{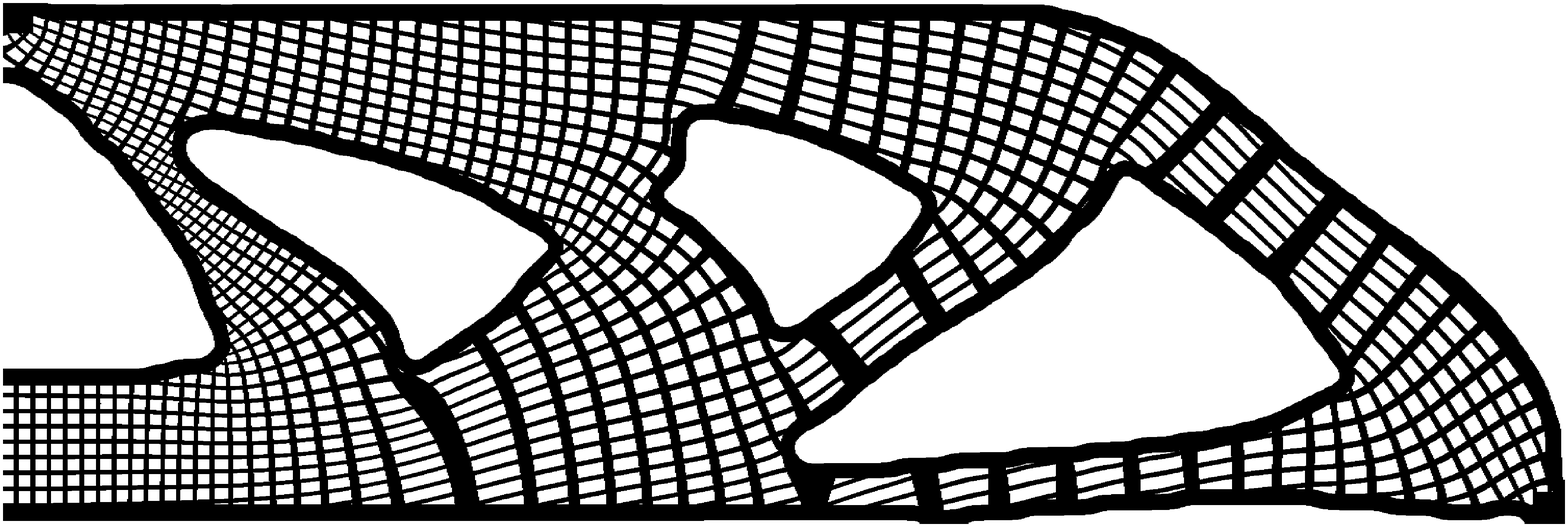}} \quad
\subfloat[$\varepsilon=30h^{f}$, $\gamma=10$,  $\mathcal{J}^{\phi}=316.48$ and $V_{\phi}=0.397$.]{\includegraphics[width=0.3\textwidth]{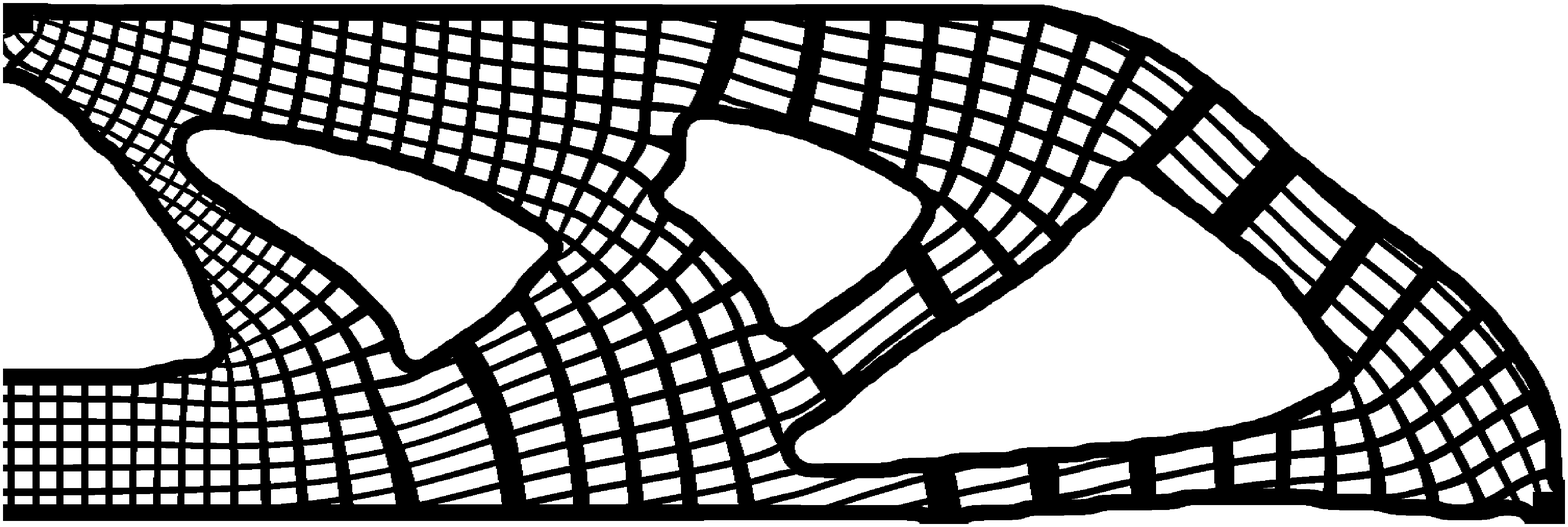}} \quad
\subfloat[$\varepsilon=40h^{f}$, $\gamma=10$, $\mathcal{J}^{\phi}=319.62$ and $V_{\phi}=0.396$.]{\includegraphics[width=0.3\textwidth]{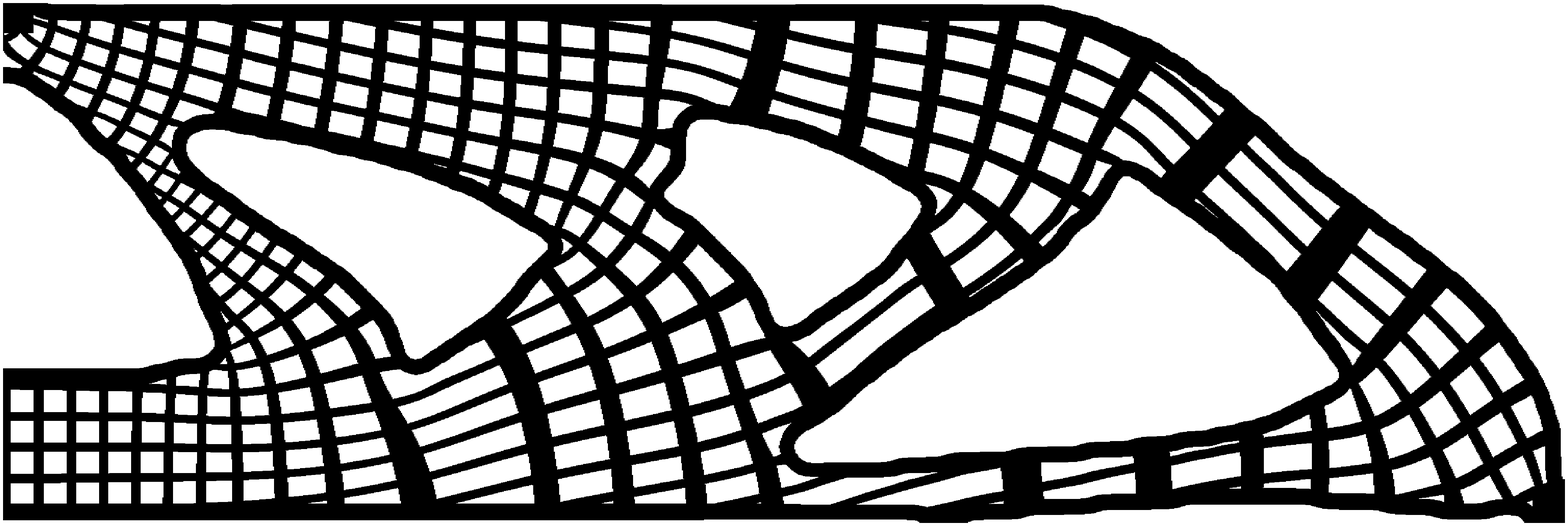}}  \\
\subfloat[$\varepsilon=20h^{f}$, $\gamma=10^3$, $\mathcal{J}^{\phi}= 307.93$ and $V_{\phi}=0.397$.]{\includegraphics[width=0.3\textwidth]{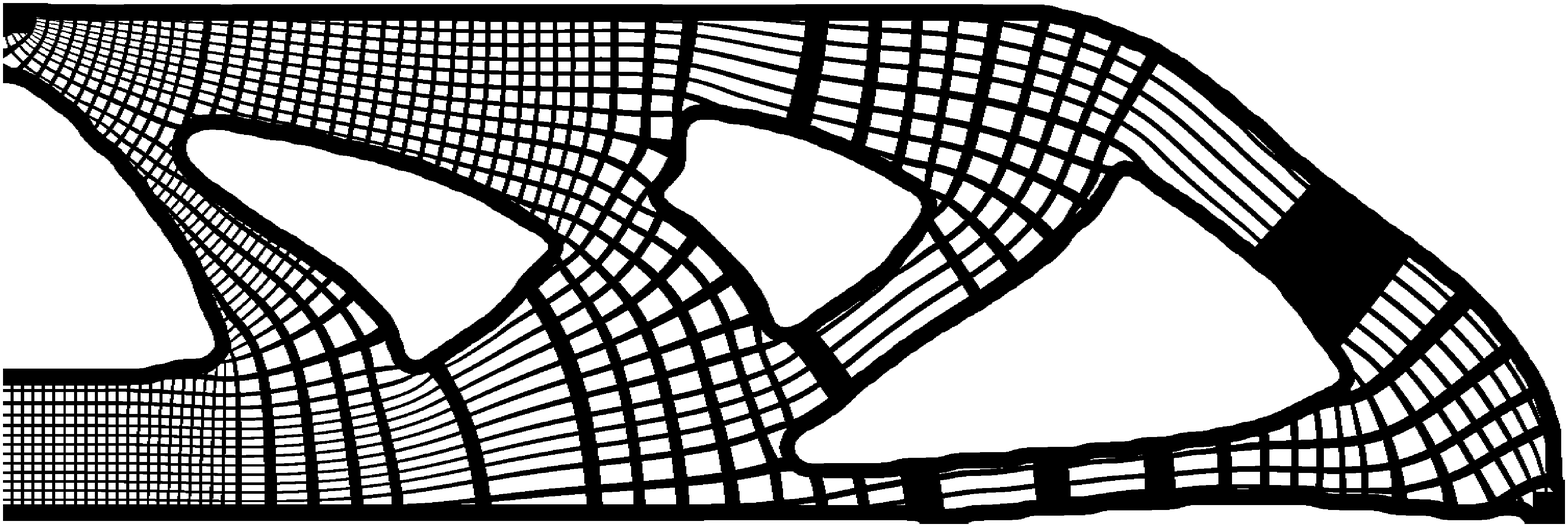}} \quad
\subfloat[$\varepsilon=30h^{f}$, $\gamma=10^3$, $\mathcal{J}^{\phi}= 304.79$ and $V_{\phi}=0.400$.]{\includegraphics[width=0.3\textwidth]{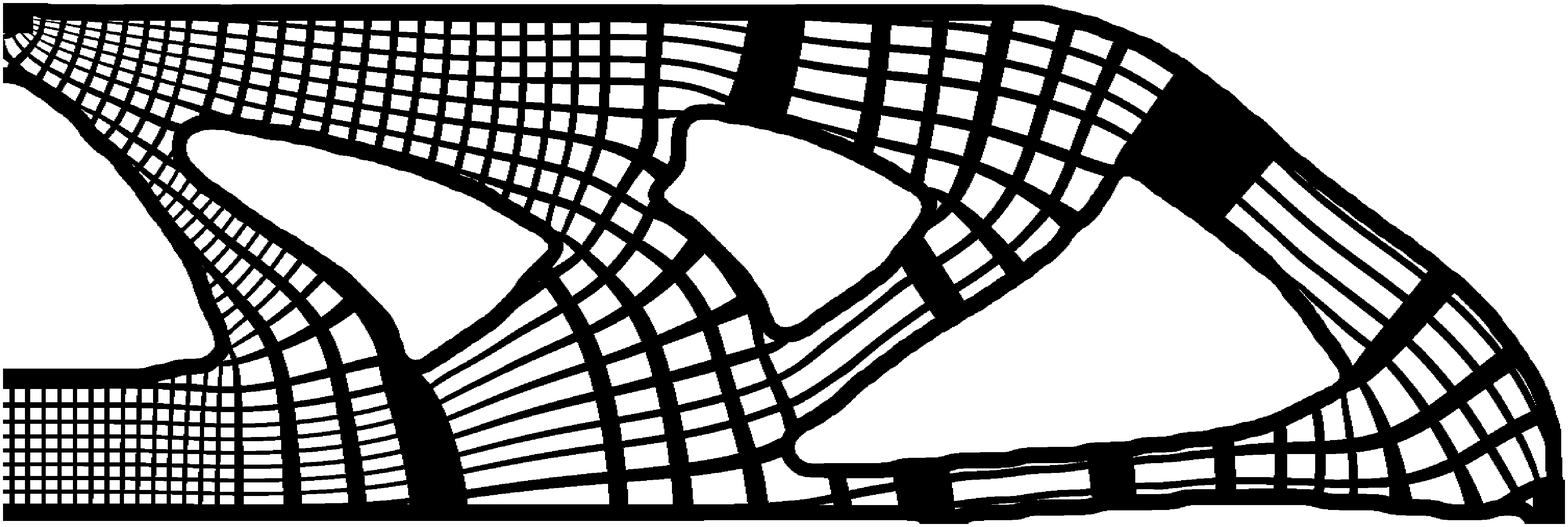}} \quad
\subfloat[$\varepsilon=40h^{f}$, $\gamma=10^3$, $\mathcal{J}^{\phi}=301.99$ and $V_{\phi}=0.411$.]{\includegraphics[width=0.3\textwidth]{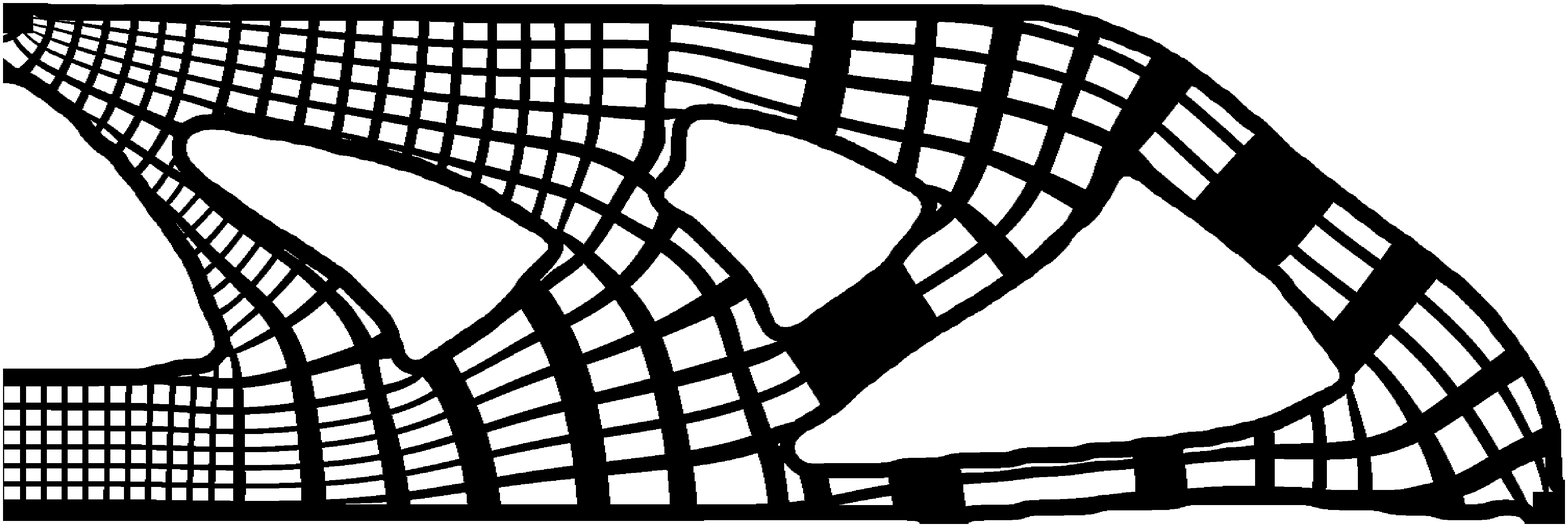}} 
\caption{Projection of the MBB-beam example for a problem of type 1, using $m^{I}=0.5$, on a fine mesh of $3000\times1000$ elements, where $\mathcal{J}^{f}= 305.32$. No adaptive periodicity is used, results are shown for various values of constraint importance $\gamma$ and $\varepsilon$.} 
\label{Fig:NumExp.proj.2}
\end{figure}

To demonstrate the performance of the projection method, we consider the MBB-beam example optimized using problem form 1, for $m^{I}=0.5$. The structure is projected on a fine mesh ($3000\times1000$ elements), without the adaptive periodicity approach. We perform the projection for two different values of $\gamma$, $\gamma=10$ and $\gamma=10^3$ to demonstrate the effect of the constraint enforcement. Furthermore, we project the structure for 3 different average unit-cell sizes $\varepsilon$. The corresponding projected structures, the compliance $\mathcal{J}^{\phi}$ and the volume of the projected structures $V_{\phi}$ are shown in Figure~\ref{Fig:NumExp.proj.2}.

From the bottom row it can be observed that a strong constraint enforcement ($\gamma = 10^{3}$) leads to structures performing very close to $\mathcal{J}^{f}$, $i.e.$ within $1\%$! Unfortunately however, we can identify a locally very distorted periodicity. These stretched unit-cells can lead to $V_{\phi}$ exceeding imposed volume constraint $V_{max}$, and do not ensure a uniform infill. Lowering the constraint enforcement to $\gamma = 10$, as is shown in the top row, yields much more regular infill patterns; however, at the cost of a slightly reduced performance.

\begin{figure}[h!]
\centering
\subfloat[$\varepsilon=20h^{f}$, $\gamma=10^{3}$, $\mathcal{J}^{\phi}=307.58$ and $V_{\phi}=0.398$.]{\includegraphics[width=0.3\textwidth]{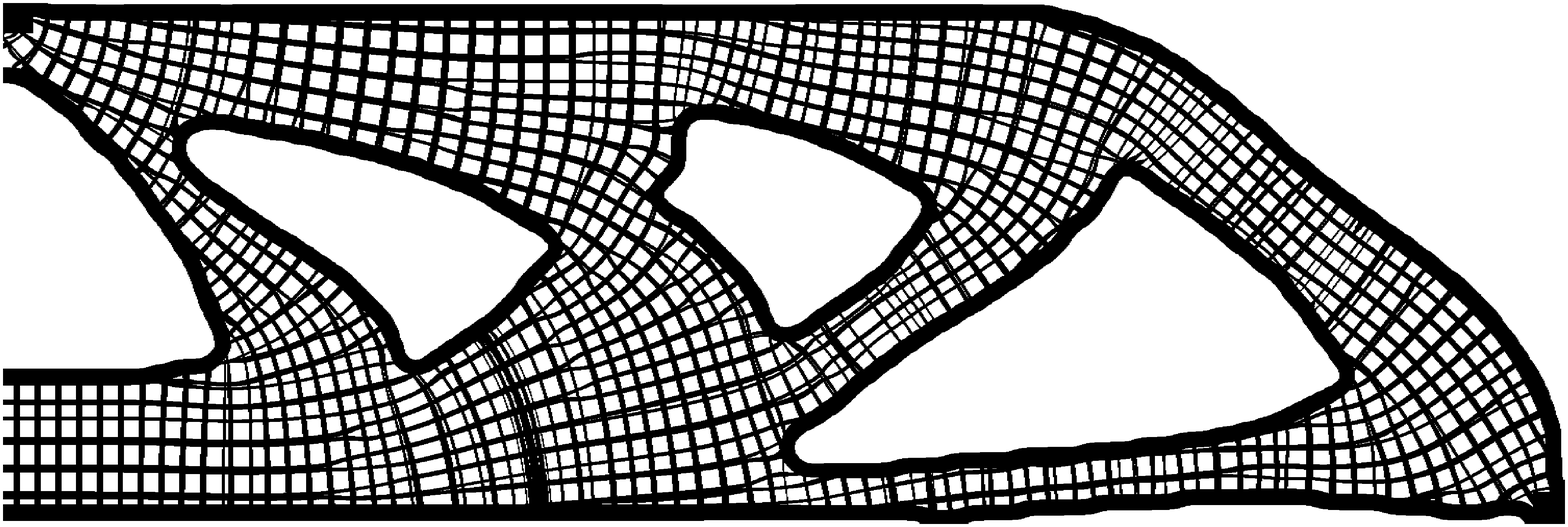}} \quad
\subfloat[$\varepsilon=30h^{f}$, $\gamma=10^{3}$,  $\mathcal{J}^{\phi}=306.41$ and $V_{\phi}=0.400$.]{\includegraphics[width=0.3\textwidth]{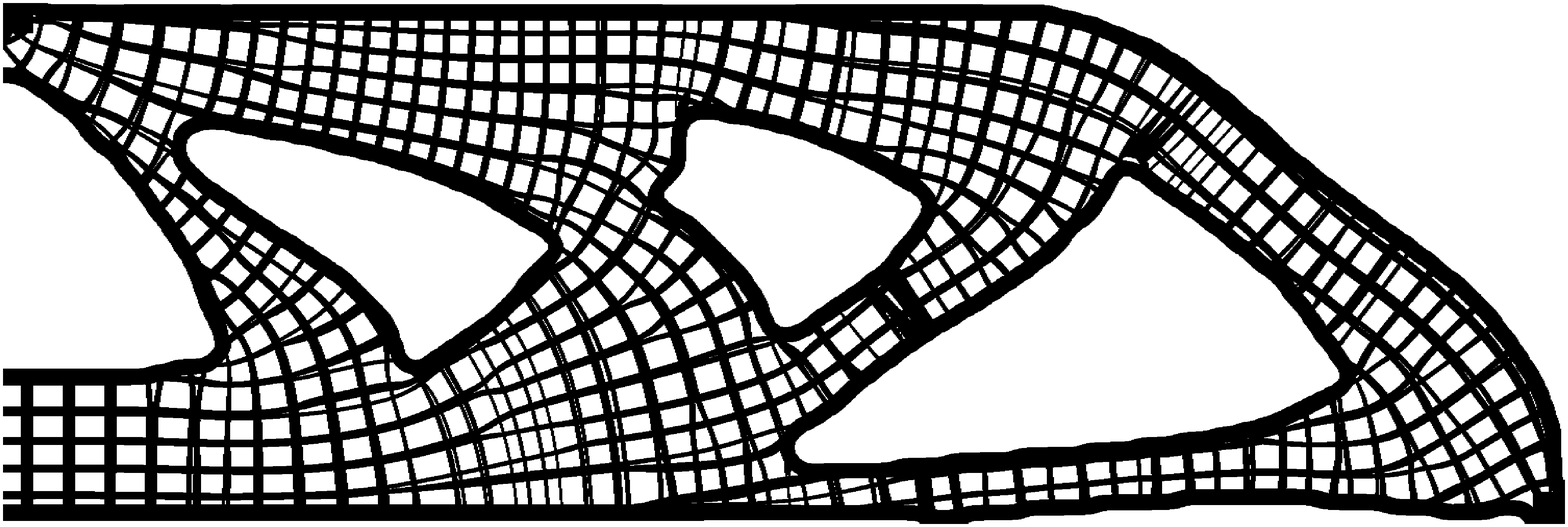}} \quad
\subfloat[$\varepsilon=40h^{f}$, $\gamma=10^{3}$, $\mathcal{J}^{\phi}=301.23$ and $V_{\phi}=0.406$.]{\includegraphics[width=0.3\textwidth]{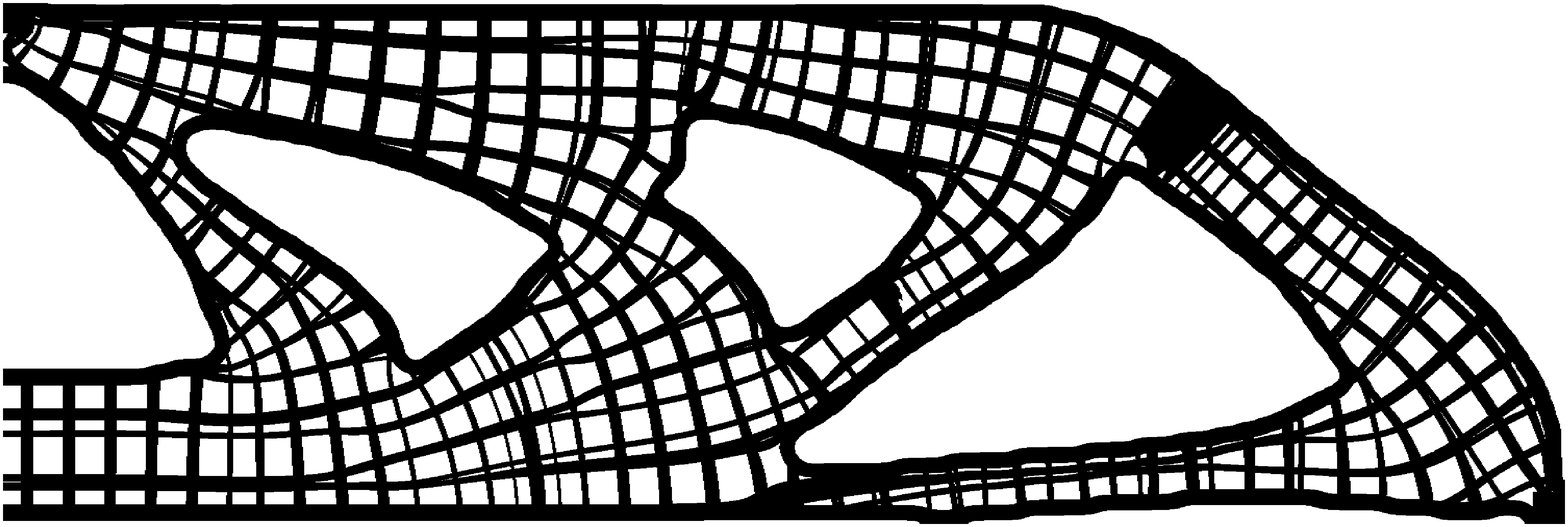}}  \\
\subfloat[$\varepsilon=30h^{f}$, $\gamma=10^{2}$, $\mathcal{J}^{\phi}=304.58$ and $V_{\phi}=0.400$.]{\includegraphics[width=0.3\textwidth]{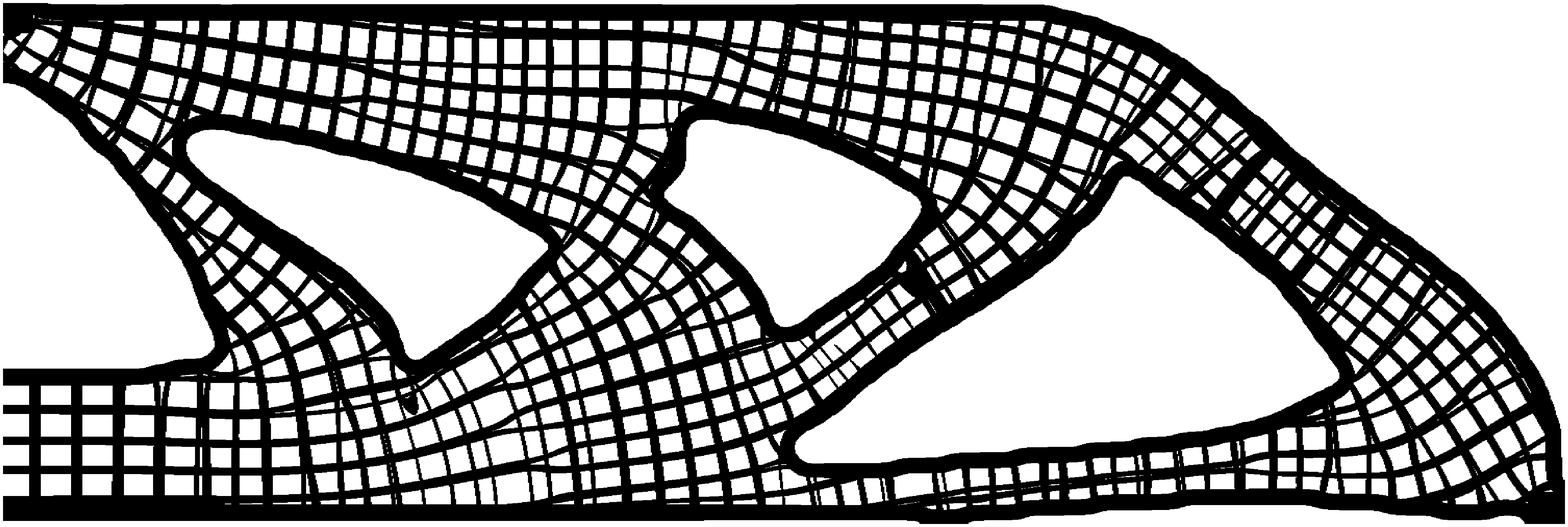}} \quad
\subfloat[$\varepsilon=30h^{f}$, $\gamma=10^{1.5}$,  $\mathcal{J}^{\phi}=308.67$ and $V_{\phi}=0.399$.]{\includegraphics[width=0.3\textwidth]{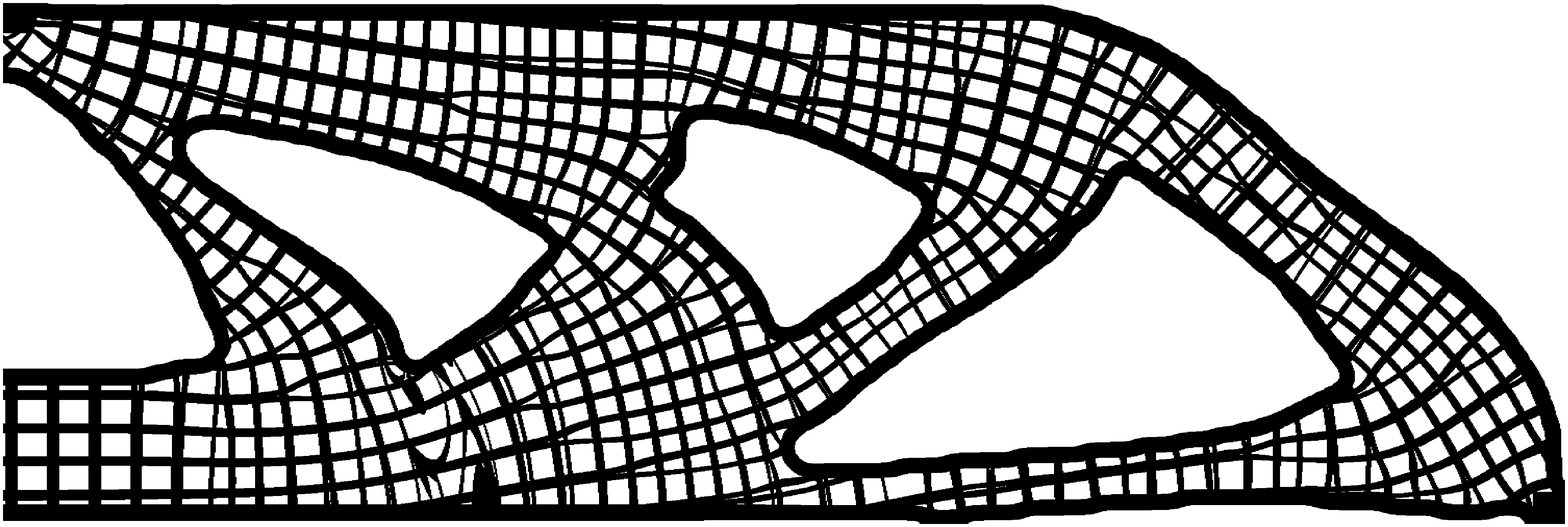}} \quad
\subfloat[$\varepsilon=30h^{f}$, $\gamma=10$, $\mathcal{J}^{\phi}=320.22$ and $V_{\phi}=0.399$.]{\includegraphics[width=0.3\textwidth]{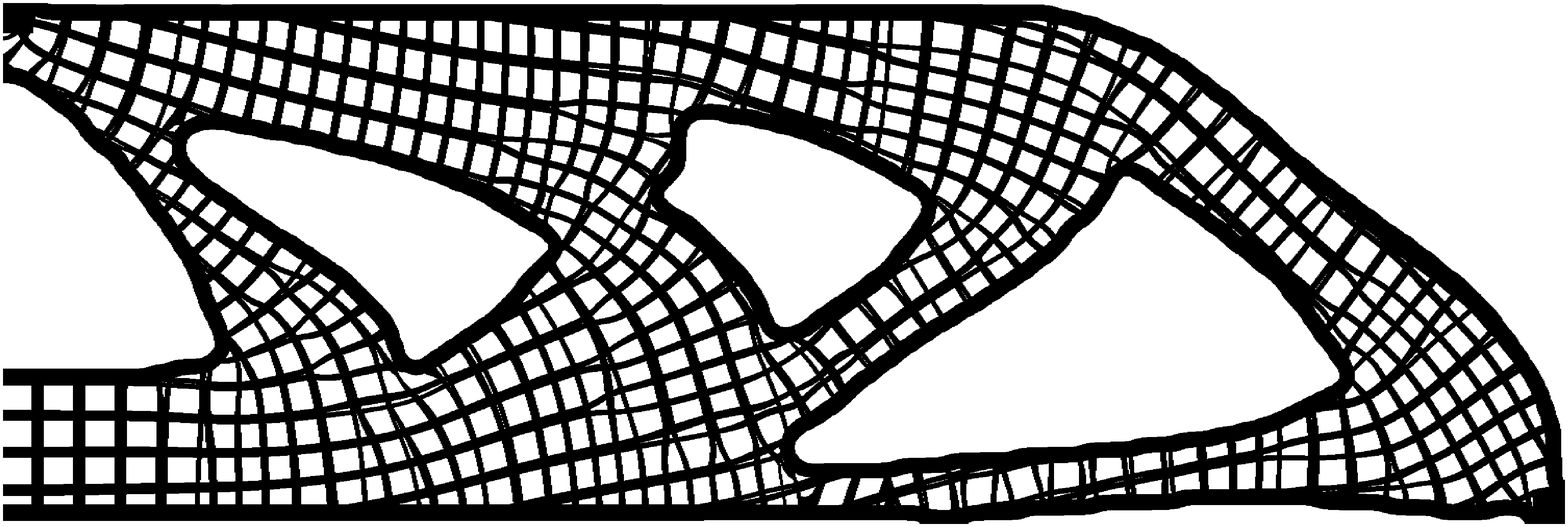}}  
\caption{Projection of the MBB-beam example for a problem of type 1, using $m^{I}=0.5$, on a fine mesh of $3000\times1000$ elements, where $\mathcal{J}^{f}= 305.32$. Adaptive periodicity projection is used, results are shown for various values of constraint importance $\gamma$ and $\varepsilon$.} 
\label{Fig:NumExp.proj.3}
\end{figure}

To use the best of both worlds, (exact angle enforcement, and regular unit-cells), we use the adaptive periodicity projection approach as is proposed in the previous section. The projected structures for $\gamma=10^{3}$ and different spacings can be seen in Figures~\ref{Fig:NumExp.proj.3}(a)-(c). 

It can be seen that values for $\mathcal{J}^{\phi}$ are again within $1\%$ of $\mathcal{J}^{f}$, while a more uniform infill is maintained. Unfortunately, the proposed adaptive periodicity mapping procedure does not work perfectly yet when $\tilde{\lambda_{i}}$ is rapidly changing. This can be best seen in the top right of the MBB-beam shown in Figure~\ref{Fig:NumExp.proj.3}(c), where the periodicity is locally undergoing a large change, such that there is no space to form nice and clear branches between different periodicities. 

To prevent large jumps in periodicity the angle enforcement can be relaxed; therefore, the projection using $\varepsilon = 30h^{f}$ is shown for various values of $\gamma$ in Figures~\ref{Fig:NumExp.proj.3}(d)-(f). 
Here it can be seen that $\gamma=10$, is the only value that completely prevents any of these local effects.
Despite being slightly misaligned with the optimal orientation, the effect on the performance is small and the projected structures perform still within $5\%$ of the homogenization-based designs.

To conclude the proposed adaptive periodicity projection approach in combination with using $\gamma=10^{3}$ shows a clear potential. The structures perform almost identically to the homogenization-based designs; while the projected structures consist of near-regular infill.  
In a future work, different formulations for adapting the periodicity in regions with large changes in periodicity will be further investigated, to improve this new and promising method even more.

\subsection{Comparison with coated structures optimized using SIMP}
\begin{figure}[h!]
\centering
\subfloat[Infill design using SIMP.]{\includegraphics[width=0.4\textwidth]{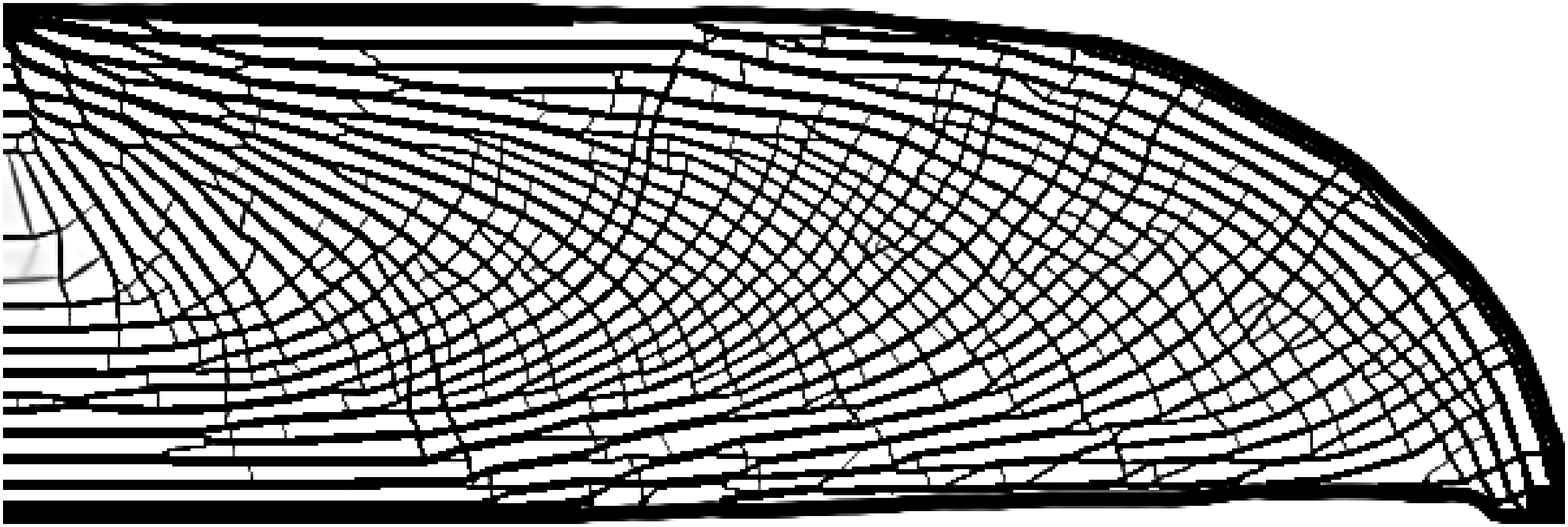}} \quad
\subfloat[Proposed projection procedure.]{\includegraphics[width=0.4\textwidth]{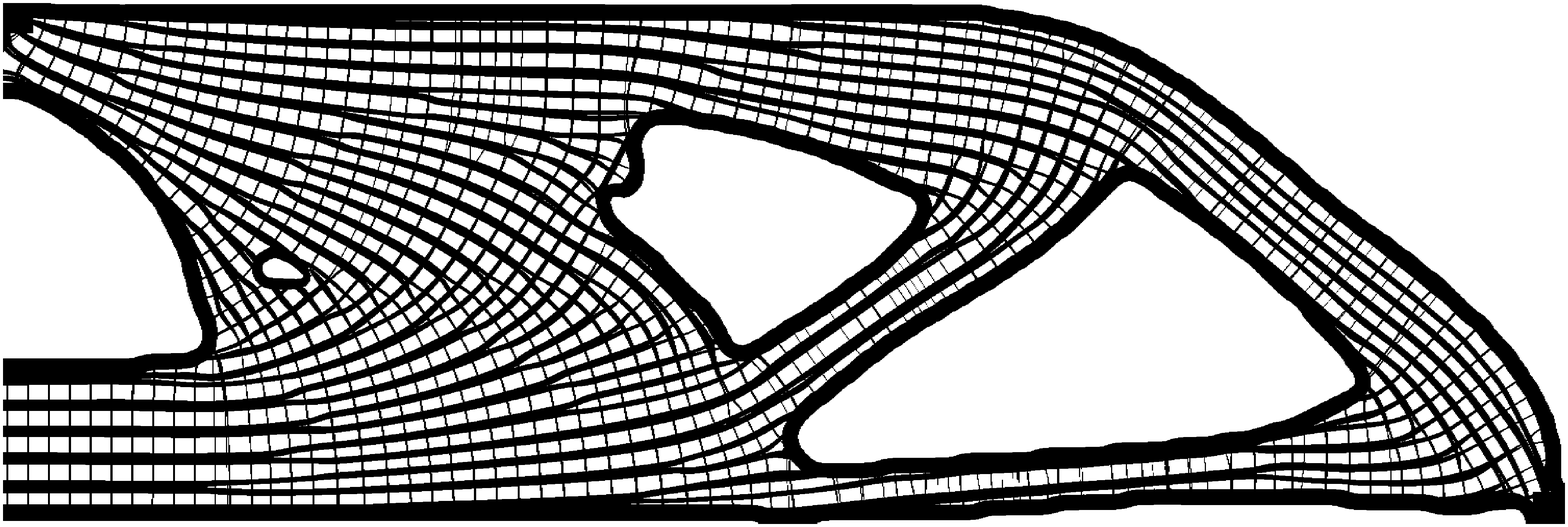}}
\caption{Comparison of the MBB-beam example using a SIMP-based approach~\citep{Bib:WuClausenSigmund2017}, and the proposed projection based approach for the infill. Uniform infill density with $m^{I}=0.5$ is used while the microstructure is allowed to vary (problem form 2).} 
\label{Fig:NumExp.proj.4}
\end{figure}

To demonstrate the performance of the approach proposed in this work, we compare it to the work of~\citet{Bib:WuClausenSigmund2017}.
In this approach a coated structure is created, where the infill is optimized using a density-based approach as can be seen for the MBB-beam example in Figure~\ref{Fig:NumExp.proj.4} (a). Here, a discretization of $600\times200$ elements is used, with $R_{1} = 0.075~L$ and $t_{ref}=0.015~L$; furthermore, $V_{max} = 0.4$, $m^{I} = 0.5$.
  
The method produces coated structures, where the infill has a locally uniform density, with no restriction on the freedom of the microstructure (close to an optimization problem of form 2). The optimization required a total time $T_{tot}=292$ minutes using $600$ design iterations. Hence, a more detailed design on a finer mesh is deemed computationally too expensive. The proposed projection procedure does not have a strong restriction on the level of detail of the projected shape, since the topology optimization can be performed on a relatively coarse mesh. The design optimized for problem form 2, with $m^{I}=0.5$ and $a_{u} = 0.9$ is projected on a fine mesh of $3000\times1000$ elements, using $\gamma=10^{3}$, $\varepsilon = 20h^{f}$ and the adaptive periodicity projection approach as can be seen in Figure~\ref{Fig:NumExp.proj.4}(b). 

For a fair comparison the structure optimized using the approach of~\citet{Bib:WuClausenSigmund2017} is mapped on a fine mesh of $3000\times1000$ elements using nearest-neighbor interpolation. The compliance values for both approaches, for various infill densities $m^{I}$ are shown in Table~\ref{Tab:NumExp.proj.2} and Table~\ref{Tab:NumExp.proj.3}. Furthermore, a breakdown of the computational cost in optimization time $T_{opt}$, mapping time $T_{\phi}$ as well as the total time $T_{tot}$ is shown. It has to be noted that all simulations are done using a single processor MATLAB code on a standard PC running Windows 7.
\begin{table}[ht!]
\centering
\caption{Compliance values for optimization mesh $\mathcal{J}^{c}$ and for fine mesh $\mathcal{J}^{f}$, as well as volume fraction $V$ and optimization time $T_{opt}$ (shown in [hh:mm:ss]), for the MBB-beam example using the approach by~\citet{Bib:WuClausenSigmund2017}.}
\label{Tab:NumExp.proj.2}
\begin{tabular}{ccccc}
\hline
$m^{I}$ & $\mathcal{J}^{c}$ & $\mathcal{J}^{f}$& $V$ & $T_{tot}$ \\ \hline
0.5 & 270.21 & 281.95 & 0.394 & 04:52:00 \\
0.6 & 241.36 & 247.54 & 0.400 & 04:58:00 \\
0.7 & 227.84 & 231.94 & 0.400 & 05:00:00 \\
 \end{tabular}
\end{table}

\begin{table}[ht!]
\centering
\caption{Compliance values for optimization mesh $\mathcal{J}^{c}$, fine mesh $\mathcal{J}^{f}$ and projected design $\mathcal{J}^{\phi}$, as well as volume fraction $V_{\phi}$ and time breakdown (shown in [hh:mm:ss]), for the MBB-beam example optimized using problem form 2 and projected using adaptive periodicity.}
\label{Tab:NumExp.proj.3}
\begin{tabular}{ccccccccc}
\hline
$m^{I}$ & $\varepsilon$ & $\mathcal{J}^{c}$ & $\mathcal{J}^{f}$ & $\mathcal{J}^{\phi}$ & $V_{\phi}$& $T_{opt}$ & $T_{\phi}$ & $T_{tot}$ \\ \hline
0.5  & $20h^{f}$ & $247.52$ & 256.31 & 261.96 & 0.397& 00:23:00& 00:00:20 & 00:23:20 \\
0.5  & $30h^{f}$ & $247.52$ & 256.31 & 260.68 &0.401 & 00:23:00 & 00:00:25 &00:23:25 \\
0.5 & $40h^{f}$ & $247.52$ & 256.31 & 257.61 & 0.409 & 00:23:00 & 00:00:30 & 00:23:30 \\ \hline
0.6 & $20h^{f}$ & $234.30$ & 242.53 & 242.03 & 0.400 & 00:21:19 & 00:00:20 & 00:21:39 \\
0.6 & $30h^{f}$ & $234.30$ & 242.53 & 239.29 & 0.402 & 00:21:19 & 00:00:25 & 00:21:44 \\
0.6 & $40h^{f}$ & $234.30$ & 242.53 & 239.21 &0.404 & 00:21:19& 00:00:30 & 00:21:49\\ \hline
0.7 & $20h^{f}$ & $227.02$ & $230.83$ & 233.92 & 0.400 & 00:21:11 & 00:00:19&00:21:30 \\
0.7 & $30h^{f}$ & $227.02$ & $230.83$ & 231.26 & 0.404 & 00:21:11 & 00:00:24 & 00:21:35\\
0.7 & $40h^{f}$ & $227.02$ & $230.83$ & 230.26 & 0.405& 00:21:11 & 00:00:29 & 00:21:40\\
 \end{tabular}
\end{table}
It can be seen that the compliance values of the structures optimized using the mapping approach are in general lower than the compliance values obtained using the approach of~\citet{Bib:WuClausenSigmund2017}. The larger $m^{I}$, the smaller the difference between the compliances obtained using both methods. Furthermore, it can be observed that a larger value of $\varepsilon$ results in a slightly larger volume of the projected structures, which results in a lower compliance. Another benefit of the projection approach, is that it results in clear solid and void structures, while the infill optimized with SIMP can remain of elements with intermediate density~\citep{Bib:WuClausenSigmund2017}.

However, the most important result is the computational efficiency of the proposed mapping method. Coated designs of high resolution (3 million elements!) can be obtained in less than half an hour, thanks to the proposed coarse scale homogenization-based optimization. While, the method by~\citet{Bib:WuClausenSigmund2017} already requires close to 5 hours on a relatively coarse mesh of $(600\times200)$ elements. A reduction in computational cost of at least an order of magnitude can thus be obtained by the proposed approach, which allows for topology optimization as a more integrated part of the structural design process.

\subsection{Room for improvement}
\begin{figure}[h!]
\centering
\subfloat[Problem form 1, $\mathcal{J}^{f} = 31.0874$, $\mathcal{J}^{\phi} = 31.7504$ and $V_{\phi} = 0.196$.]{\includegraphics[width=0.4\textwidth]{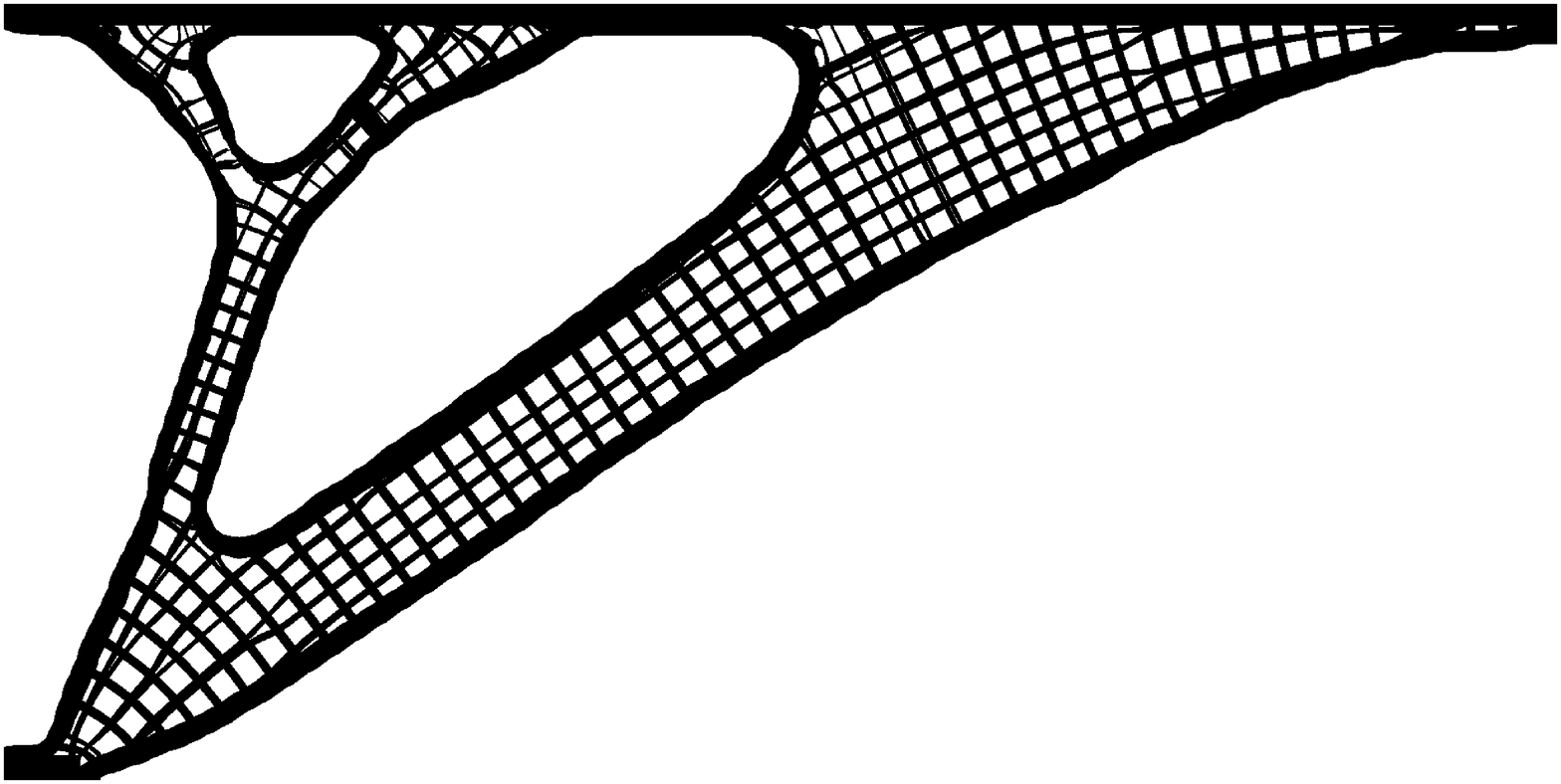}} \quad
\subfloat[Problem form 3, $\mathcal{J}^{f} = 26.4330$, $\mathcal{J}^{\phi} = 26.8968$ and $V_{\phi} = 0.198$.]{\includegraphics[width=0.4\textwidth]{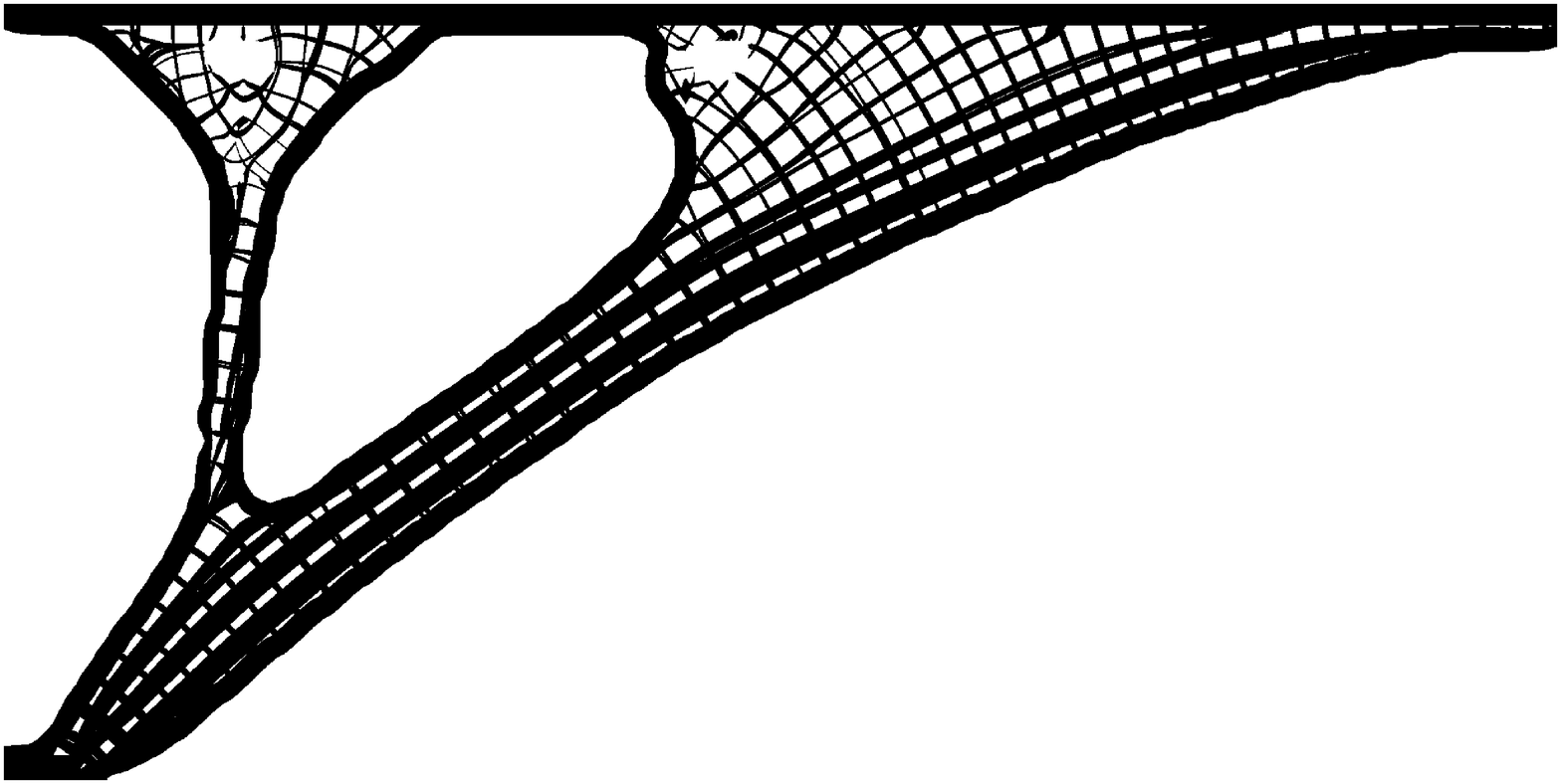}}
\caption{Projected structures of the bridge example on a fine mesh of $2000\times1000$, elements using the adaptive periodicity approach, $\varepsilon = 20h^{f}$ and $\gamma = 10^{3}$.} 
\label{Fig:NumExp.proj.5}
\end{figure}

As is discussed in the previous section, the adaptive periodicity projection approach will restrict the local unit-cell spacing to the interval $[\varepsilon2^{-1/2}, \varepsilon2^{1/2}]$, except in the transition zone. When the angle changes slowly the microstructure is spaced in a very regular manner, as can be seen in the main load carrying member of Figure~\ref{Fig:NumExp.proj.5}(a). 

In the top left of the bridge-example the angle field is rapidly changing. Hence, there is no room for a smooth transition through all the periodicities and the spacing is slightly less regular as in the rest of the domain. A more worrying effect can be seen in the projected figure optimized for problem form 3, shown in~\ref{Fig:NumExp.proj.5}(b). Here there is a singularity in the angle field below the void in the top left corner. Although the angle constraint ensures that most bars are well aligned with $\theta$, and the performance is within $2\%$ of the homogenization-based performance, it can be seen that there is a jump in orientation of mapping functions $\phi_{i}$. We note that the observed singular points have low or negligible stresses, making the appearance less critical in terms of objective value. However, to avoid large changes in orientation and prevent the occurrence of singularities the angle field can possibly be regularized, e.g. using the approach presented by~\citet{Bib:GDondersAllairePantz2018}. We are confident that such a regularization will get $\mathcal{J}^{\phi}$ even closer to $\mathcal{J}^{f}$ and result in even more regular mapped designs. 

%% file: Conclusion.tex
\section{Concluding remarks}
\label{Sec:Conclusion} 
An efficient approach to perform topology optimization of coated structures with orthotropic infill has been presented. Performing homogenization-based topology optimization allows for the modeling of designs with complex microstructures on a relatively coarse mesh, thus resulting in low computational cost. Furthermore, the double filter approach ensures in almost all cases a clear distinction between coating, infill and void. 

In the second part of the work, a projection approach is presented to map the coated designs from the assumption of infinite periodicity on a fine but realizable scale. A novel method to adaptively refine the periodicity is presented to allow for a regular spacing of the infill. Numerical experiments demonstrate that the projected designs, despite a lack of separation of scales, are very close (within 1-2$\%$) to the homogenization-based performance. Furthermore, a comparison with~\citep{Bib:WuClausenSigmund2017} where the infill is optimized using a density based method, shows that the projection procedure yields similar or even better performing designs at a finer resolution and at a computational cost which is at least 10 times lower, and potentially more in a case of mutual refinement.

This overall promising approach allows for extension of the method to 3D or to more complex loading situations. The main challenge here will lie in finding a parameterization that allows for smoothly varying microstructures through the domain. We are confident that such a parameterization can and will be found.

\section*{Acknowledgments}
The authors acknowledge the financial support from the Villum Foundation (InnoTop VILLUM investigator project) and DTU Mechanical Engineering. Furthermore, the authors would like to express their gratitude to Anders Clausen for sharing his code on the optimization of coated structures. Finally, the authors wish to thank Krister Svanberg for the Matlab MMA code.